\documentclass[amsmath,amssymb,aps,prx,reprint,floatfix]{revtex4-2}

\usepackage{graphicx}
\usepackage{dcolumn}
\usepackage{bm}
\usepackage{tikz,lipsum,lmodern}
\usepackage[most]{tcolorbox}

\usepackage{xcolor}
\definecolor{myBlue}{HTML}{5AA4CD}
\definecolor{myYellow}{HTML}{E9B65C}
\definecolor{myGreen}{HTML}{5AC0A4}

\usepackage{glossaries}
\usepackage[per-mode=symbol]{siunitx}
\usepackage[version=3]{mhchem}
\usepackage{enumitem}

\usepackage[    unicode,    colorlinks=true,    urlcolor=black,    linkcolor=black,    citecolor=black]{hyperref}

\newacronym{aic}{AIC}{Akaike information criterion}
\newacronym{ardr}{ARDR}{automatic relevance detection regression}
\newacronym{bic}{BIC}{Bayesian information criterion}
\newacronym{ce}{CE}{cluster expansion}
\newacronym{cv}{CV}{cross-validation}
\newacronym{dft}{DFT}{density-functional theory}
\newacronym{eci}{ECI}{effective cluster interaction}
\newacronym{fcc}{FCC}{face-centered cubic}
\newacronym{lspr}{LSPR}{localized surface plasmon resonance}
\newacronym{lasso}{LASSO}{least absolute shrinkage and selection operator}
\newacronym{mc}{MC}{Monte Carlo}
\newacronym{nn}{NN}{nearest neighbor}
\newacronym{np}{NP}{nanoparticle}
\newacronym{ols}{OLS}{ordinary least squares}
\newacronym{pbc}{PBC}{periodic boundary conditions}
\newacronym{rmse}{RMSE}{root mean square error}
\newacronym{rfe}{RFE}{recursive feature elimination}
\newacronym{sa}{SA}{simulated annealing}
\newacronym{sgc}{SGC}{semi-grand canonical}
\newacronym{vcsgc}{VCSGC}{variance-constrained semi-grand canonical}

\newcommand{\icet}{\textsc{icet}}

\newcommand{\sklearn}{\textsc{scikit-learn}}

\DeclareSIUnit\angstrom{\text {Å}}
\DeclareSIUnit\site{\text{site}}
\DeclareSIUnit\atom{\text{atom}}
\renewcommand{\vec}[1]{\ensuremath\boldsymbol{#1}}

\newcommand{\phys}{
    Department of Physics,
    Chalmers University of Technology,
    SE 412~96 Gothenburg, Sweden
}

\begin{document}

\title{Construction and sampling of alloy cluster expansions --- A tutorial}

\author{Pernilla Ekborg-Tanner}
\author{Petter Rosander}
\author{Erik Fransson}
\author{Paul Erhart}
\email{erhart@chalmers.se}
\affiliation{\phys}

\date{\today}

\begin{abstract}
Crystalline alloys and related mixed systems make up a large family of materials with high tunability which have been proposed as the solution to a large number of energy related materials design problems. 
Due to the presence of chemical order and disorder in these systems, neither experimental efforts nor ab-initio computational methods alone are sufficient to span the inherently large configuration space.
Therefore, fast and accurate models are necessary.
To this end, \glsentrylongpl{ce} have been widely and successfully used for the past decades. 
\Glsentrylongpl{ce} are generalized Ising models designed to predict the energy of any atomic configuration of a system after training on a small subset of the available configurations. 
Constructing and sampling a \glsentrylong{ce} consists of multiple steps that have to be performed with care. 
In this tutorial, we provide a comprehensive guide to this process, highlighting important considerations and potential pitfalls. 
The tutorial consists of three parts, starting with \glsentrylong{ce} construction for a relatively simple system, continuing with strategies for more challenging systems such as surfaces and closing with examples of \glsentrylong{mc} sampling of \glsentrylongpl{ce} to study order-disorder transitions and phase diagrams.
\end{abstract}

\maketitle

\section{Introduction}
Materials engineering proposes a solution to many of the energy related problems our society is faced with.  
Tremendous research efforts are currently put towards optimizing green technologies such as solar cells, batteries, catalysts and fuel cells.
The performance of these technologies is dictated by the properties of the materials involved, meaning that optimization is achieved by improving the materials design, which in turn requires tunability of the materials.
A common strategy to expand the tunability is to consider multicomponent systems, since the larger chemical space enables tailoring via chemical composition and ordering. 
Recent examples can be found in the fields of photovoltaics \cite{MaLutZhe09, XiaLiYao13, LuFer13, FeuReiAva16, LiCaiVal22}, batteries \cite{BalLuoQiu16, LaoZhaLuo17, SonLiuMi19}, catalysis \cite{SinXu13, HanGiaFly20, NakFur22}, fuel cells \cite{Ant03, RenLvLie20}, nanophotonics and nanoplasmonics \cite{GonLyuPal20, DarKhaTom21}, construction and manufacturing \cite{DurSou14, GarColBlo19, TorCam20, ZhaZhaWu22} as well as two-dimensional materials such as MXenes \cite{AnaLukGog17, TanGuoWu18, GogAna19, YinLiYao21}. 

The increased tunability comes at the cost of added complexity in the materials design process and computational methods are often needed to efficiently span the composition space. 
While computational efforts are ideally based on ab-initio methods such as \gls{dft} calculations, such methods are typically computationally too expensive for sampling the relevant composition space.
To exemplify, a binary system consisting of $N$ atoms corresponds to approximately $2^N$ indistinguishable atomic configurations, meaning that already at system sizes of \num{100} atoms one would need to examine roughly $10^{30}$ atomic configurations. 
For this reason, more effective models are necessary.
For crystalline materials, \glspl{ce} are ideal candidate models as indicated by the large number of successful applications including phase diagram prediction for metals \cite{AstMcCFon93, OzoWolZun98a,Meng2009,  RahLöfFraErh21, GreFraAngErhWah2021} and semiconductors \cite{VanAydCed98, CedVenMar00, ZhoMaxCed06, BecVan18, LinRahErh22}, ordering phenomena \cite{KimKavTho10, BurWalSto11, BurWal12, AngLinErh16, AngErh17, TroRigDra17, GunPucVan18} as well as the properties of surfaces \cite{DraReiFah01, SluKaw03, WelWieKer10, SteHamHwa10, CheSchSch11, CaoMue15, HerBraSch15, FraGreWah2021, FraGreLarWah2021, EkbErh21, XieJia23} and nanoparticles \cite{MueCed10, CheButWal10, Yug11, Mue12, TanWanJoh12, WanTanJoh14, LiRacPu18, CaoLiMue18}. 

\begin{figure}[b!]
\centering
\includegraphics[scale=0.94]{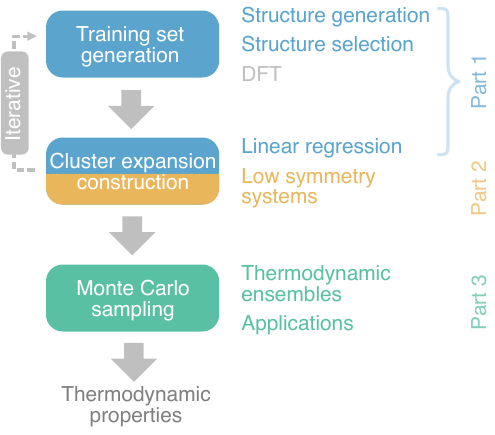}
\caption{
    \textbf{Constructing and sampling a \gls{ce}.}
    An overview of the typical procedure of constructing and sampling a \gls{ce} and a connection to the three main parts of this tutorial. 
}
\label{fig:overview}
\end{figure}

\Glspl{ce} are generalized Ising models that can in principle predict the energy (or any other function of the configuration) of any atomic configuration after training on only a small subset of the available atomic configurations (\autoref{fig:overview}).
They are typically sampled in \gls{mc} simulations to extract thermodynamic information about the system and study thermodynamic observables such as chemical ordering, free energy and heat capacity. 
While we focus on energy prediction in this tutorial we note that \glspl{ce} have also been used to model, for example, activation barriers \cite{VanCedAst01}, vibrational properties \cite{MorWalCed00, WalCed02b}, chemical expansion \cite{AngErh17} and transport properties \cite{AngLinErh16, BroZhaPal21}, which can serve as suitable starting points for interested readers.
There are various software packages that implement the \gls{ce} framework and thermodynamic sampling via \gls{mc} sampling, including, e.g., \textsc{atat} \cite{ATAT}, \textsc{uncle} \cite{LerWieHar09}, \textsc{clease} \cite{CLEASE}, \textsc{casm} \cite{CASM}, \icet{} \cite{Angqvist2019} and \textsc{smol} \cite{SMOL}.

In this tutorial, we present an overview of practical aspects to consider when constructing and sampling \glspl{ce}.
It is accompanied by a set of jupyter notebooks available online (\url{https://ce-tutorials.materialsmodeling.org} \cite{url-link, zenodo-link}) that use the \icet{} package \cite{Angqvist2019}.
The tutorial is organized as follows. 
First, the \gls{ce} formalism is briefly outlined in \autoref{sect:method} followed by practical considerations for constructing a \gls{ce} in \autoref{sect:constructing-cluster-expansions}. 
Then, we review the \gls{ce} construction process for a relatively simple system (\ce{Mo_{1-x}V_{x}C_{1 - y}\square_y}) in \autoref{sect:basic-example}, with emphasis on training set generation and training procedure. 
Next, we discuss \gls{ce} construction for a low symmetry system, namely a \ce{Au_{x}Pd_{1-x}} surface, in \autoref{sect:low-symmetry}, which allows us to discuss the use of local symmetries, Bayesian priors and constraints for improving \gls{ce} performance.
Lastly, in \autoref{sect:monte-carlo-sampling} we introduce sampling of \glspl{ce} via \gls{mc} simulations to obtain various thermodynamic observables for \ce{Au_3Pd} as well as the \ce{Ag_{x}Pd_{1-x}} alloy system.

\section{Cluster expansion formalism}\label{sect:method}

The theory behind \glspl{ce} has been discussed at length elsewhere \cite{SanDucGra84, CedVenMar00, Sanchez2010, Fon94, Zunger2002, Wal09, XieZhoJia22}.
In the present context, we present a short overview and focus on practical considerations. 

A \gls{ce} predicts the value of some observable $E$ as a function of the atomic configuration, represented by the occupation vector $\vec{\sigma}$, for a crystalline material, based on all involved atomic clusters $\vec{\alpha}$.
A cluster is a set of $k$ atomic sites on the crystalline lattice where $k=0, 1, 2, \ldots$ is the order of the cluster.
The observable can be expressed as  
\begin{align}\label{eq:CE-formalism-1}
    E(\vec{\sigma}) = \sum _{\vec{\alpha}} J_{\vec{\alpha}} \Pi_{\vec{\alpha}}(\vec{\sigma}),
\end{align}
where $J_{\vec{\alpha}}$ is the contribution of cluster $\vec{\alpha}$, called the \gls{eci}, and $\Pi_{\vec{\alpha}}$ are orthogonal basis functions spanning the space of atomic configurations \cite{SanDucGra84, Wal09}.  
All symmetrically equivalent clusters have the same \gls{eci} and can be grouped into what we will refer to as an orbit (originating from group theory and used in \cite{Angqvist2019}), which is also referred to as a representative cluster in the \gls{ce} literature. 
Exploiting this fact simplifies Eq.~\eqref{eq:CE-formalism-1} to
\begin{align}\label{eq:CE-formalism-2}
    E(\vec{\sigma}) = \sum _{\vec{\beta}} m_{\vec{\beta}} J_{\vec{\beta}} \left<\Pi_{\vec{\alpha}}(\vec{\sigma})\right>_{\vec{\beta}},
\end{align}
where $m_{\vec{\beta}}$ is the multiplicity of orbit $\vec{\beta}$ and the $\left<\ldots\right>_{\vec{\beta}}$ bracket indicates that the basis function $\Pi_{\vec{\alpha}}$ is averaged over all clusters $\vec{\alpha}$ in the orbit $\vec{\beta}$.
The sum in Eq.~\eqref{eq:CE-formalism-2} runs over all orbits up to some cutoff criterion, typically defined by a cutoff radius $r$ for each cluster order $k$.

Training a \gls{ce} involves finding the optimal \glspl{eci} based on a set of atomic structures for which the value of the observable $E$ is known, called the training set. 
In most cases, the observable of interest $E$ is the energy or a variant thereof, say the mixing energy or a migration barrier, with reference data obtained from from \gls{dft} calculations. 
Equation \eqref{eq:CE-formalism-2} can be cast as a linear problem,
\begin{align}\label{eq:CE-formalism-mat}
    \vec{y} = \vec{X}\vec{w},
\end{align} 
where $\vec{y}$ is a vector comprising the target values of the observable $E$ for the training set, $\vec{w}$ is an unknown vector containing the \glspl{eci}, and $\vec{X}$ is the so-called design matrix, where each row correspond to the values of the basis functions $\Pi_{\vec{\alpha}}$ for a structure in the training set.
Note that in this form the multiplicities are absorbed into either $\vec{w}$ or $\vec{X}$.
The number of \glspl{eci}, i.e., the size of $\vec{w}$ and accordingly the number of columns in $\vec{X}$, are determined by the cutoff radii used to select the allowed orbits of different order.

The optimal \glspl{eci} $\vec{w}_\text{opt}$ can be found using linear regression techniques (see \autoref{sect:constructing-cluster-expansions}).
Once the optimal \glspl{eci} are found, the \gls{ce} can be used to predict $E$ for any atomic configuration representable by a supercell of any size and concentration.

The \gls{ce} formalism applies to ideal lattices.
In most cases, however, relaxed structures are more thermodynamically relevant than structures with atoms residing precisely on the lattice sites.
\Glspl{ce} are therefore often used to model the energy of relaxed structures mapped to the closest corresponding occupation on the ideal lattice. 
This means that the \glspl{eci} effectively include variations in the interactions due to volume changes and atomic relaxations.

The validity of a \gls{ce} employed over a concentration range has been debated in the literature \cite{Sanchez2010, Mueller2010, Sanchez2017}.
In practice, for rare cases, it can become difficult to model the full concentration range with a single \gls{ce}, especially if the system undergoes sharp transitions at some specific concentrations.
This can (but need not) occur, e.g., in materials with electronic band gaps if the charge state of a species changes with composition (e.g., $\ce{Mn^{2+}}\rightarrow\ce{Mn^{3+}}\rightarrow\ce{Mn^{4+}}$ \cite{BarZhoYan22}) or if the species that are being mixed are aliovalent (e.g., \ce{Si^{4+}} and \ce{Al^{3+}} \cite{FanAngHel21}).
In these situations one needs to consider the overall charge balance as the total energy becomes in principle dependent on the Fermi level (i.e., the electron chemical potential).
This issue has been circumvented, e.g., in the case of zeolites by constructing \glspl{ce} exclusively on charge balanced configurations and then using only \gls{mc} trial moves that maintain charge balance \cite{FanAngHel21}, while a more general approach for ionic materials has been described in Ref.~\citenum{BarZhoYan22}.
There are also cases were \glspl{ce} may not be sufficient to capture all relevant interactions in the system, such as long-ranged strain interactions which requires extending the \gls{ce} formalism to reciprocal space \cite{Laks1992, OzoWolZun98a, Rahm2022}.

\section{Constructing cluster expansions}
\label{sect:constructing-cluster-expansions}

A \gls{ce} model should be both accurate and transferable.
In practice this means that we aim to find an optimal set of \glspl{eci}, $\vec{w}$ in Eq.~\eqref{eq:CE-formalism-2}.
This entails choosing a regression method along with related hyper-parameters, cutoff parameters (i.e., the size of the model) as well as composing a set of training structures.
In the following we first briefly review regression methods (\autoref{sect:regression-methods}), followed by a short discussion of how to assess model performance (\autoref{sect:assessing-model-performance}) as well as the role of cutoffs and hyper-parameters (\autoref{sect:parameter-selection-in-practice}).
The impact of the composition of the training structure set will be discussed through practical examples in the next section (\autoref{sect:structure-generation}).

\subsection{Regression methods}
\label{sect:regression-methods}

There are many approaches to solving the linear regression problem in Eq.~\eqref{eq:CE-formalism-mat} \cite{NelOzoRee13, Zhou2014, Angqvist2019, FraEriErh20, Cheng2020}.
Here we provide a short introduction to several common methods.

The solution of the linear problem, Eq.~\eqref{eq:CE-formalism-mat}, with some typical regularization terms can be written as
\begin{equation}\label{eq:linear_regression}
    \boldsymbol{w}_\text{opt} = \underset{\boldsymbol{w}}{\text{argmin}} \left\{ \Vert \boldsymbol{X}\boldsymbol{w} - \boldsymbol{y} \Vert^2_2 +  \boldsymbol{w}^T\boldsymbol{\Lambda} \boldsymbol{w} + \Vert \boldsymbol{M}\boldsymbol{w} \Vert_1 \right \},
\end{equation}
where $\boldsymbol{w}_\text{opt}$ is the solution vector, $\boldsymbol{\Lambda}$ is the $\ell_2$-regularization matrix and $\boldsymbol{M}$ is the $\ell_1$ regularization matrix.
Setting $\boldsymbol{\Lambda}=\boldsymbol{M}=\boldsymbol{0}$ yields the \acrfull{ols} solution.
\Gls{ols} is, however, prone to overfitting, meaning that the model captures noise in the training data and therefore performs poorly on new unseen data.
For $\boldsymbol{M}=0$ and $\boldsymbol{\Lambda}=\alpha \boldsymbol{1}$, Eq.~\eqref{eq:linear_regression} reduces to ridge regression with regularization parameter $\alpha$, and with $\boldsymbol{\Lambda}=0$ and $\boldsymbol{M}=\alpha \boldsymbol{1}$ one obtains the expression used in the \gls{lasso}.

The regularized regression methods generally assume that the design matrix, $\boldsymbol{X}$, is standardized.
Therefore, it is common practice to rescale the columns of $\boldsymbol{X}$ to have unit variance before solving the problem, and afterwards apply the inverse procedure to obtain the unscaled parameters.
This rescaling needs to be taken into consideration if manually choosing values for the regularization matrices $\boldsymbol{\Lambda}$ and $\boldsymbol{M}$ (for instance for Bayesian \glspl{ce}, see \autoref{sect:bayesian-ce}).
Additionally, one commonly trains \glspl{ce} for the mixing (or formation) energy, rather than the total energy in order to avoid very large target values, $\boldsymbol{y}$.

Feature selection techniques can be used to reduce the number of nonzero \glspl{eci} in the solution vector. 
This can lead to more transferable models and faster predictions (e.g., in \gls{mc} simulations).
\Gls{rfe} is an iterative algorithm where in each iteration, one solves the linear problem (typically with \gls{ols}) and the least important (smallest) \glspl{eci} are pruned (set to zero).
This is iterated until a desired number of nonzero \glspl{eci} is reached.
\Gls{ardr} is based on the Bayesian ridge regression technique where the regularization matrix is diagonal with elements  $\Lambda_{ii}=\lambda_i$.
The individual regularization strengths $\lambda_i$ for each parameter are updated throughout the optimization, and parameters are pruned (set to zero) if $\lambda_i$ increases above a threshold ($\lambda$-threshold) \cite{MacKay1992}.
These linear regression methods can readily be employed for \glspl{ce} using \icet{} via \sklearn{} \cite{scikit-learn} or even more directly via the \textsc{trainstation} interface to the former \cite{FraEriErh20}.

\subsection{Model performance and learning curves}
\label{sect:assessing-model-performance}

In general, one wants to construct models that require as little training data and as few non-zero \glspl{eci} as possible.
Less training data means fewer (usually computationally demanding) reference calculations, while fewer non-zero \glspl{eci} translate to reduced model complexity, a feature that is often associated with better transferability, i.e., such models perform more reliably on unseen data.
These aspects need to be taken into account when constructing \glspl{ce}.

Techniques such as ridge regression, \gls{rfe}, \gls{ardr} or \gls{lasso} involve hyper-parameters, e.g., the regularization parameter $\alpha$ in ridge regression, the number of features in \gls{rfe} or the $\lambda$-threshold in the case of \gls{ardr}.
In addition one must make select the cutoffs that determine the range of the summation in Eq.~\eqref{eq:CE-formalism-2}.
These parameters directly affect model performance in terms of accuracy, transferability as well as data efficiency, i.e., the amount of reference data needed to obtain a well converged model.

In practice one usually determines optimal parameters through so-called \emph{learning curves}, which show a suitable performance indicator (see below) as a function of, e.g., hyper-parameters, cutoff values or training set size.
We will exemplify this approach throughout this tutorial (see, e.g., Figs.~\ref{fig:fit_methods}, \ref{fig:merge_learning_curves}, and \ref{fig:bayesian_rmse}).

The most widely used measure for model performance is the \gls{rmse} score calculated over a \emph{validation} set, since the \gls{rmse} over the \emph{training} set (i.e., the first term on the right-hand side of Eq.~\eqref{eq:linear_regression}) is a poor estimate of how a model will perform on unseen data points.
The validation \gls{rmse} is commonly calculated via \gls{cv}.

As we already noted above simpler models tend to exhibit better transferability.
This principle is approximately represented through information criteria such as \gls{aic} and \gls{bic} \cite{Aka74, Sch78, AhoDerPet14, Mur13}, which weigh validation \gls{rmse} vs model size.
These measures can hence be useful when selecting between models which have similar \gls{rmse} values.
We note, however, that in our experience these information criteria are often not sufficiently conclusive as a general measure when constructing \gls{ce} models.

Lastly, we emphasize that none of these performance measures are perfect when it comes to assessing the general performance of a \gls{ce} and that they will always reflect the choice of training data to some degree.
We therefore recommend to always study the property or properties of interest (e.g., phase diagram, heat capacity, or order parameters) throughout the \gls{ce} construction process to make sure that convergence is achieved.

\subsection{Cutoff selection vs regularization}
\label{sect:parameter-selection-in-practice}

When it comes to selecting cutoffs, the conventional approach, especially when using \gls{ols} (and thus no regularization, i.e., $\boldsymbol{\Lambda}=\boldsymbol{M}=\boldsymbol{0}$ in Eq.~\eqref{eq:linear_regression}), is to start with a set of small cutoffs and low orders and iteratively increase the size (and complexity) of the model.
Too small cutoffs (and thus a small number of \glspl{eci}) lead to underfitting, whereas too long cutoffs (and thus a large number of \glspl{eci}) lead to overfitting.
A good starting point for the length of the cutoff is on the order of the lattice parameter, and cutoffs longer than three lattice parameters are very rarely needed.
The interaction strength decreases with the order of the cluster, meaning that including terms up to third or fourth order is sufficient in most cases.

When using regularization and feature selection approaches, one can in principle choose a large initial number of orbits using both larger cutoffs and higher-order orbits, which is then reduced by the regression method of choice.
We note that \gls{ardr} is usually performing very well in such situations, see, e.g., Ref. \citenum{FraEriErh20, LinRahErh22} and examples below.
This means that with regularization, the importance of cutoff selection decreases, as long as the cutoffs are large enough. 
In practice, however, one can run into problems if cutoffs are selected too large.
In our experience, the best performance is therefore often found when combining \gls{ardr} with cutoff selection.

Lastly, we note that one can also use a Bayesian method, where physical intuition is encoded in a regularization matrix $\boldsymbol{\Lambda}$ (see, e.g., Ref.~\citenum{MueCed09} and \autoref{sect:bayesian-ce}), to for instance enforce higher importance of low-order and/or short-ranged orbits.
This approach could be useful for certain complex systems, but requires significantly more work to set up than, for instance, \gls{ardr}.

%
\section{Part 1: A first example}\label{sect:basic-example}

\begin{tcolorbox}[colback=myBlue!15!white,colframe=myBlue, title=\textbf{Key takeaways}]
\begin{itemize}[leftmargin=*]
\item
    Simple linear regression techniques such as \gls{ols} often lead to overfitting and large validation errors.
    Use regularization and/or feature selection to improve model performance. 
\item
    Similarly, naive structure selection schemes such as randomization can lead to poorly performing models, while more advanced schemes generally yield better models and require fewer training structures.
\item
    We recommend generating structures via either condition number minimization or uncertainty maximization in conjunction with \gls{ardr}.
\end{itemize}
\end{tcolorbox}

In the first part of this tutorial, we illustrate the construction of a \gls{ce} with an emphasis on the choice of regression techniques (\autoref{sect:comparison-of-regression-methods}) and approaches for structure selection (\autoref{sect:structure-generation}).
To this end, we consider a simple carbide \ce{Mo_{1-x}V_{x}C_{1 - y}\square_y} with two sublattices on a rocksalt lattice.
On the metal sublattice we consider mixing between Mo and  V, and on the carbon sublattice we consider C atoms and vacancies ($\square$), with a maximum of 30\% carbon vacancies.
We employ cutoffs of \qty{9}{\angstrom} and \qty{5}{\angstrom} for two and three-body clusters, respectively.
This yields a total number of 52 \glspl{eci}, including two singlets, 25 pairs, 24 triplets as well as a constant, sometimes referred to as the zerolet.
A detailed analysis of the cutoff selection can be found in the notebooks accompanying this tutorial \cite{url-link, zenodo-link}.
We use reference data from \gls{dft} calculations available online \cite{zenodo-link}.
The computational details concerning these calculations can be found in Appendix \ref{sect:dft-details}.

\subsection{Comparison of regression methods}
\label{sect:comparison-of-regression-methods}

\begin{figure}
\centering
\includegraphics{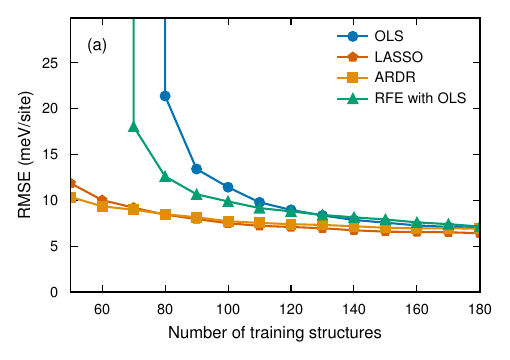}
\includegraphics{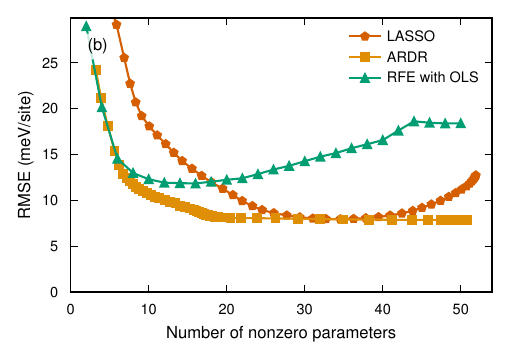}
\caption{
    \textbf{\Gls{rmse} using different regression methods in the \ce{Mo_{1-x}V_{x}C_{1 - y}\square_y} system.}
    (a) The validation \gls{rmse} based on the random training set using different regression methods as a function of the number of structures included in the training set.
    (b) The validation \gls{rmse} when using 80 training structures as a function of the number of nonzero parameters obtained after regression and feature selection.
    Here, the number of nonzero parameters obtained is controlled by varying the hyperparameter controlling the sparsity of the solution continuously.
    For \gls{ardr} this is the $\lambda$-threshold parameter, for \gls{lasso} it is $\alpha$ (see Eq.~\eqref{eq:linear_regression}) and for \gls{rfe} it is the number of features.
}
\label{fig:fit_methods}
\end{figure}

Let us now explore some of the above mentioned regression methods for solving Eq.~\eqref{eq:CE-formalism-mat} in order to illuminate some of their differences.
For the sake of simplicity we consider a set of 200 ``random'' training structures, where each structure has a randomized supercell size (up to maximum 50 atoms), random Mo/V and C/vacancy concentrations, and random occupation of the lattice.
This set of structures will be referred to as the random set of structures from here on out.
The resulting validation error as a function of number of training structures is shown in \autoref{fig:fit_methods}a).
First, we note that all four regression methods yield similar validation errors when using a large training set.
For \gls{ols} the validation error increases rapidly when decreasing the number of training structures, which is also to lesser extent the case for \gls{rfe}, due to overfitting.
Both \gls{ardr} and \gls{lasso} perform significantly better than \gls{ols} and \gls{rfe} when using a small number of training structures.

\Gls{ardr}, \gls{rfe}, and \gls{lasso} all have one hyperparameter that controls the sparsity (number of nonzero parameters in the solution), which needs to be chosen.
This is typically done by scanning over the hyperparameter value and selecting the value yielding the smallest validation error.
Here, we use this procedure each time a model is trained.

\autoref{fig:fit_methods}b shows an example for the variation of the validation error with the resulting number of features during a hyper parameter scan.
\Gls{lasso} has a minimum at about 35 nonzero-parameters whereas \gls{ardr} achieves the same validation error with only 20 parameters.
The tendency of \gls{lasso} to over-select is known \cite{Zou06} and we have observed this behavior previously for both \glspl{ce} \cite{Angqvist2019} and in other applications such as force constant expansions \cite{FraEriErh20}.
\Gls{rfe} has a minimum at about 15 nonzero-parameters but with slightly higher validation error compared to the other two regression methods.

The trends described above are, in our experience, general, and we have found \gls{ardr} to be the most performant approach in most cases.
Therefore, we recommend using \gls{ardr} as a starting point for \gls{ce} construction.

\subsection{Training set generation methods}
\label{sect:structure-generation}

\begin{figure}
\centering
\includegraphics{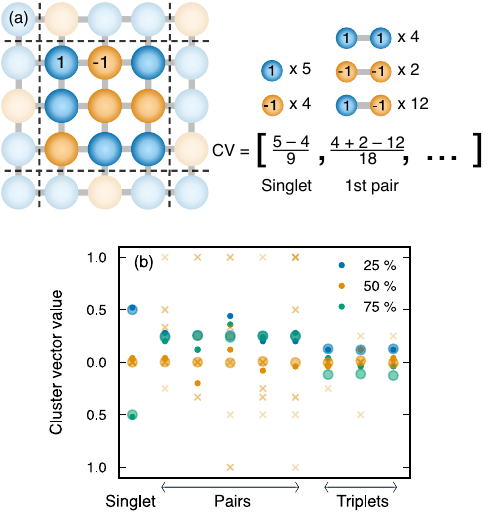}
\caption{
    \textbf{Cluster vectors for random structures.}
    (a) An example of how the first elements of a cluster vector are calculated for a 2D structure.
    (b) Cluster vectors for randomized atomic configurations of a 2D square binary system. 
    The concentration is indicated by the color. 
    The large and small circles represents large (\num{1e4} atoms) and small (25 atoms) random structures, respectively.
    The crosses represents 10 randomly selected structures from a pool of enumerated structures with up to 8 atoms (and concentration 50\%) for comparison.
  }
\label{fig:rnd_cv_correlations}
\end{figure}

For simplicity in \autoref{sect:comparison-of-regression-methods} we used randomized structures.
Let us now discuss more informed approaches to generating or selecting training structures.

Recall that we are trying to solve the linear problem Eq.~\eqref{eq:CE-formalism-mat}, with the design matrix $\vec{X}$ operating on the solution vector $\vec{w}$, i.e., the \glspl{eci}.
Commonly we would like to find solutions in large cluster spaces (long cutoffs and more bodies), expecting though that only relatively few \glspl{eci} are significant due to the near-sightedness of physical interactions.
This means $\vec{w}$ should be sparse.
Under these circumstances it can be shown that optimal design matrices $\vec{X}$ should be nearly orthonormal, a feature that is characterized by the restricted isometry property \cite{CanTao05}.
Heuristically, this can be thought of as setting up the training set, with as much variability in the cluster vectors as possible.

At first it might be tempting to use randomized structures, since two random configurations will very rarely be identical and the cluster vector is a function of the configuration. 
If one considers how the cluster vector is composed (\autoref{fig:rnd_cv_correlations}a), it becomes, however, quickly evident that this is a poor choice.
Let us imagine a simple binary system. 
After the zerolet, the first cluster vector item is the singlet, which reflects the overall composition.
Then we have the pairs which reflect the proportion of pairs between alike (A--A, B--B) and unalike (A--B) pairs. 
For a large random structure, there is no tendency for ordering which means that the proportion of alike and unalike pairs will be determined by the overall composition and take on the same value for all orders (\autoref{fig:rnd_cv_correlations}b).  
Smaller random structures allow for some variation around the large structure limit, but these variations are small compared to what can be achieved with other generation methods.
A similar argument can be made for higher order clusters. 
This example demonstrates that while random structures are rarely identical, they have similar cluster vectors.
For contrast, we can compare the cluster vectors of large and small random structures as well as enumerated structures (\autoref{fig:rnd_cv_correlations}b), which shows that by going beyond random structures, we can achieve a much larger spread of cluster vectors.

\begin{figure}
\centering
\includegraphics[scale=0.94]{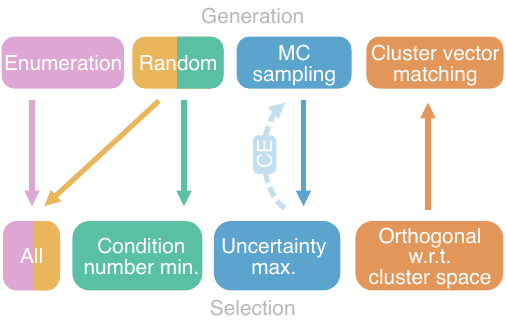}
\caption{
    \textbf{Structure generation and selection schemes.} 
    The process of generating and selecting a set of structures for the training set can be designed in multiple ways. 
    In this tutorial, we focus on the five approaches represented by the arrows in this schematic overview. 
}
\label{fig:schematic_structure_generation}
\end{figure}

There are many methods for generating a set of training structures, each having their own pros and cons.
Strictly speaking, one should discriminate between structure \textit{generation} and \textit{selection}:

In this tutorial, we consider four ways of \textit{generating} structures; enumeration of all possible structures up to some maximum structure size, randomized structures (in terms of structure size, composition and/or configuration), \gls{mc} sampling (see \autoref{sect:monte-carlo-sampling}) of existing \glspl{ce} and selecting one (or a set of) target cluster vector(s) and finding the closest matching structure(s).

We also consider four ways of \textit{selecting} structures. 
The most straightforward choice is to select all the generated structures, which leaves little control of the quality of the training set. 
More advanced selection methods can be applied with the aim of optimizing some aspect of the training set, such as minimizing the condition number of the design matrix, selecting structures with the largest uncertainty or trying to achieve a set of structures with orthogonal cluster vectors.

These generation and selection schemes can be combined and modified to create a sheer endless number of different training sets. 
Here, we consider five approaches which we refer to as uncertainty maximization, condition number minimization, structure orthogonalization, structure enumeration as well as the random set discussed in \autoref{sect:comparison-of-regression-methods}
(\autoref{fig:schematic_structure_generation}).
In the following, we will present and discuss these approaches.

\textbf{Uncertainty maximization} is a form of \textit{active learning} which is a common training set generation approach for model construction in general, including \gls{ce} construction \cite{WalCed02, Sek2009, Kleiven2021}.
Here, the model is iteratively trained and new structures are selected at each iteration based on the model uncertainty prediction for a pool of structures.
The uncertainty of structures can be estimated from, for example, an ensemble of \glspl{ce} from bootstrap sampling \cite{Angqvist2019, Kleiven2021}.
This means that a collection of \glspl{ce} trained in identical fashion but based on different training sets is generated.
The different training sets are generated by resampling the original training set with replacement (meaning one structure can appear multiple times in a training set).
Structures can, e.g., be generated using \gls{mc} simulations in the canonical ensemble (see \autoref{sect:monte-carlo-sampling}) and the ones with the largest uncertainties are selected to be included in the training set.
The benefit of this structure generation approach is that the training structures added during each iteration are structures for which the current ensemble of models predicts large uncertainty, and therefore will lead to large accuracy improvements in each iteration.
Another benefit of this approach is that structures are generated from \gls{mc} simulations with the desired conditions (concentrations, temperatures etc.) meaning the structures selected will be thermodynamically relevant ones.
On the other hand, this also limits the pool of structures to select from in terms of how they span the cluster vector space.
Furthermore, this approach requires a relatively large effort due to the iteration process and the fact that one needs to define a set of structures for which the uncertainty is predicted.

\textbf{Condition number minimization} refers to the process of selecting structures such that the condition number of the linear problem in Eq.~\eqref{eq:CE-formalism-mat} is minimized.
The condition number $c$ is defined as \cite{GolGenLoaCha13}
\begin{align}
    c = \frac{\text{max}(\Gamma)}{\text{min}(\Gamma)},
\end{align}
where $\Gamma$ are the singular values of the design matrix.
The condition number describes how well conditioned the linear problem is, with smaller values indicating better conditioning.
It can be thought of as a measure of how sensitive the fitting result is with respect to changes or errors in the input data.

This approach starts by generating a large pool containing on the order of millions of randomly generated structures.
An initial training set of $N$ structures is randomly drawn from the pool, where $N$ is on the order of hundreds.
Next, a simulated annealing \gls{mc} simulation (see \autoref{sect:monte-carlo-sampling}) is carried out where structures from the large random pool are randomly swapped in and out of the training set with probability $P=\mathrm{e}^{-\Delta c/\Theta}$ where $\Theta$ is an artificial temperature and $\Delta c$ is the change in condition number when swapping two structures.
The resulting training set will be an approximate solution to the problem of selecting a subset of $N$ structures from the large pool with the lowest condition number.

\begin{figure}
\centering
\includegraphics{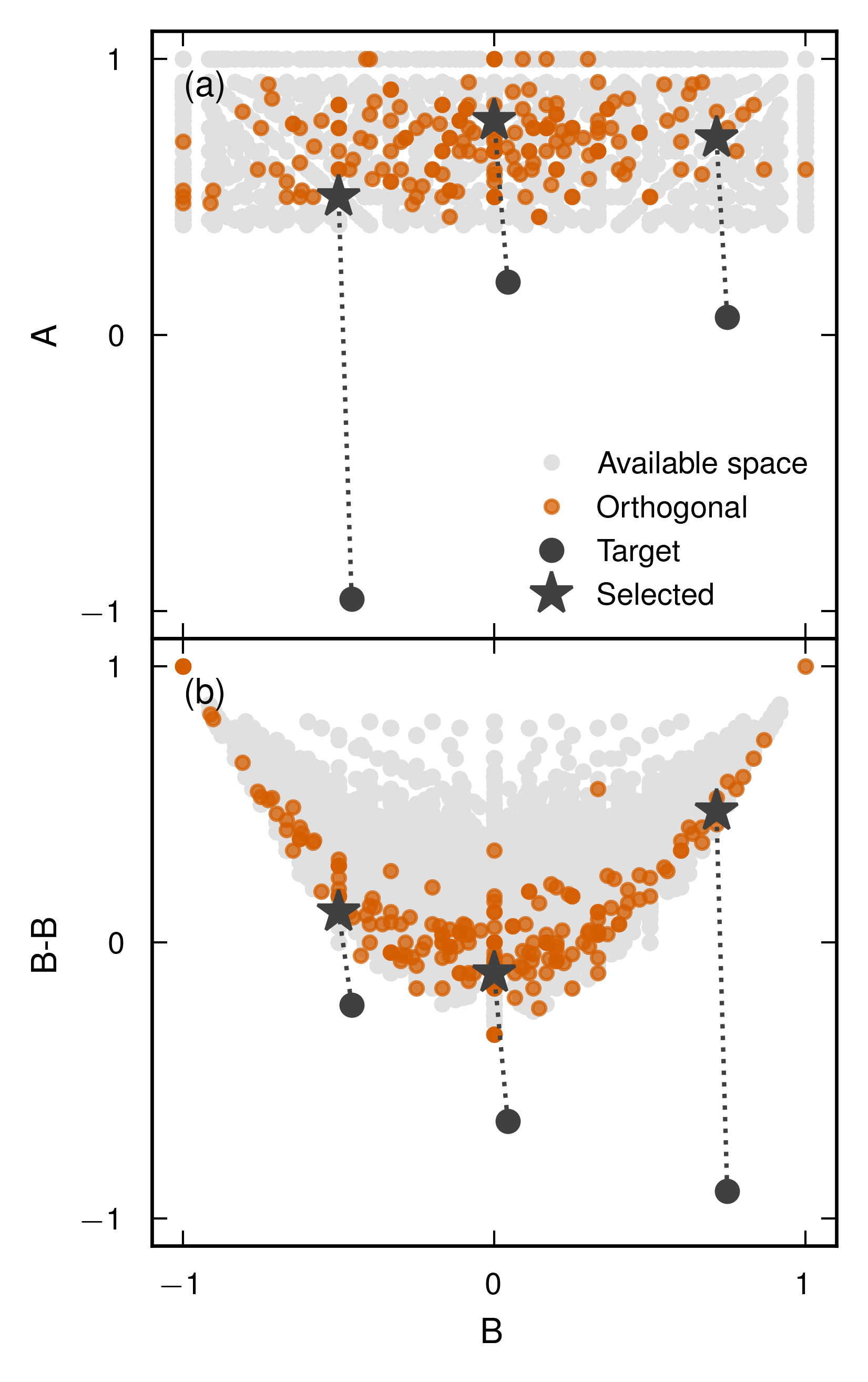}
\caption{
    \textbf{Correlations between orbits}.
    Correlations between (a) the singlets of the carbon (A) and metal (B) sublattices and (b) the metal singlet and nearest neighbor.
    The light grey area indicates the available space represented by all the structures from the large random pool used for condition number minimization. 
    The orthogonal data set is shown in orange. 
    The orthogonalization procedure is visualized by the initial (circle) and final (star) positions of three example structures.  
  }
\label{fig:cluster_correlations}
\end{figure}

\textbf{Structure orthogonalization} aims at providing a set of structures with cluster vectors orthogonal to each other \cite{NelHarZho13, NelOzoRee13}.
It is related to the condition number approach described above in the sense that both methods aim to span the cluster vector space.
The procedure starts from a structure with a random cluster vector. 
New structures are then added iteratively by finding a cluster vector that is orthogonal to the rest and identifying the closest matching structure. 

A benefit with this approach is that, in principle, one can quickly produce a training set without the need for any iterative models (as in the uncertainty maximization) or large pool of structures (as in the condition number minimization). 
The latter depends, however, on the ability to find a matching structure for a target cluster vector without, e.g., a pool of structures. 
In \icet{}, this is possible due to an implemented method based on simulated annealing. 

There are, however, two major drawbacks to this approach. 
First, the entire cluster vector space is not available due to correlations between the cluster vector elements (\autoref{fig:cluster_correlations}).
For instance, the number of possible A--B nearest neighbor pairs is limited by the value of the singlet. 
We show this in \autoref{fig:cluster_correlations} for the entire available space (represented by the large random pool used for condition number minimization) and the data set produced using this approach. 
Clearly, both the fact that the concentration range for the carbon sublattice is restricted and that the singlet and pairs are correlated significantly limits the available space. 
The orthogonalization procedure will more often than not ask for a structure outside of this space and the resulting structure (with the closest matching cluster vector) will not be orthogonal to the other structures.

Second, the number of orthogonal cluster vectors is limited by the number of elements in the cluster vector (which in turn is determined by the cutoffs).
This means that the training set size is limited by the cluster cutoffs when using this approach. 

\textbf{Structure enumeration} is a method that generates all symmetrically inequivalent structures that are permissible given a certain lattice, under the constraint that the number of atoms in each structure must be smaller than some given number \cite{HarFor08}. 
The benefit of structure enumeration is that it requires no other input than the maximum number of atoms in a supercell and will systematically generate high-symmetry and ground-state structures.
One drawback is that the maximum number of atoms needs to be quite small (typically less than 15 atoms) in order for the number of structures to be computationally feasible, and this can lead to the training set not spanning long-ranged interactions.
For our model system, \ce{Mo_{1-x}V_{x}C_{1 - y}\square_y} (with the chosen cutoffs), enumeration up to 12 atoms leads to about 650 structures, but produces an ill-conditioned design matrix and should thus not be used for training.
For this reason, the enumerated training set is here only used for testing purposes.
Additionally, for systems with larger and/or complex primitive cells, it may be unfeasible to use altogether as the number of enumerated structures grows exponentially with the size of the primitive cell.

\begin{figure}
\centering
\includegraphics{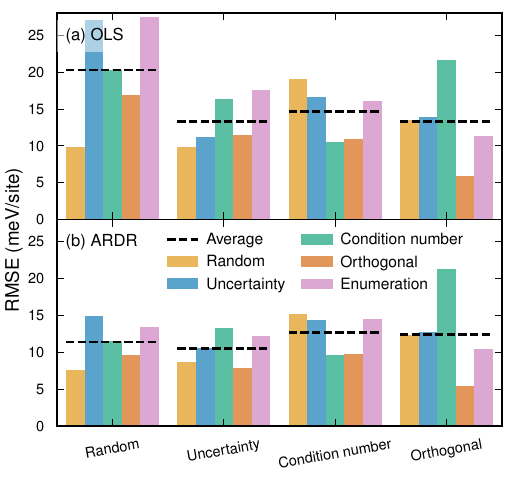}
\caption{
    \textbf{Comparison of structure generation schemes for the \ce{Mo_{1-x}V_{x}C_{1 - y}\square_y} system}.
    Here, \glspl{ce} were constructed using four different training sets (as indicated on the x-axis) consisting of \num{110} structures and trained using either (a) \gls{ols} or (b) \gls{ardr}.
    The \glspl{rmse} were evaluated using the different training sets as test sets (as indicated by the bar color) including a set of enumerated structures.
    The dashed line indicates the average over the five test sets.
  }
\label{fig:validation_barplot}
\end{figure}

Before we compare these approaches, let us note that the validation error is not a suitable measure to evaluate the quality of a training set generation scheme, because a procedure generating very similar (or identical) structures would lead to models with low validation errors but poor predictions on structures outside the training set.
Therefore, we evaluate each \gls{ce} on all the structures sets generated  (\autoref{fig:validation_barplot}).

In \autoref{fig:validation_barplot}a the validation error is shown over all sets of structures for \glspl{ce} trained with each training set using \gls{ols}.
Here, we see that the random training set leads to larger errors across all structure sets, whereas training using the three other structure sets yields lower validation errors across the board.
In \autoref{fig:validation_barplot}b the same analysis is carried out for training with \gls{ardr}.
Here, we see that the validation errors obtained with a random training set are on the same level as the other training sets, demonstrating the efficacy of \gls{ardr} even with a poor choice of training structures.
We also note that training with the structure orthogonalization structures leads to a large error over the condition number structures, yet very small validation error, indicating poor transferability.

\subsection{Conclusions}
In the first part of this tutorial, we have demonstrated the \gls{ce} construction process for a relatively simple system, focusing on linear regression and structure generation and selection. 
We have found that \gls{ols} with a training set consisting of randomized structures leads to large validation errors, and that
this can be prevented by using better structure selection schemes as well as regularization and feature selection. 
We note that the material considered here presents a rather simple system with high symmetry and for more complex cases, optimizing the training set and regression method typically yields a larger benefit.

Based on these findings, our general recommendation is to generate structures by either uncertainty minimization (if it seems worthwhile to put in the effort) or condition number minimization (if you want to avoid an iterative process) in conjunction with \gls{ardr} when solving the linear problem Eq.~\eqref{eq:CE-formalism-mat}.
Lastly we again emphasize that the training set generation and selection methods can be combined in multiple ways and the procedures outlined here do not need to be followed strictly.
It might, for instance, be beneficial to start with a fast method (enumeration or orthogonalization) and extend the training set with an iterative method. 
In practice, it is also often beneficial to include thermodynamically relevant structures, in particular ground-state structures, in the training set.

\section{Part 2: Low-symmetry systems}
\label{sect:low-symmetry}

\begin{tcolorbox}[colback=myYellow!15!white,colframe=myYellow, title=\textbf{Key takeaways}]
  \begin{itemize}[leftmargin=*]
      \item Low symmetry system generally have a large number of \acrshortpl{eci} which makes \acrshort{ce} construction challenging. 
      \item Local symmetries and Bayesian inference can be used to couple similar orbits. 
      \item Weighted constraints can be applied to ensure certain properties are represented more accurately by the \acrshort{ce}.  
  \end{itemize}
\end{tcolorbox}

\begin{figure}
\centering
\includegraphics[scale=0.88]{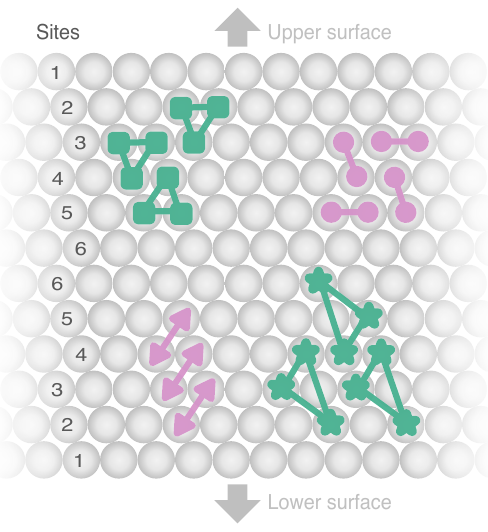}
\caption{
    \textbf{Merging of orbits.}
    Side view of a 12-layer \gls{fcc}-(111) surface slab. 
    The numbers on the left row of atoms indicate the symmetrically inequivalent sites. 
    Examples of orbits that can be merged if only site 1 is considered a surface site are marked with pink for pairs and green for triplets and different symbols for inequivalent orbits (note the difference in length between sets of pairs). 
}
\label{fig:merge_orbits}
\end{figure}

In the second part of this tutorial, we consider \gls{ce} construction for a low-symmetry system, specifically a surface, which comes with new challenges. 
The number of orbits for a given set of cutoffs grows with the number of symmetrically inequivalent sites that make up the material.
For simple bulk systems, such as the one in \autoref{sect:basic-example}, the unit cell typically comprises one or a few atoms which constitute the maximum number of inequivalent sites. 
For \textit{low-symmetry} systems such as surfaces and nanoparticles, on the other hand, the unit cell is generally larger. 
For instance, the number of inequivalent sites for a surface slab is at least the number of layers divided by two (\autoref{fig:merge_orbits}).
Consequentially, the number of orbits increases rapidly with the number of layers and the \gls{ce} construction procedure outlined in \autoref{sect:basic-example} can be insufficient.

In the following, we discuss three approaches to improve \gls{ce} construction for low-symmetry systems. 
The first two approaches are based on grouping similar orbits and either explicitly merging them or coupling them in a Bayesian framework. 
The third approach consists of adding weighted constraints to ensure that specific properties are more accurately predicted. 
These approaches can be combined and are not restricted to low-symmetry systems.  

To showcase these approaches we will construct \glspl{ce} for a 12-layer \ce{Au_xPd_{1-x}} \gls{fcc}-(111) surface slab. 
The cutoffs used for \gls{ce} construction are \qty{6}{\angstrom} and \qty{3}{\angstrom} for two and three-body clusters, respectively, resulting in 76 orbits. 
The training set consists of the pure Au and Pd slabs and 201 structures generated using the orthogonal cluster space approach described in \autoref{sect:structure-generation} (using slightly larger cutoffs to avoid underdetermined systems during training).
Details on the \gls{dft} calculations can be found in Ref.~\citenum{EkbErh21}.

\begin{figure}
\centering
\includegraphics[]{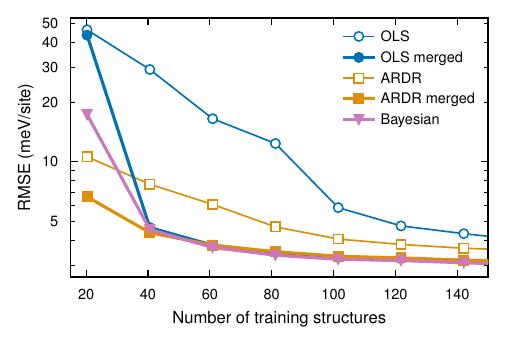}
\caption{
    \textbf{Learning curves for a surface \gls{ce}.}
    The validation error as a function of the training set size for different linear regression methods, with and without merging of orbits. 
}
\label{fig:merge_learning_curves}
\end{figure}

We begin by training \glspl{ce} as  in \autoref{sect:basic-example}, using plain \gls{ols} and \gls{ardr} to fit the \glspl{eci}. 
As expected, \gls{ardr} outperforms \gls{ols} in terms of the validation error as well as with respect to the number of structures necessary to reach convergence (\autoref{fig:merge_learning_curves}). 

The number of structures needed to reach convergence is, however, relatively large for both fitting methods. 

\begin{figure*}
\centering
\includegraphics[]{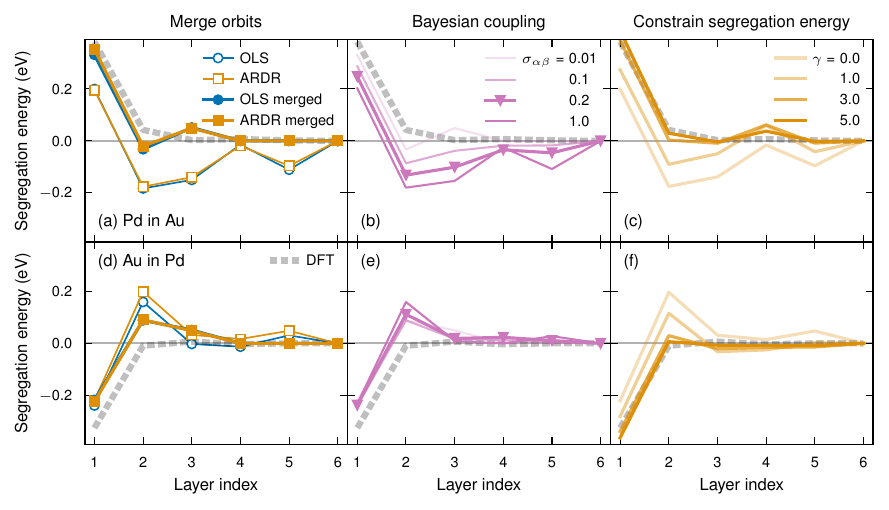}
\caption{
    \textbf{Segregation energy predictions for surface \glspl{ce}.}
    Segregation energies for a single Pd atom in Au (upper panel) and a single Au atom in Pd (lower panel) positioned in each atomic layer calculated using \gls{dft} and predicted by different \glspl{ce}.
    The layer index is counting from the surface, as in \autoref{fig:merge_orbits}.
    (a, d) \glspl{ce} trained using \gls{ols} and \gls{ardr} with and without merging of orbits.
    (b, e) \glspl{ce} trained using Bayesian coupling of similar orbits. The \gls{ce} with the optimal coupling parameter $\sigma_{\alpha\beta}$ from \autoref{fig:bayesian_rmse} is highlighted with triangles. 
    (c, f) \glspl{ce} trained with (unmerged) \gls{ardr} and an added constraint enforcing correct segregation energy prediction with increasing weight $\gamma$.   
}
\label{fig:segregation_energy}
\end{figure*}

The validation error is not always a sufficient measure for \gls{ce} accuracy. 
In the present example of a surface, we find that although the validation energy has converged, the segregation energy prediction is associated with large errors, which will ultimately result in an erroneous predictions of the surface segregation. 
The segregation energy is the energy difference caused by moving, e.g., a single Pd atom in an otherwise pure Au slab from the middle of the slab (i.e., the bulk) to a site in the surface region.
(Here, we construct \numproduct{2x2x12} slabs for this purpose.)
This calculation requires the prediction of two configurations which mean it is the relative error between these two configuration that matters. 
In \autoref{fig:segregation_energy}a and d, we find that the segregation energies predicted by the plain \gls{ols} and \gls{ardr} \glspl{ce} have large errors and systematically overestimate the energy gain of moving Pd (in Au) to the surface and the energy cost of moving Au (in Pd) to the surface.
There are also unphysical oscillations of the segregation energy in the inner layers whereas the \gls{dft} results indicate that the segregation energy only varies in the outer 2 to 3 layers. 
This suggest that the models predict too large differences between the atoms in the inner layers, as discussed further in the following section. 

\subsection{Merging of orbits}

At a sufficient distance from the surface, the atomic interactions should reach the bulk limit. 
For example, the energetic contribution from a nearest neighbour pair in layer 5 should be very similar to the one in layer 6.
We can use this idea to define so-called \emph{local} symmetries.
Orbits that belong to the same local symmetry should have the same \gls{eci}, which means that they can be merged into a single orbit. 
This process is analogous to when the individual clusters are grouped into orbits in Eq.~\eqref{eq:CE-formalism-2}. 
Note that in this approach, we now longer rely on the \emph{global} symmetries, which can be rigorously derived from the underlying lattice.
Instead the specification of the local symmetries is up to the person constructing the \gls{ce}, and based on additional \emph{physical} knowledge about the system, such as the range of the interactions and the similarity of the local environments.

Note that it is important to be aware of the treatment of multiplicities when merging orbits.
In many implementations of the \gls{ce} method, including \icet{}, the default behavior is to include the multiplicities in the \gls{eci} vector $\vec{w}$.
This will lead to incorrect results if any of the merged orbits have different multiplicities. 
Instead, the multiplicities should be included in the design matrix $\vec{X}$, which will lead to correct averaging of the multiplicities when merging.

In the following, we choose to consider only the outermost layer as surface and treat the remaining layers as bulk. 
We then merge all orbits with the same order and radius that consist exclusively of bulk sites (\autoref{fig:merge_orbits}), reducing the number of orbits from 76 to 20.
Other options include extending the surface region to several layers, introducing a subsurface region treated differently and treating the local symmetries different depending on cluster order. 

The present, rather aggressive, merging strategy leads to a significant reduction of the number of training structures necessary to reach a certain accuracy while actually improving the final validation error (\autoref{fig:merge_learning_curves}). 
Except for the smallest training set considered, there is almost no difference between \gls{ols} and \gls{ardr}. 
This is because the lack of regularization in \gls{ols} often leads to overfitting, which is avoided here due to the small number of features. 
We also find that the segregation energy predictions are significantly better for the merged \glspl{ce} (\autoref{fig:segregation_energy}a and d).
This is a result of restricting the differences between the \glspl{eci} in the inner layers. 

\subsection{Bayesian coupling of orbits}\label{sect:bayesian-ce}

\begin{figure}
\centering
\includegraphics{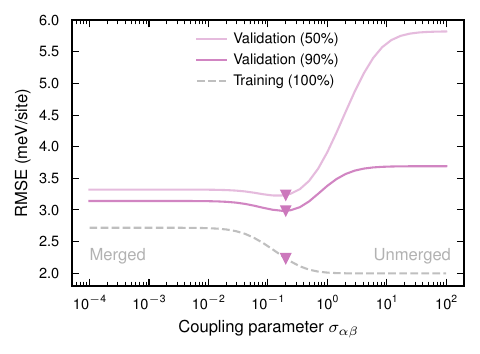}
\caption{
    \textbf{Performance of Bayesian surface \glspl{ce} with varying coupling strength.}
    Validation and training error with varying coupling parameter $\sigma_{\alpha \beta}$ for similar orbits.
    The validation error is shown for two different training set sizes (50 and 90\%), and the training error is obtained using the full training set (100\%). 
    The optimal coupling parameter, $\sigma_{\alpha \beta} = 0.2$, is highlighted with triangles. 
}
\label{fig:bayesian_rmse}
\end{figure}

The assumption that similar orbits should have similar energetic contribution to the total energy is an example of a physical insight about the system. 
If such insights can be formulated in the form of Bayesian priors, they can be used to construct improved \glspl{ce} within a Bayesian framework \cite{MueCed09, CocWal10, NelOzoRee13}.
Here, we follow the approach first presented by Mueller and Ceder in Ref.~\citenum{MueCed09}.

We assume Gaussian priors for the \glspl{eci} such that
\begin{equation}
    P(\vec{w}|\vec{X}) \propto \prod_\alpha e^{-w_\alpha^2/2\sigma_\alpha^2} \prod_{\alpha, \beta\neq\alpha} e^{-(w_\alpha-w_\beta)^2/2\sigma_{\alpha\beta}^2},
\end{equation}
where $P(\vec{w}|\vec{X})$ is the posterior.
Here, the first product controls the magnitude of the \glspl{eci} via $\sigma_{\alpha}$ which is the standard deviation of the prior and should be chosen to roughly correspond to the expected value of the respective \gls{eci}.
If one, for instance, wanted to achieve smaller \glspl{eci} for large clusters, one could define $\sigma_{\alpha}$ such that it decreases with orbit size. 
The second product controls the coupling between orbits via $\sigma_{\alpha\beta}$ which is the inverse coupling strength between orbits $\alpha$ and $\beta$.

Going back to the linear problem formulated in Eq.~\eqref{eq:linear_regression}, the Bayesian priors are introduced via the $\ell_2$ regularization matrix $\Lambda$ which has diagonal elements $\Lambda_{\alpha\alpha} = \frac{\sigma^2}{\sigma_\alpha^2} + \sum_{\beta\neq\alpha}\frac{\sigma^2}{\sigma_{\alpha\beta}^2}$ and off-diagonal elements $\Lambda_{\alpha\beta}=\Lambda_{\beta\alpha}=-\frac{\sigma^2}{\sigma_{\alpha\beta}^2}$,
where $\sigma$ reflects the typical error of the model.
Then, the maximum posterior estimate for the \glspl{eci} is given by
\begin{equation}
    \vec{w}_\text{opt} = (\vec{X}^T\vec{X} + \vec{\Lambda})^{-1}\vec{X}^T\vec{y}.
\end{equation}

Similar to the merging approach, one has to be careful with the treatment of multiplicities when using Bayesian coupling of similar orbits. 
Coupling orbits means that we expect the values of their \gls{eci} to be similar. 
It is therefore crucial that the multiplicities are included in the design matrix and not the \gls{eci} vector when applying this approach. 
In addition, in linear regression one typically employs standardization, i.e., rescaling of the design matrix to improve the linear regression. 
If this is the case, one has to employ the same scaling of the Bayesian regularization matrix.

In this tutorial, we apply a rather simple Bayesian approach using the same definition of local symmetries as in \autoref{sect:low-symmetry} and couple the orbits belonging to the same local symmetry using a single coupling parameter $\sigma_{\alpha\beta}$ for all coupled orbits. 
The \gls{eci} magnitude is controlled by a single value of $\sigma_{\alpha}$ for all orbits as well. 
We emphasize that the Bayesian priors can have a much more intricate design than this to allow for greater tunability \cite{MueCed09}.
For example, one can take into account the order and size of orbits and define complex criteria for the similarity of orbits. 

In \autoref{fig:bayesian_rmse} we show how the validation and training error varies with the coupling parameter $\sigma_{\alpha\beta}$ while keeping $\sigma_{\alpha}=10$.
For large $\sigma_{\alpha\beta}$, the unmerged \gls{ce} is obtained with a high validation error but low training error (indicating overfitting).
For small $\sigma_{\alpha\beta}$, the merged \gls{ce} is recovered and the training and validation errors approach each other with a significant reduction of the validation compared to the large $\sigma_{\alpha\beta}$ limit.
The latter is particularly prominent for smaller training sets. 
In between these to extremes, at about $\sigma_{\alpha\beta} = 0.2$, the validation error has a minimum. 
This indicates that the optimal compromise between the freedom of unmerged orbits and reduced feature space of merged orbits is found. 
\autoref{fig:segregation_energy}b and e show the segregation energy prediction for various values of $\sigma_{\alpha\beta}$.

The identified optimal $\sigma_{\alpha\beta} = 0.2$ does provide some improvement in segregation energy prediction compared to the unmerged \glspl{ce}, but decreasing $\sigma_{\alpha\beta}$ further towards the merged limit results in significantly better segregation energy prediction, in particular for the case of Pd in Au. 
This again shows that the validation \gls{rmse} is not always sufficient to find the most physically sound \gls{ce}.

\subsection{Adding constraints and weights}

A \gls{ce} can also be manipulated by introducing constraints and/or weights to ensure that certain properties are reproduced more accurately. 
The property of interest depends on the purpose of the model.
If, for example, the model is going to be used to study surface segregation phenomena, a natural choice is the segregation energy. 
Calculating the segregation energy requires \gls{dft} calculations of surface slabs with a single Pd atom in a Au slab, respectively positioned in each atomic layer, and vice versa resulting in a total of 12 structures.
A straightforward approach to improve the segregation energy prediction is to include these structures in the training set.
Additionally, one could give these structures a higher weight by simply multiplying the corresponding rows of the design matrix $\vec{X}$ and elements in the solution vector $\vec{y}$ with a suitably chosen factor.
By doing so, the prediction error will shrink for these specific structures, effectively reducing the segregation energy error. 

Another option, which we demonstrate in this tutorial, is to explicitly enforce better predictions of the segregation energy as a constraint. 
This entails reformulating the linear problem as
\begin{align}
\begin{split}
    \vec{w}_\text{opt} = \underset{\vec{w}}{\text{argmin}}
    \big\{
    &\Vert \vec{X}\vec{w} - \vec{y} \Vert^2_2
    \\
    &+ \gamma \Vert (\vec{X}_\text{S} - \vec{X}_\text{B})\vec{w} - \vec{E}_\text{seg} \Vert^2_2
    \big\},
\end{split}
\label{eq:constrained}
\end{align}
excluding any regularization terms (see Eq.~\eqref{eq:linear_regression}).
Here, $\vec{X_\text{S}}$ contains the cluster vectors of the structures with a single atom positioned in different atomic layers.
Similarly, $\vec{X_\text{B}}$ contains the cluster vectors of the corresponding structures with a single atom in the bulk position (i.e., the innermost atomic layer).
$\vec{E_\text{seg}}$ is the corresponding segregation energies and $\gamma$ is a weight factor that dictates how strongly the constraint is enforced. 
In practice, this is achieved by adding rows to the design matrix $\vec{X}$ corresponding to $\gamma  (\vec{X}_\text{S} - \vec{X}_\text{B})$ and the target vector $\vec{y}$ corresponding to $\vec{E_\text{seg}}$.

In \autoref{fig:segregation_energy}c and f, we show the segregation energy predictions for different weight factors $\gamma$. 
As expected, the segregation energy predictions move closer to the \gls{dft} line with increasing weight. 
The improvement in segregation energy prediction comes at the cost of a slightly increased validation error, from \qty{2.8}{\milli\electronvolt\per\site} for $\gamma=0$ to \qty{3.4}{\milli\electronvolt\per\site} for $\gamma=5$.

The strategy of introducing constraints and/or weights is not restricted to low symmetry systems and can be applied to any property that can be expressed as a function of the cluster vector. 
For example, one could constrain the mixing energy of the pure Au and Pd structures to be 0, give higher weight to structures close to some specific composition or use weights to compensate for an uneven sampling of the configuration space.

\subsection{Conclusions}
In this section, we have presented three ways to adapt the \gls{ce} construction process in order to achieve accurate models for complex systems, such as surfaces. 
These approaches should be seen as tools that can be used on their own (as shown here) or in conjunction and can be modified to provide a tailored solution for the problem at hand.

\section{Part 3: Monte Carlo sampling}
\label{sect:monte-carlo-sampling} 

\begin{tcolorbox}[colback=myGreen!15!white,colframe=myGreen, title=\textbf{Key takeaways}]
  \begin{itemize}[leftmargin=*]
      \item \acrshort{mc} simulations can be used to sample a \acrshort{ce} in different thermodynamical ensembles.
      \item \acrshort{vcsgc} can be used for sampling across miscibility gaps, but not
      \acrshort{sgc} since the chemical potential maps to two different concentrations.
      \item Phase transitions can be identified from different thermodynamic properties such as heat capacity as well as short and long range order parameters.
  \end{itemize}
\end{tcolorbox}

In the third and last part of this tutorial, we show how the configuration space described by a \gls{ce} can be sampled via \gls{mc} simulations.
This allows for calculation of thermodynamic observables such as free energies, order parameters and heat capacities.
We will begin this section with a short introduction to \gls{mc} simulations and the most common thermodynamic ensembles, and then illustrate these concepts using two examples.

\subsection{Monte Carlo simulations}

Thermodynamic sampling is usually carried out via Metropolis \gls{mc} simulations \cite{Fre23} which are generally executed as follows.
Starting from some arbitrary initial state, a so-called trial step is suggested, which consists of a change in the atomic configuration. 
This step is either accepted (i.e., implemented) or rejected according to the Metropolis criterion with a probability given by
\begin{equation*}
    \mathcal{P} = \min{\left\{1,  \mathrm{exp}\left(- \Delta\psi / k_B T\right)\right\}},
\end{equation*}
where $\Delta \psi$ is the change in the relevant thermodynamic potential (which will be introduced further in Appendix \ref{sect:ensembles}).
If the trial step is rejected, the system remains in its current state.
The procedure is repeated until convergence is achieved.

\subsection{Thermodynamic ensembles}
\label{sect:ensembles}

\Gls{mc} simulations can be executed in different thermodynamic ensembles. 
The choice of which depends on the goal of the simulation and the properties to be extracted.

\textbf{The canonical ensemble} models a situation where the total number of particles ($N$) as well as their concentrations ($c_i$) are kept constant along with the temperature ($T)$.
It thus represents a system free to exchange heat with a reservoir at temperature $T$.

Strictly speaking, in the canonical ensemble (and similarly in the \gls{sgc} and \gls{vcsgc} ensembles, see below) the volume ($V$) is constant.
When constructing a \gls{ce} one, however, commonly trains against configurations that have been relaxed with respect to both atomic positions and volume/cell shape.
\Gls{ce} models trained in this fashion therefore incorporate the strain energy term that separates the canonical ($Nc_iVT$) from the isobaric-isothermal ensemble ($Nc_iTp$).
It is therefore more sensible to interpret the results of such \gls{ce}-\gls{mc} simulations in terms of the latter ensemble.
In keeping with the literature, we will, however, use the terms for the constant-volume ensembles in the following.

The trial steps used to sample the canonical ensemble need to preserve the conserved quantities, specifically the concentrations of the different species ($c_i$).
This can be accomplished by (simultaneously) swapping the occupancy of two sites.
The change of the thermodynamic potential, $\psi$, is then given by the energy difference between the two states, i.e., $\Delta\psi = E_\text{new} - E_\text{old} = \Delta E$.
In the context of this tutorial, $E$ refers to the energy predicted by the \gls{ce}.

It is also possible to gradually reduce the temperature in an \gls{mc} simulation in the canonical ensemble, an approach that is referred to as \textit{simulated annealing}.
This procedure can be used as a general optimization algorithm, for example to find the lowest energy (ground-state) structures or in structure selection as implemented in \autoref{sect:structure-generation}.

\textbf{The \gls{sgc} ensemble} refers to a case where the total number of particles ($N$) is fixed while the concentration of the different species is controlled via the relative chemical potentials ($\Delta \mu_i$), again at constant temperature ($T$).
The \gls{sgc} ensemble thus represents a system in connection with both a heat reservoir ($T$) and one or several particle reservoirs ($\Delta\mu_i$) under the constraint of a constant number of particles ($N$).

In the \gls{ce} literature it is not uncommon for the \gls{sgc} ensemble to be referred to as the grand canonical ensemble, which is, however, a misnomer.
The actual \emph{grand canonical} ensemble represents an open system that has no constraint on the total number of particles with the absolute chemical potentials ($\mu_i$) as variables \cite{Pathria22}.
By contrast, the \emph{semi-grand canonical} ensemble represents a semi-open system, where the relative proportion of particles of different species but not their total number may change.
In cases involving at least one sublattice with vacancies (such as the example discussed in \autoref{sect:basic-example}), one can interpret \gls{ce}-based \gls{mc} simulations in the \gls{sgc} or \gls{vcsgc} ensembles in terms of the grand canonical ensemble by imposing suitable thermodynamic boundary conditions.
This approach is described and demonstrated in, e.g., Ref.~\citenum{RahLöfFraErh21}, which also discusses para and full equilibria in this context.

As before, in the case of the canonical ensemble, trial moves used for sampling the \gls{sgc} ensemble need to respect the respective constraints.
A suitable trial move is therefore to change the occupation of a single site.
The change in the thermodynamic potential is given by $\Delta\psi = \Delta E - N\sum_{i>1}\Delta c_i\Delta \mu_i$, where $N$ is the total number of sites, $c_i$ is the concentration of species $i$ and $\Delta\mu_i = \mu_i - \mu_1$ is the chemical potential difference of species $i$ relative to the first species.

A big advantage of the \gls{sgc} ensemble compared to the canonical ensemble is that it directly connects to the derivative of the free energy per atom with respect to the concentration(s)
\begin{equation*}
    \frac{1}{N}\frac{\partial F}{\partial c_i} = -\Delta \mu_i,
\end{equation*}
which can be integrated along the concentration axis (or axes) to yield the free energy.
Inside multi-phase regions, the same chemical potentials map, however, to multiple different concentrations, which renders it impossible to integrate over (and in fact sample inside) miscibility gaps.

\textbf{The \gls{vcsgc} ensemble} can, in contrast to the \gls{sgc} ensemble, be used for sampling inside and across miscibility gaps \cite{SadErh12}.
Similar to the \gls{sgc} ensemble it imposes a flexible constraint on the mean concentration(s) but unlike the \gls{sgc} ensemble it also constrains their fluctuation(s).
This is achieved via two variables, $\bar{\phi}$ and $\bar{\kappa}$, which control the mean and the variance of the concentration(s), respectively.
The corresponding thermodynamic potential is given by $\Delta\psi = \Delta E + Nk_BT\bar{\kappa}(\Delta c + \bar{\phi} / 2)^2$, and the ensemble can be sampled using the same kind of moves as those used for the \gls{sgc} ensemble.

The free energy derivative per atom with respect to the concentration is given by
\begin{equation}
    \frac{1}{N}\frac{\partial F}{\partial c} = - 2k_BT\bar{\kappa}\left( \left<c\right> + \bar{\phi} / 2 \right)
    \label{eq:vcsgc-free-energy}
\end{equation}
As a result of the constraint on the variance of the concentration, one can also stabilize concentrations inside of miscibility gaps.
This allows one to obtain the free energy as a continuous function of concentration, and thus enables free energy integration.

The choice of $\bar{\kappa}$ affects the strength of the variance constraint, corresponding to the inverse variance of the expected concentration.
Larger values enforce smaller fluctuations, which reduces the acceptance ratios, requiring more sampling.
Smaller values on the other hand, can cause the fluctuations to become too large to allow sampling two-phase regions as the \gls{vcsgc} ensemble then approaches the \gls{sgc} ensemble \cite{SadErh12}.
Empirically, we have found that $\bar\kappa=200$ strikes a good balance between sampling efficiency and quality for most systems.

We can get an idea for how to select suitable values for the parameter $\bar{\phi}$ by considering Eq.~\eqref{eq:vcsgc-free-energy}.
For typical temperatures and values of $\bar{\kappa}$ the terms in front of the bracket on the right-hand side of Eq.~\eqref{eq:vcsgc-free-energy} are much larger than the variation of the free energy on the left-hand side.
As a result, we can assume that $2\left<c\right> + \bar{\phi} \approx 0$.
It then follows that the concentration limits $\left<c\right>\rightarrow 0$ and $\left<c\right>\rightarrow 1$ are reached by setting $\bar{\phi}\approx -2-\delta$ and $\bar{\phi}\approx \delta$, respectively.
Here, $\delta$ indicates that one needs to sample somewhat beyond those limits in order to cover the full composition range.
In practice, a value of $\delta$ of about 0.3 is sufficient in most cases.

\subsection{Ordering in \texorpdfstring{\ce{Au3Pd}}{AuPd3} using the canonical ensemble}
\label{sect:mc-AuPd}

\begin{figure}
\centering
\includegraphics{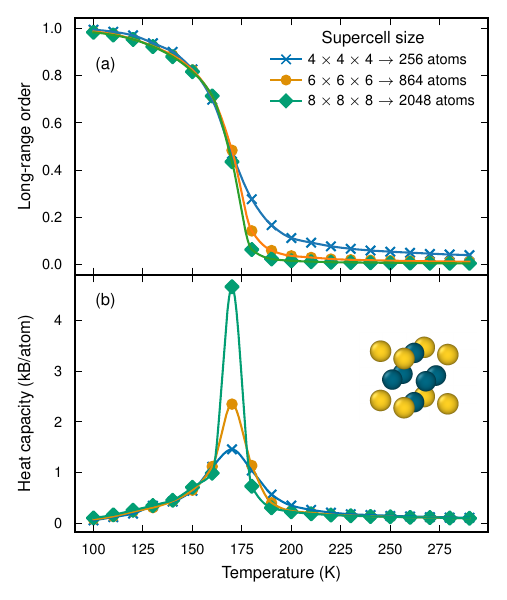}
\caption{
    \textbf{Order-disorder transition in \ce{AuPd3} from \gls{mc} simulations.}
    (a) Long-range order parameter and (b) heat capacity as a function of temperature and system size, where the solid lines are guides for the eye.
    The system size is given in multiples of the primitive (4-atom) unit cell.
    The long-range order is represented by the partial static structure factor between Pd-atoms.
    The inset shows the ordered low temperature phase.
}
\label{fig:AuPd:cv-lro}
\end{figure}

As the first illustration we consider the order-disorder transition in \ce{AuPd3}, which can be conveniently accessed via simulations in the canonical ensemble.
Here, we use a \gls{ce} developed in Ref.~\citenum{RahLöfFraErh21} for \ce{Au_{1-x}Pd_x} on the \gls{fcc} lattice.
In \autoref{fig:AuPd:cv-lro} we show the resulting long-range order parameter and heat capacity as a function of temperature.
The long-range order is here represented by the partial static structure factor calculated between Pd atoms at $\mathbf{q} = \frac{2\pi}{a_0}[1, 0, 0]$ (see Eq.~\eqref{eq:LRO}), while the heat capacity is calculated via Eq.~\eqref{eq:heat-capacity}.

The sharp peak in the heat capacity at approximately \qty{175}{\kelvin} indicates that the material undergoes a continuous phase transition from a disordered to an ordered phase (see inset in \autoref{fig:AuPd:cv-lro}b) with decreasing temperature.
The phase transition can also be seen in the long-range order parameter, as this is increasing sharply at the phase transition.

All simulations indicate a phase transition around \qty{175}{\kelvin}, but there the sharpness of the transition is sensitive to supercell size, with larger systems yielding sharper transitions.
This behavior is typical for continuous phase transitions, as the correlation length of the fluctuations in the systems diverges at the critical temperature \cite{Ma_StatPhys}.
In order to achieve convergence one therefore needs to consider a range of system sizes, and extrapolate to the infinite size limit if possible.

\subsection{Phase diagram of \texorpdfstring{\ce{Ag_{x}Pd_{1-x}}}{Ag(x)Pd(1-x)} via the \texorpdfstring{\gls{sgc}}{SGC} and \texorpdfstring{\gls{vcsgc}}{VCSGC} ensembles}

Next, we illustrate the construction of a phase diagram for a system with a miscibility gap, namely \gls{fcc} \ce{Ag_{x}Pd_{1-x}}, using a \gls{ce} from Ref.~\citenum{Angqvist2019}, a \numproduct{10x10x10} supercell and \gls{mc} simulations in both the \gls{sgc} and \gls{vcsgc} ensembles.

In the case of the \gls{sgc}-\gls{mc} simulations, the chemical potential was sampled using \num{105} evenly spaced points between \num{-1.04} and \qty{1.04}{\electronvolt\per\atom}.
For the \gls{vcsgc}-\gls{mc} simulations the variance constraint parameter $\bar{\kappa}$ was set to \num{200}, while the average constraint parameter $\bar{\phi}$ was varied between \num{-2.3} and \num{0.3} with \num{105} evenly spaced points.

\begin{figure}
\centering
\includegraphics{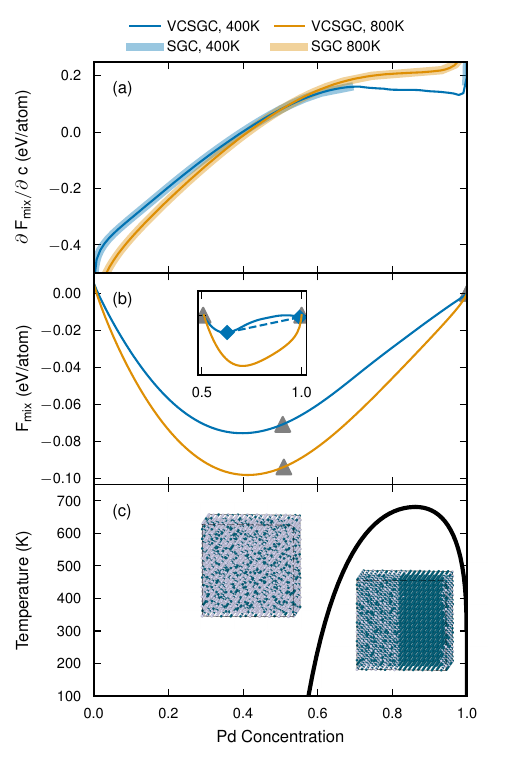}
\caption{
    \textbf{\Gls{mc} sampling of the \ce{Ag_xPd_{1-x}} system.}
    (a) Derivative of the free energy of mixing with respect to composition (i.e., the chemical potential), (b) free energy of mixing and (c) phase diagram.
    The inset in (b) shows the zoomed-in and tilted free energy of mixing in the vicinity of the miscibility gap (indicated by the blue diamonds).
    Due to the concave free energy at \qty{400}{\kelvin}, the \gls{sgc}-\gls{mc} ensemble cannot form a stable solution.
    The miscibility gap is extracted from the free energy landscape of the \gls{vcsgc} ensemble.
}
\label{fig:AgPd:phase-diagram}
\end{figure}

\textbf{Miscibility gap.}
Figure~\ref{fig:AgPd:phase-diagram}a shows the free energy derivative obtained from sampling in both ensembles at a temperature below (\qty{400}{\kelvin}) and above (\qty{800}{\kelvin}) the miscibility gap.
Here, the miscibility gap separates a Pd-rich phase containing almost no Ag, from a mixed phase with up to approximately 60\% to 70\% Pd.

Outside the miscibility gap, simulations in the \gls{sgc} and \gls{vcsgc} ensembles yields numerically identical results.
In the \gls{sgc}-\gls{mc} simulations at \qty{400}{\kelvin} it is, however, apparent that the miscibility gap, i.e., the two-phase region between approximately 60\% and 99\%, is not accessible as the mapping between $\partial F/\partial c$ and the concentration $c$ is one-to-many (i.e., non injective), and the variance of the concentration diverges.
In fact, here the identification of the free energy derivative with the relative chemical potential breaks down, as the latter is only meaningfully defined in single-phase regions.

When using \gls{sgc}-\gls{mc} simulations the free energy in the single-phase regions can still be obtained through thermodynamic integration starting from known limits such as in the low or high-temperature expansions \cite{WalAst02}.
Furthermore it is possible to obtain the phase diagram by tracing the phase boundaries \cite{WalAst02}.
To mitigate hysteresis effects, this requires iteratively sampling across two-phase regions as well as careful consideration of the convergence with respect to the numerical resolution of the chemical potential.

To counteract the divergence of the concentration in miscibility gaps, in the \gls{vcsgc} ensemble one includes a term representing a \emph{constraint} on the variance in the thermodynamic potential.
This allows one to access also compositions inside the miscibility gap and sample the free energy derivative as a continuous function of composition (\autoref{fig:AgPd:phase-diagram}a).
This in turn enables integrating the free energy over the entire concentration range (\autoref{fig:AgPd:phase-diagram}b).
Finally, via the convex hull construction (inset in \autoref{fig:AgPd:phase-diagram}b) one can then readily extract the phase boundaries (\autoref{fig:AgPd:phase-diagram}c).

Thanks to the possibility to compute the free energy across two-phase regions one can furthermore extract \emph{excess} free energies.
Thereby, it is possible to compute, e.g., interface free energies as a function of interface orientation.
The interested reader can find more information on this topic in, e.g., Refs.~\citenum{SadErh12} and \citenum{Rahm2022}.

Finally, the \gls{vcsgc} ensemble naturally enables an even sampling of the concentration axis as equidistant values of $\bar{\phi}$ leads to an approximately equidistant concentrations.
This avoids adaptive sampling in the vicinity of phase transitions that are necessary when using the \gls{sgc} ensemble.

\begin{figure}
\centering
\includegraphics{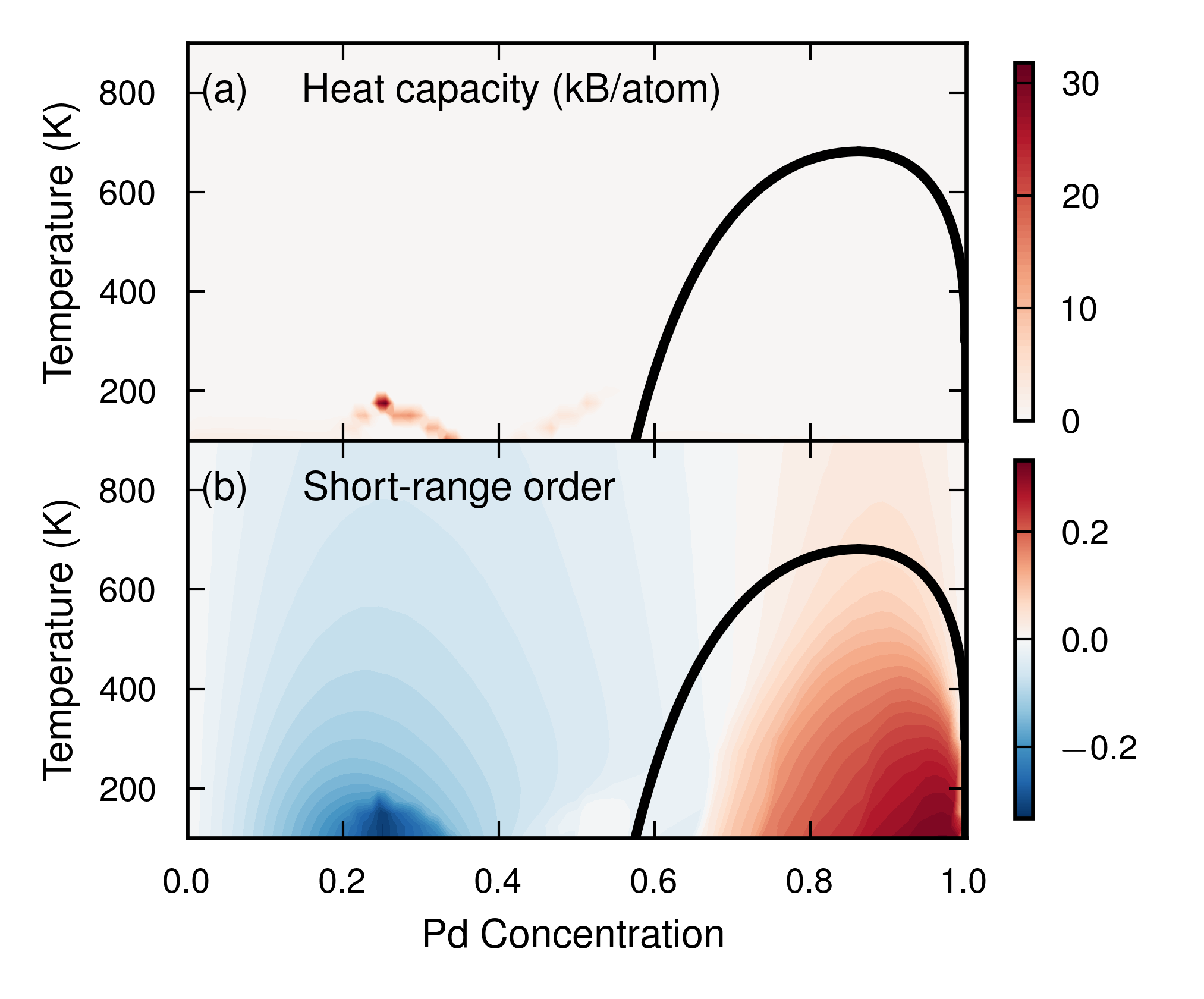}
\caption{
    \textbf{Order-disorder transitions in \ce{Ag_xPd_{1-x}} from \gls{vcsgc}-\gls{mc} simulations.}
    (a) The heat capacity map reveals a continuous phase transition between a disordered and an ordered phase around 25\% Pd.
    (b) The short-range order parameter [Eq.~\eqref{eq:SRO}] indicates the existence of a miscibility gap, however, it is hard to identify the exact boundaries. It also shows a sharp change at the order-disorder phase transition.
}
\label{fig:AgPd:cv-sro}
\end{figure}

\textbf{Secondary phase transitions.}
By mapping out the free energy, we were able to locate the miscibility gap in the Ag--Pd system.
This phase boundary corresponds to a first-order phase transition as evident from the sudden (and discontinuous) transition in the derivative of the free energy with respect to composition.
It is, however, very difficult (if not practically impossible for numerical reasons) to obtain information about continuous phase transitions.
To this end, as we already saw above (\autoref{sect:mc-AuPd}) the heat capacity and order parameters are better suited.

Analysis of the heat capacity as a function of temperature and composition does indeed reveal a transition around 25\% Pd and \qty{200}{\kelvin} (\autoref{fig:AgPd:cv-sro}a).
This transition closely resembles the one in \ce{Au3Pd} that we analyzed before but here we also obtain the composition dependence of the transition.
At 25\% Pd the transition is very sharp, while both the heat capacity and the transition temperature drop with either decreasing or increasing composition.
Note that the heat capacity shows no discernible features around the miscibility gap.

The order-disorder transition at around 25\% Pd is also apparent from the short-range order parameter computed according to Eq.~\eqref{eq:SRO} (\autoref{fig:AgPd:cv-sro}b).
The short-range order map (unlike the heat capacity), however, also exhibits structure across the miscibility gap, which is the result of the variation in the proportions of the two different phases (Pd-rich and mixed).
It is, however, clearly apparent that the short-range order is not suited for identifying the actual phase boundaries as the map is smooth and distinct features only emerge well inside the miscibility gap.

\subsection{Conclusions}
In the final part of this tutorial, we demonstrated the utility of \gls{mc} simulations in different thermodynamic ensembles for extracting thermodynamic observables and identifying phase boundaries.
Which ensemble is best suited depends on the task at hand.
Generally speaking the canonical ensemble is simple to use (and understand) but limited in so far as it does not directly enable one to observe a first derivative of the free energy.
It can still be used for thermodynamic integration (which was not covered here).
The \gls{sgc} and \gls{vcsgc} ensembles both provide access to the first derivative of the free energy with respect to concentration and thereby can be used more readily for obtaining phase diagrams.
The \gls{vcsgc} ensemble additionally allows one to handily extract excess free energies.
Here, we did not discuss acceptance ratios but in closing point out that while the canonical ensemble tends to yield higher acceptance ratios for small concentrations, the other two ensembles achieve higher ratios at higher concentrations \cite{SadErhStu12}.

\section{Outlook}

In this tutorial we have provided a short practical introduction into the topic of constructing and sampling \glspl{ce} for the study of many-component systems.
These techniques can be used to investigate a huge range of materials, including in particular materials with relevance to energy applications.
In the field of battery research, for example, \glspl{ce} have been applied to study chemical ordering \cite{VanCed04}, voltage curves \cite{CheSaiHua17} and migration barriers \cite{YuLinTho14, ChaJorLof21}, whereas with respect to photovoltaic materials, successful applications of \glspl{ce} include, e.g., studies of the phase stability of perovskites \cite{YamIikYam17, BecVan18}, bandgap engineering \cite{XuJia19, HanYeuYe22} as well as chemical order and the effect of vacancies \cite{YuCar16}.
Catalysis is another field where \glspl{ce} have proven to be useful, for instance when studying the effect of nanoparticle size and shape \cite{EomKimLim18} and composition \cite{AiChaTyg23a}, as well as for comparing active sites \cite{YanTanSai20} and performing high-throughput compositional screening of candidate materials \cite{AiChaTyg23b}.
In the field of thermoelectrics, \gls{ce} have been used to study the impact of chemical order \cite{AngLinErh16} and composition \cite{ChaReeDon10, TroRigDra17} on phase stability and even transport properties, for materials such as SiGe nanowires \cite{ChaReeDon10}, clathrates \cite{AngLinErh16, TroRigDra17, BroZhaPal21} and skutterudites \cite{IsaWol19}.
Lastly, \glspl{ce} are of importance for studying various high performance alloys, including superalloys \cite{MaiHofMul16}, high-entropy alloys \cite{ShaSinJoh16} and intermetallic compounds \cite{AliKleAko20, NatVan17}.
These materials often exhibit favorable thermal and mechanical properties and high tunability, and have applications in for instance light-weight construction and high temperature environments. 

Most applications of \glspl{ce} consist of \gls{ce} construction and \gls{mc} sampling which means that the contents of Part 1 and 3 of this tutorial are relevant in most cases. 
The structure selection and training strategies from Part 1 is of particular importance for more complex systems, such as the multicomponent high-performance alloys mentioned above, where these aspects are especially challenging due to the large number of components and/or larger unit cells.
Similarly, the content of Part 3 are of particular interest for studies with an emphasis on phase diagrams, phase transitions and chemical order, including, e.g., thermoelectrics and high performance alloys. 
In addition, understanding the thermodynamic ensembles, in particular the \gls{sgc} and \gls{vcsgc} ensembles, is crucial when the chemical potential is of interest, for instance for battery voltage curves and systems in contact with gases.
Lastly, the strategies in Part 2 are useful for studies of surfaces and nanoparticles, which is almost always the case in the field of catalysis, but can also be applied for other challenging systems where the strategies from Part 1 are not sufficient. 

Finally, we would like to emphasize that there are various more advanced topics related to \glspl{ce} that are beyond the scope of this tutorial that we invite so-inclined readers to pursue.
With regard to the \emph{construction} of \glspl{ce} this includes, e.g., managing sublattices \cite{RahLöfFraErh21}, strain \cite{Laks1992, OzoWolZun98a, Rahm2022}, handling thermodynamic constraints through nullspaces \cite{EriFraErh19, RahLöfFraErh21} or quadratic programming \cite{HuaUrbRon17} as well as alternative regression approaches \cite{ZhoCheBar22}.
In terms of \emph{applications}, one can mention, e.g., ground-state finding \cite{LarJacSch18, HuaKit2016}, materials under pressure \cite{AliKleAko20}, precipitate formation \cite{KleAko20}, the description of migration barriers \cite{VanCedAst01} and \gls{ce}-based kinetic \gls{mc} \cite{DenMisXie23}.

\appendix
\section{Computational details}\label{sect:dft-details}
The \gls{dft} calculations for the simple carbide system, \ce{Mo_{1-x}V_{x}C_{1 - y}\square_y}, were carried out using the projector augmented method \cite{Blo94, KreJou99} as implemented in \textsc{vasp} \cite{KreHaf93, KreFur1996-1, KreFur1996-2}.
We used the van-der-Waals density functional method with consistent exchange \cite{DioRydSch04, BerHyl2014}  for the exchange-correlation potential.
The Brillouin zone was sampled with a $\Gamma$ centered $\boldsymbol{k}$-point grid, with the smallest allowed spacing between two k-grid points being $\SI{0.25}{\angstrom^{-1}}$ and the plane wave cutoff energy $\SI{520}{eV}$.
Both the positions and cell were allowed to relax, the largest allowed residual forces were set to $\SI{0.02}{eV\angstrom^{-1}}$.

\section{Thermodynamic properties}
In addition to the properties mentioned above, one can obtain several other thermodynamic properties of interest from \gls{mc} simulations.

The \textbf{heat capacity} is readily obtained via
\begin{equation}
    C_V(T) = \frac{\mathrm{Var}[E(T)]}{k_B T^2},
    \label{eq:heat-capacity}
\end{equation}
where $\mathrm{Var}[E(T)]$ is the variance of the energy.

\textbf{Long-range order} can be assessed from the partial static structure factors, which are defined for atom types A and B as
\begin{equation}
    S_\text{AB}(\vec{q}) \propto \sum_{j}^{N_\text{A}} \sum_{k}^{N_\text{B}} \mathrm{exp}\left[-\mathrm{i}\vec{q}\cdot(\vec{R}_j - \vec{R}_k) \right], \label{eq:LRO}
\end{equation}
where $\vec{q}$ is a reciprocal lattice point and $\vec{R}_k$ the position of atom $k$.

\textbf{Short-range order} is often measured in terms of the Warren-Cowley short-range order parameter, which determines to which degree a binary (A-B) system mixes or segregates \cite{Cowley1950}.
For an atom $i$ of type A it is defined as
\begin{equation}
    \alpha_i = 1 - \frac{Z_\text{B}}{Z_\text{tot} c_\text{B}}, \label{eq:SRO}
\end{equation}
where $Z_\text{B}$ is the number of B neighbors in the first neighbor shell, $Z_\text{tot}$ is the total number of neighbors in the first shell and $c_\text{B}$ is the concentration of B atoms.
For a random mixture we obtain $\alpha=0$, for a phase separated system we get $\alpha > 0$, while $\alpha < 0$ indicates ordering.

\section*{Acknowledgments}

We gratefully acknowledge funding from the Swedish Research Council (Nos.~2020-04935 and 2021-05072), the Swedish Energy Agency (grant No. 45410-1), the Excellence Initiative Nano at Chalmers, and the Chalmers Initiative for Advancement of Neutron and Synchrotron Techniques. 
The computations were enabled by resources provided by the National Academic Infrastructure for Supercomputing in Sweden (NAISS) at C3SE, NSC, and PDC partially funded by the Swedish Research Council through grant agreements no. 2022-06725 and no. 2018-05973.


\begin{thebibliography}{142}%
\makeatletter
\providecommand \@ifxundefined [1]{%
 \@ifx{#1\undefined}
}%
\providecommand \@ifnum [1]{%
 \ifnum #1\expandafter \@firstoftwo
 \else \expandafter \@secondoftwo
 \fi
}%
\providecommand \@ifx [1]{%
 \ifx #1\expandafter \@firstoftwo
 \else \expandafter \@secondoftwo
 \fi
}%
\providecommand \natexlab [1]{#1}%
\providecommand \enquote  [1]{``#1''}%
\providecommand \bibnamefont  [1]{#1}%
\providecommand \bibfnamefont [1]{#1}%
\providecommand \citenamefont [1]{#1}%
\providecommand \href@noop [0]{\@secondoftwo}%
\providecommand \href [0]{\begingroup \@sanitize@url \@href}%
\providecommand \@href[1]{\@@startlink{#1}\@@href}%
\providecommand \@@href[1]{\endgroup#1\@@endlink}%
\providecommand \@sanitize@url [0]{\catcode `\\12\catcode `\$12\catcode
  `\&12\catcode `\#12\catcode `\^12\catcode `\_12\catcode `\%12\relax}%
\providecommand \@@startlink[1]{}%
\providecommand \@@endlink[0]{}%
\providecommand \url  [0]{\begingroup\@sanitize@url \@url }%
\providecommand \@url [1]{\endgroup\@href {#1}{\urlprefix }}%
\providecommand \urlprefix  [0]{URL }%
\providecommand \Eprint [0]{\href }%
\providecommand \doibase [0]{https://doi.org/}%
\providecommand \selectlanguage [0]{\@gobble}%
\providecommand \bibinfo  [0]{\@secondoftwo}%
\providecommand \bibfield  [0]{\@secondoftwo}%
\providecommand \translation [1]{[#1]}%
\providecommand \BibitemOpen [0]{}%
\providecommand \bibitemStop [0]{}%
\providecommand \bibitemNoStop [0]{.\EOS\space}%
\providecommand \EOS [0]{\spacefactor3000\relax}%
\providecommand \BibitemShut  [1]{\csname bibitem#1\endcsname}%
\let\auto@bib@innerbib\@empty
\bibitem [{\citenamefont {Ma}\ \emph {et~al.}(2009)\citenamefont {Ma},
  \citenamefont {Luther}, \citenamefont {Zheng}, \citenamefont {Wu},\ and\
  \citenamefont {Alivisatos}}]{MaLutZhe09}%
  \BibitemOpen
  \bibfield  {author} {\bibinfo {author} {\bibfnamefont {W.}~\bibnamefont
  {Ma}}, \bibinfo {author} {\bibfnamefont {J.~M.}\ \bibnamefont {Luther}},
  \bibinfo {author} {\bibfnamefont {H.}~\bibnamefont {Zheng}}, \bibinfo
  {author} {\bibfnamefont {Y.}~\bibnamefont {Wu}},\ and\ \bibinfo {author}
  {\bibfnamefont {A.~P.}\ \bibnamefont {Alivisatos}},\ }\bibfield  {title}
  {\bibinfo {title} {Photovoltaic devices employing ternary \ce{PbS_xSe_{1-x}}
  nanocrystals},\ }\href {https://doi.org/10.1021/nl900388a} {\bibfield
  {journal} {\bibinfo  {journal} {Nano Letters}\ }\textbf {\bibinfo {volume}
  {9}},\ \bibinfo {pages} {1699–1703} (\bibinfo {year} {2009})}\BibitemShut
  {NoStop}%
\bibitem [{\citenamefont {Xiao}\ \emph {et~al.}(2013)\citenamefont {Xiao},
  \citenamefont {Li}, \citenamefont {Yao}, \citenamefont {Deng}, \citenamefont
  {Ding}, \citenamefont {Wu}, \citenamefont {Yang}, \citenamefont {Li},
  \citenamefont {Dong}, \citenamefont {Liu}, \citenamefont {Zhang},\ and\
  \citenamefont {Zhao}}]{XiaLiYao13}%
  \BibitemOpen
  \bibfield  {author} {\bibinfo {author} {\bibfnamefont {Z.-Y.}\ \bibnamefont
  {Xiao}}, \bibinfo {author} {\bibfnamefont {Y.-F.}\ \bibnamefont {Li}},
  \bibinfo {author} {\bibfnamefont {B.}~\bibnamefont {Yao}}, \bibinfo {author}
  {\bibfnamefont {R.}~\bibnamefont {Deng}}, \bibinfo {author} {\bibfnamefont
  {Z.-H.}\ \bibnamefont {Ding}}, \bibinfo {author} {\bibfnamefont
  {T.}~\bibnamefont {Wu}}, \bibinfo {author} {\bibfnamefont {G.}~\bibnamefont
  {Yang}}, \bibinfo {author} {\bibfnamefont {C.-R.}\ \bibnamefont {Li}},
  \bibinfo {author} {\bibfnamefont {Z.-Y.}\ \bibnamefont {Dong}}, \bibinfo
  {author} {\bibfnamefont {L.}~\bibnamefont {Liu}}, \bibinfo {author}
  {\bibfnamefont {L.-G.}\ \bibnamefont {Zhang}},\ and\ \bibinfo {author}
  {\bibfnamefont {H.-F.}\ \bibnamefont {Zhao}},\ }\bibfield  {title} {\bibinfo
  {title} {Bandgap engineering of \ce{Cu2Cd_xZn_{1-x}SnS4} alloy for
  photovoltaic applications: {A} complementary experimental and
  first-principles study},\ }\href {https://doi.org/10.1063/1.4829457}
  {\bibfield  {journal} {\bibinfo  {journal} {Journal of Applied Physics}\
  }\textbf {\bibinfo {volume} {114}},\ \bibinfo {pages} {183506} (\bibinfo
  {year} {2013})}\BibitemShut {NoStop}%
\bibitem [{\citenamefont {Lu}\ and\ \citenamefont {Ferguson}(2013)}]{LuFer13}%
  \BibitemOpen
  \bibfield  {author} {\bibinfo {author} {\bibfnamefont {N.}~\bibnamefont
  {Lu}}\ and\ \bibinfo {author} {\bibfnamefont {I.}~\bibnamefont {Ferguson}},\
  }\bibfield  {title} {\bibinfo {title} {{III}-nitrides for energy production:
  photovoltaic and thermoelectric applications},\ }\href
  {https://doi.org/10.1088/0268-1242/28/7/074023} {\bibfield  {journal}
  {\bibinfo  {journal} {Semiconductor Science and Technology}\ }\textbf
  {\bibinfo {volume} {28}},\ \bibinfo {pages} {074023} (\bibinfo {year}
  {2013})}\BibitemShut {NoStop}%
\bibitem [{\citenamefont {Feurer}\ \emph {et~al.}(2016)\citenamefont {Feurer},
  \citenamefont {Reinhard}, \citenamefont {Avancini}, \citenamefont {Bissig},
  \citenamefont {L\"{o}ckinger}, \citenamefont {Fuchs}, \citenamefont {Carron},
  \citenamefont {Weiss}, \citenamefont {Perrenoud}, \citenamefont
  {Stutterheim}, \citenamefont {Buecheler},\ and\ \citenamefont
  {Tiwari}}]{FeuReiAva16}%
  \BibitemOpen
  \bibfield  {author} {\bibinfo {author} {\bibfnamefont {T.}~\bibnamefont
  {Feurer}}, \bibinfo {author} {\bibfnamefont {P.}~\bibnamefont {Reinhard}},
  \bibinfo {author} {\bibfnamefont {E.}~\bibnamefont {Avancini}}, \bibinfo
  {author} {\bibfnamefont {B.}~\bibnamefont {Bissig}}, \bibinfo {author}
  {\bibfnamefont {J.}~\bibnamefont {L\"{o}ckinger}}, \bibinfo {author}
  {\bibfnamefont {P.}~\bibnamefont {Fuchs}}, \bibinfo {author} {\bibfnamefont
  {R.}~\bibnamefont {Carron}}, \bibinfo {author} {\bibfnamefont {T.~P.}\
  \bibnamefont {Weiss}}, \bibinfo {author} {\bibfnamefont {J.}~\bibnamefont
  {Perrenoud}}, \bibinfo {author} {\bibfnamefont {S.}~\bibnamefont
  {Stutterheim}}, \bibinfo {author} {\bibfnamefont {S.}~\bibnamefont
  {Buecheler}},\ and\ \bibinfo {author} {\bibfnamefont {A.~N.}\ \bibnamefont
  {Tiwari}},\ }\bibfield  {title} {\bibinfo {title} {Progress in thin film
  {CIGS} photovoltaics – {Research} and development, manufacturing, and
  applications},\ }\href {https://doi.org/10.1002/pip.2811} {\bibfield
  {journal} {\bibinfo  {journal} {Progress in Photovoltaics: Research and
  Applications}\ }\textbf {\bibinfo {volume} {25}},\ \bibinfo {pages}
  {645–667} (\bibinfo {year} {2016})}\BibitemShut {NoStop}%
\bibitem [{\citenamefont {Li}\ \emph {et~al.}(2022)\citenamefont {Li},
  \citenamefont {Cai}, \citenamefont {Valdes}, \citenamefont {Wang},
  \citenamefont {Shafarman}, \citenamefont {Wei},\ and\ \citenamefont
  {Janotti}}]{LiCaiVal22}%
  \BibitemOpen
  \bibfield  {author} {\bibinfo {author} {\bibfnamefont {W.}~\bibnamefont
  {Li}}, \bibinfo {author} {\bibfnamefont {X.-F.}\ \bibnamefont {Cai}},
  \bibinfo {author} {\bibfnamefont {N.}~\bibnamefont {Valdes}}, \bibinfo
  {author} {\bibfnamefont {T.}~\bibnamefont {Wang}}, \bibinfo {author}
  {\bibfnamefont {W.}~\bibnamefont {Shafarman}}, \bibinfo {author}
  {\bibfnamefont {S.-H.}\ \bibnamefont {Wei}},\ and\ \bibinfo {author}
  {\bibfnamefont {A.}~\bibnamefont {Janotti}},\ }\bibfield  {title} {\bibinfo
  {title} {\ce{In2Se3}, \ce{In2Te3}, and \ce{In2(Se, Te)3} alloys as
  photovoltaic materials},\ }\href
  {https://doi.org/10.1021/acs.jpclett.2c02975} {\bibfield  {journal} {\bibinfo
   {journal} {The Journal of Physical Chemistry Letters}\ }\textbf {\bibinfo
  {volume} {13}},\ \bibinfo {pages} {12026–12031} (\bibinfo {year}
  {2022})}\BibitemShut {NoStop}%
\bibitem [{\citenamefont {Balogun}\ \emph {et~al.}(2016)\citenamefont
  {Balogun}, \citenamefont {Luo}, \citenamefont {Qiu}, \citenamefont {Liu},\
  and\ \citenamefont {Tong}}]{BalLuoQiu16}%
  \BibitemOpen
  \bibfield  {author} {\bibinfo {author} {\bibfnamefont {M.-S.}\ \bibnamefont
  {Balogun}}, \bibinfo {author} {\bibfnamefont {Y.}~\bibnamefont {Luo}},
  \bibinfo {author} {\bibfnamefont {W.}~\bibnamefont {Qiu}}, \bibinfo {author}
  {\bibfnamefont {P.}~\bibnamefont {Liu}},\ and\ \bibinfo {author}
  {\bibfnamefont {Y.}~\bibnamefont {Tong}},\ }\bibfield  {title} {\bibinfo
  {title} {A review of carbon materials and their composites with alloy metals
  for sodium ion battery anodes},\ }\href
  {https://doi.org/10.1016/j.carbon.2015.09.091} {\bibfield  {journal}
  {\bibinfo  {journal} {Carbon}\ }\textbf {\bibinfo {volume} {98}},\ \bibinfo
  {pages} {162–178} (\bibinfo {year} {2016})}\BibitemShut {NoStop}%
\bibitem [{\citenamefont {Lao}\ \emph {et~al.}(2017)\citenamefont {Lao},
  \citenamefont {Zhang}, \citenamefont {Luo}, \citenamefont {Yan},
  \citenamefont {Sun},\ and\ \citenamefont {Dou}}]{LaoZhaLuo17}%
  \BibitemOpen
  \bibfield  {author} {\bibinfo {author} {\bibfnamefont {M.}~\bibnamefont
  {Lao}}, \bibinfo {author} {\bibfnamefont {Y.}~\bibnamefont {Zhang}}, \bibinfo
  {author} {\bibfnamefont {W.}~\bibnamefont {Luo}}, \bibinfo {author}
  {\bibfnamefont {Q.}~\bibnamefont {Yan}}, \bibinfo {author} {\bibfnamefont
  {W.}~\bibnamefont {Sun}},\ and\ \bibinfo {author} {\bibfnamefont {S.~X.}\
  \bibnamefont {Dou}},\ }\bibfield  {title} {\bibinfo {title} {Alloy‐based
  anode materials toward advanced sodium‐ion batteries},\ }\href
  {https://doi.org/10.1002/adma.201700622} {\bibfield  {journal} {\bibinfo
  {journal} {Advanced Materials}\ }\textbf {\bibinfo {volume} {29}},\ \bibinfo
  {pages} {1700622} (\bibinfo {year} {2017})}\BibitemShut {NoStop}%
\bibitem [{\citenamefont {Song}\ \emph {et~al.}(2019)\citenamefont {Song},
  \citenamefont {Liu}, \citenamefont {Mi}, \citenamefont {Chou}, \citenamefont
  {Chen},\ and\ \citenamefont {Shen}}]{SonLiuMi19}%
  \BibitemOpen
  \bibfield  {author} {\bibinfo {author} {\bibfnamefont {K.}~\bibnamefont
  {Song}}, \bibinfo {author} {\bibfnamefont {C.}~\bibnamefont {Liu}}, \bibinfo
  {author} {\bibfnamefont {L.}~\bibnamefont {Mi}}, \bibinfo {author}
  {\bibfnamefont {S.}~\bibnamefont {Chou}}, \bibinfo {author} {\bibfnamefont
  {W.}~\bibnamefont {Chen}},\ and\ \bibinfo {author} {\bibfnamefont
  {C.}~\bibnamefont {Shen}},\ }\bibfield  {title} {\bibinfo {title} {Recent
  progress on the alloy‐based anode for sodium‐ion batteries and
  potassium‐ion batteries},\ }\href {https://doi.org/10.1002/smll.201903194}
  {\bibfield  {journal} {\bibinfo  {journal} {Small}\ }\textbf {\bibinfo
  {volume} {17}},\ \bibinfo {pages} {1903194} (\bibinfo {year}
  {2019})}\BibitemShut {NoStop}%
\bibitem [{\citenamefont {Singh}\ and\ \citenamefont {Xu}(2013)}]{SinXu13}%
  \BibitemOpen
  \bibfield  {author} {\bibinfo {author} {\bibfnamefont {A.~K.}\ \bibnamefont
  {Singh}}\ and\ \bibinfo {author} {\bibfnamefont {Q.}~\bibnamefont {Xu}},\
  }\bibfield  {title} {\bibinfo {title} {Synergistic catalysis over bimetallic
  alloy nanoparticles},\ }\href {https://doi.org/10.1002/cctc.201200591}
  {\bibfield  {journal} {\bibinfo  {journal} {ChemCatChem}\ }\textbf {\bibinfo
  {volume} {5}},\ \bibinfo {pages} {652–676} (\bibinfo {year}
  {2013})}\BibitemShut {NoStop}%
\bibitem [{\citenamefont {Hannagan}\ \emph {et~al.}(2020)\citenamefont
  {Hannagan}, \citenamefont {Giannakakis}, \citenamefont
  {Flytzani-Stephanopoulos},\ and\ \citenamefont {Sykes}}]{HanGiaFly20}%
  \BibitemOpen
  \bibfield  {author} {\bibinfo {author} {\bibfnamefont {R.~T.}\ \bibnamefont
  {Hannagan}}, \bibinfo {author} {\bibfnamefont {G.}~\bibnamefont
  {Giannakakis}}, \bibinfo {author} {\bibfnamefont {M.}~\bibnamefont
  {Flytzani-Stephanopoulos}},\ and\ \bibinfo {author} {\bibfnamefont
  {E.~C.~H.}\ \bibnamefont {Sykes}},\ }\bibfield  {title} {\bibinfo {title}
  {Single-atom alloy catalysis},\ }\href
  {https://doi.org/10.1021/acs.chemrev.0c00078} {\bibfield  {journal} {\bibinfo
   {journal} {Chemical Reviews}\ }\textbf {\bibinfo {volume} {120}},\ \bibinfo
  {pages} {12044–12088} (\bibinfo {year} {2020})}\BibitemShut {NoStop}%
\bibitem [{\citenamefont {Nakaya}\ and\ \citenamefont
  {Furukawa}(2022)}]{NakFur22}%
  \BibitemOpen
  \bibfield  {author} {\bibinfo {author} {\bibfnamefont {Y.}~\bibnamefont
  {Nakaya}}\ and\ \bibinfo {author} {\bibfnamefont {S.}~\bibnamefont
  {Furukawa}},\ }\bibfield  {title} {\bibinfo {title} {Catalysis of alloys:
  {Classification}, principles, and design for a variety of materials and
  reactions},\ }\href {https://doi.org/10.1021/acs.chemrev.2c00356} {\bibfield
  {journal} {\bibinfo  {journal} {Chemical Reviews}\ }\textbf {\bibinfo
  {volume} {123}},\ \bibinfo {pages} {5859–5947} (\bibinfo {year}
  {2022})}\BibitemShut {NoStop}%
\bibitem [{\citenamefont {Antolini}(2003)}]{Ant03}%
  \BibitemOpen
  \bibfield  {author} {\bibinfo {author} {\bibfnamefont {E.}~\bibnamefont
  {Antolini}},\ }\bibfield  {title} {\bibinfo {title} {Formation of
  carbon-supported {PtM} alloys for low temperature fuel cells: a review},\
  }\href {https://doi.org/10.1016/s0254-0584(02)00389-9} {\bibfield  {journal}
  {\bibinfo  {journal} {Materials Chemistry and Physics}\ }\textbf {\bibinfo
  {volume} {78}},\ \bibinfo {pages} {563–573} (\bibinfo {year}
  {2003})}\BibitemShut {NoStop}%
\bibitem [{\citenamefont {Ren}\ \emph {et~al.}(2020)\citenamefont {Ren},
  \citenamefont {Lv}, \citenamefont {Liu}, \citenamefont {Liu}, \citenamefont
  {Wang}, \citenamefont {Liu},\ and\ \citenamefont {Wu}}]{RenLvLie20}%
  \BibitemOpen
  \bibfield  {author} {\bibinfo {author} {\bibfnamefont {X.}~\bibnamefont
  {Ren}}, \bibinfo {author} {\bibfnamefont {Q.}~\bibnamefont {Lv}}, \bibinfo
  {author} {\bibfnamefont {L.}~\bibnamefont {Liu}}, \bibinfo {author}
  {\bibfnamefont {B.}~\bibnamefont {Liu}}, \bibinfo {author} {\bibfnamefont
  {Y.}~\bibnamefont {Wang}}, \bibinfo {author} {\bibfnamefont {A.}~\bibnamefont
  {Liu}},\ and\ \bibinfo {author} {\bibfnamefont {G.}~\bibnamefont {Wu}},\
  }\bibfield  {title} {\bibinfo {title} {Current progress of {Pt} and
  {Pt}-based electrocatalysts used for fuel cells},\ }\href
  {https://doi.org/10.1039/c9se00460b} {\bibfield  {journal} {\bibinfo
  {journal} {Sustainable Energy \& Fuels}\ }\textbf {\bibinfo {volume} {4}},\
  \bibinfo {pages} {15–30} (\bibinfo {year} {2020})}\BibitemShut {NoStop}%
\bibitem [{\citenamefont {Gong}\ \emph {et~al.}(2020)\citenamefont {Gong},
  \citenamefont {Lyu}, \citenamefont {Palm}, \citenamefont {Memarzadeh},
  \citenamefont {Munday},\ and\ \citenamefont {Leite}}]{GonLyuPal20}%
  \BibitemOpen
  \bibfield  {author} {\bibinfo {author} {\bibfnamefont {T.}~\bibnamefont
  {Gong}}, \bibinfo {author} {\bibfnamefont {P.}~\bibnamefont {Lyu}}, \bibinfo
  {author} {\bibfnamefont {K.~J.}\ \bibnamefont {Palm}}, \bibinfo {author}
  {\bibfnamefont {S.}~\bibnamefont {Memarzadeh}}, \bibinfo {author}
  {\bibfnamefont {J.~N.}\ \bibnamefont {Munday}},\ and\ \bibinfo {author}
  {\bibfnamefont {M.~S.}\ \bibnamefont {Leite}},\ }\bibfield  {title} {\bibinfo
  {title} {Emergent opportunities with metallic alloys: {From} material design
  to optical devices},\ }\href {https://doi.org/10.1002/adom.202001082}
  {\bibfield  {journal} {\bibinfo  {journal} {Advanced Optical Materials}\
  }\textbf {\bibinfo {volume} {8}},\ \bibinfo {pages} {2001082} (\bibinfo
  {year} {2020})}\BibitemShut {NoStop}%
\bibitem [{\citenamefont {Darmadi}\ \emph {et~al.}(2021)\citenamefont
  {Darmadi}, \citenamefont {Khairunnisa}, \citenamefont {Tomeček},\ and\
  \citenamefont {Langhammer}}]{DarKhaTom21}%
  \BibitemOpen
  \bibfield  {author} {\bibinfo {author} {\bibfnamefont {I.}~\bibnamefont
  {Darmadi}}, \bibinfo {author} {\bibfnamefont {S.~Z.}\ \bibnamefont
  {Khairunnisa}}, \bibinfo {author} {\bibfnamefont {D.}~\bibnamefont
  {Tomeček}},\ and\ \bibinfo {author} {\bibfnamefont {C.}~\bibnamefont
  {Langhammer}},\ }\bibfield  {title} {\bibinfo {title} {Optimization of the
  composition of {PdAuCu} ternary alloy nanoparticles for plasmonic hydrogen
  sensing},\ }\href {https://doi.org/10.1021/acsanm.1c01242} {\bibfield
  {journal} {\bibinfo  {journal} {ACS Applied Nano Materials}\ }\textbf
  {\bibinfo {volume} {4}},\ \bibinfo {pages} {8716–8722} (\bibinfo {year}
  {2021})}\BibitemShut {NoStop}%
\bibitem [{\citenamefont {Dursun}\ and\ \citenamefont
  {Soutis}(2014)}]{DurSou14}%
  \BibitemOpen
  \bibfield  {author} {\bibinfo {author} {\bibfnamefont {T.}~\bibnamefont
  {Dursun}}\ and\ \bibinfo {author} {\bibfnamefont {C.}~\bibnamefont
  {Soutis}},\ }\bibfield  {title} {\bibinfo {title} {Recent developments in
  advanced aircraft aluminium alloys},\ }\href
  {https://doi.org/10.1016/j.matdes.2013.12.002} {\bibfield  {journal}
  {\bibinfo  {journal} {Materials \& Design (1980-2015)}\ }\textbf {\bibinfo
  {volume} {56}},\ \bibinfo {pages} {862–871} (\bibinfo {year}
  {2014})}\BibitemShut {NoStop}%
\bibitem [{\citenamefont {García}\ \emph {et~al.}(2019)\citenamefont
  {García}, \citenamefont {Collado~Ciprés}, \citenamefont {Blomqvist},\ and\
  \citenamefont {Kaplan}}]{GarColBlo19}%
  \BibitemOpen
  \bibfield  {author} {\bibinfo {author} {\bibfnamefont {J.}~\bibnamefont
  {García}}, \bibinfo {author} {\bibfnamefont {V.}~\bibnamefont
  {Collado~Ciprés}}, \bibinfo {author} {\bibfnamefont {A.}~\bibnamefont
  {Blomqvist}},\ and\ \bibinfo {author} {\bibfnamefont {B.}~\bibnamefont
  {Kaplan}},\ }\bibfield  {title} {\bibinfo {title} {Cemented carbide
  microstructures: a review},\ }\href
  {https://doi.org/10.1016/j.ijrmhm.2018.12.004} {\bibfield  {journal}
  {\bibinfo  {journal} {International Journal of Refractory Metals and Hard
  Materials}\ }\textbf {\bibinfo {volume} {80}},\ \bibinfo {pages} {40–68}
  (\bibinfo {year} {2019})}\BibitemShut {NoStop}%
\bibitem [{\citenamefont {Torralba}\ and\ \citenamefont
  {Campos}(2020)}]{TorCam20}%
  \BibitemOpen
  \bibfield  {author} {\bibinfo {author} {\bibfnamefont {J.~M.}\ \bibnamefont
  {Torralba}}\ and\ \bibinfo {author} {\bibfnamefont {M.}~\bibnamefont
  {Campos}},\ }\bibfield  {title} {\bibinfo {title} {High entropy alloys
  manufactured by additive manufacturing},\ }\href
  {https://doi.org/10.3390/met10050639} {\bibfield  {journal} {\bibinfo
  {journal} {Metals}\ }\textbf {\bibinfo {volume} {10}},\ \bibinfo {pages}
  {639} (\bibinfo {year} {2020})}\BibitemShut {NoStop}%
\bibitem [{\citenamefont {Zhang}\ \emph {et~al.}(2022)\citenamefont {Zhang},
  \citenamefont {Zhang}, \citenamefont {Wu}, \citenamefont {Ji},\ and\
  \citenamefont {Kumar}}]{ZhaZhaWu22}%
  \BibitemOpen
  \bibfield  {author} {\bibinfo {author} {\bibfnamefont {Z.-X.}\ \bibnamefont
  {Zhang}}, \bibinfo {author} {\bibfnamefont {J.}~\bibnamefont {Zhang}},
  \bibinfo {author} {\bibfnamefont {H.}~\bibnamefont {Wu}}, \bibinfo {author}
  {\bibfnamefont {Y.}~\bibnamefont {Ji}},\ and\ \bibinfo {author}
  {\bibfnamefont {D.~D.}\ \bibnamefont {Kumar}},\ }\bibfield  {title} {\bibinfo
  {title} {Iron-based shape memory alloys in construction: Research,
  applications and opportunities},\ }\href {https://doi.org/10.3390/ma15051723}
  {\bibfield  {journal} {\bibinfo  {journal} {Materials}\ }\textbf {\bibinfo
  {volume} {15}},\ \bibinfo {pages} {1723} (\bibinfo {year}
  {2022})}\BibitemShut {NoStop}%
\bibitem [{\citenamefont {Anasori}\ \emph {et~al.}(2017)\citenamefont
  {Anasori}, \citenamefont {Lukatskaya},\ and\ \citenamefont
  {Gogotsi}}]{AnaLukGog17}%
  \BibitemOpen
  \bibfield  {author} {\bibinfo {author} {\bibfnamefont {B.}~\bibnamefont
  {Anasori}}, \bibinfo {author} {\bibfnamefont {M.~R.}\ \bibnamefont
  {Lukatskaya}},\ and\ \bibinfo {author} {\bibfnamefont {Y.}~\bibnamefont
  {Gogotsi}},\ }\bibfield  {title} {\bibinfo {title} {{2D} metal carbides and
  nitrides ({MXenes}) for energy storage},\ }\href
  {https://doi.org/10.1038/natrevmats.2016.98} {\bibfield  {journal} {\bibinfo
  {journal} {Nature Reviews Materials}\ }\textbf {\bibinfo {volume} {2}},\
  \bibinfo {pages} {16098} (\bibinfo {year} {2017})}\BibitemShut {NoStop}%
\bibitem [{\citenamefont {Tang}\ \emph {et~al.}(2018)\citenamefont {Tang},
  \citenamefont {Guo}, \citenamefont {Wu},\ and\ \citenamefont
  {Wang}}]{TanGuoWu18}%
  \BibitemOpen
  \bibfield  {author} {\bibinfo {author} {\bibfnamefont {X.}~\bibnamefont
  {Tang}}, \bibinfo {author} {\bibfnamefont {X.}~\bibnamefont {Guo}}, \bibinfo
  {author} {\bibfnamefont {W.}~\bibnamefont {Wu}},\ and\ \bibinfo {author}
  {\bibfnamefont {G.}~\bibnamefont {Wang}},\ }\bibfield  {title} {\bibinfo
  {title} {{2D} metal carbides and nitrides ({MX}enes) as high‐performance
  electrode materials for lithium‐based batteries},\ }\href
  {https://doi.org/10.1002/aenm.201801897} {\bibfield  {journal} {\bibinfo
  {journal} {Advanced Energy Materials}\ }\textbf {\bibinfo {volume} {8}},\
  \bibinfo {pages} {1801897} (\bibinfo {year} {2018})}\BibitemShut {NoStop}%
\bibitem [{\citenamefont {Gogotsi}\ and\ \citenamefont
  {Anasori}(2019)}]{GogAna19}%
  \BibitemOpen
  \bibfield  {author} {\bibinfo {author} {\bibfnamefont {Y.}~\bibnamefont
  {Gogotsi}}\ and\ \bibinfo {author} {\bibfnamefont {B.}~\bibnamefont
  {Anasori}},\ }\bibfield  {title} {\bibinfo {title} {The rise of {MX}enes},\
  }\href {https://doi.org/10.1021/acsnano.9b06394} {\bibfield  {journal}
  {\bibinfo  {journal} {ACS Nano}\ }\textbf {\bibinfo {volume} {13}},\ \bibinfo
  {pages} {8491–8494} (\bibinfo {year} {2019})}\BibitemShut {NoStop}%
\bibitem [{\citenamefont {Yin}\ \emph {et~al.}(2021)\citenamefont {Yin},
  \citenamefont {Li}, \citenamefont {Yao}, \citenamefont {Wang}, \citenamefont
  {Jia}, \citenamefont {Liu}, \citenamefont {Li}, \citenamefont {Li},\ and\
  \citenamefont {He}}]{YinLiYao21}%
  \BibitemOpen
  \bibfield  {author} {\bibinfo {author} {\bibfnamefont {L.}~\bibnamefont
  {Yin}}, \bibinfo {author} {\bibfnamefont {Y.}~\bibnamefont {Li}}, \bibinfo
  {author} {\bibfnamefont {X.}~\bibnamefont {Yao}}, \bibinfo {author}
  {\bibfnamefont {Y.}~\bibnamefont {Wang}}, \bibinfo {author} {\bibfnamefont
  {L.}~\bibnamefont {Jia}}, \bibinfo {author} {\bibfnamefont {Q.}~\bibnamefont
  {Liu}}, \bibinfo {author} {\bibfnamefont {J.}~\bibnamefont {Li}}, \bibinfo
  {author} {\bibfnamefont {Y.}~\bibnamefont {Li}},\ and\ \bibinfo {author}
  {\bibfnamefont {D.}~\bibnamefont {He}},\ }\bibfield  {title} {\bibinfo
  {title} {{MX}enes for solar cells},\ }\href
  {https://doi.org/10.1007/s40820-021-00604-8} {\bibfield  {journal} {\bibinfo
  {journal} {Nano-Micro Letters}\ }\textbf {\bibinfo {volume} {13}},\ \bibinfo
  {pages} {78} (\bibinfo {year} {2021})}\BibitemShut {NoStop}%
\bibitem [{\citenamefont {Asta}\ \emph {et~al.}(1993)\citenamefont {Asta},
  \citenamefont {McCormack},\ and\ \citenamefont {de~Fontaine}}]{AstMcCFon93}%
  \BibitemOpen
  \bibfield  {author} {\bibinfo {author} {\bibfnamefont {M.}~\bibnamefont
  {Asta}}, \bibinfo {author} {\bibfnamefont {R.}~\bibnamefont {McCormack}},\
  and\ \bibinfo {author} {\bibfnamefont {D.}~\bibnamefont {de~Fontaine}},\
  }\bibfield  {title} {\bibinfo {title} {Theoretical study of alloy phase
  stability in the {Cd-Mg} system},\ }\href
  {https://doi.org/10.1103/PhysRevB.48.748} {\bibfield  {journal} {\bibinfo
  {journal} {Physical Review B}\ }\textbf {\bibinfo {volume} {48}},\ \bibinfo
  {pages} {748} (\bibinfo {year} {1993})}\BibitemShut {NoStop}%
\bibitem [{\citenamefont {Ozoli\c{n}\v{s}}\ \emph {et~al.}(1998)\citenamefont
  {Ozoli\c{n}\v{s}}, \citenamefont {Wolverton},\ and\ \citenamefont
  {Zunger}}]{OzoWolZun98a}%
  \BibitemOpen
  \bibfield  {author} {\bibinfo {author} {\bibfnamefont {V.}~\bibnamefont
  {Ozoli\c{n}\v{s}}}, \bibinfo {author} {\bibfnamefont {C.}~\bibnamefont
  {Wolverton}},\ and\ \bibinfo {author} {\bibfnamefont {A.}~\bibnamefont
  {Zunger}},\ }\bibfield  {title} {\bibinfo {title} {\ce{Cu-Au}, \ce{Ag-Au},
  \ce{Cu-Ag}, and \ce{Ni-Au} intermetallics: {First}-principles study of
  temperature-composition phase diagrams and structures},\ }\href
  {https://doi.org/10.1103/PhysRevB.57.6427} {\bibfield  {journal} {\bibinfo
  {journal} {Physical Review B}\ }\textbf {\bibinfo {volume} {57}},\ \bibinfo
  {pages} {6427} (\bibinfo {year} {1998})}\BibitemShut {NoStop}%
\bibitem [{\citenamefont {Meng}\ and\ \citenamefont {Arroyo-de
  Dompablo}(2009)}]{Meng2009}%
  \BibitemOpen
  \bibfield  {author} {\bibinfo {author} {\bibfnamefont {Y.~S.}\ \bibnamefont
  {Meng}}\ and\ \bibinfo {author} {\bibfnamefont {M.~E.}\ \bibnamefont
  {Arroyo-de Dompablo}},\ }\bibfield  {title} {\bibinfo {title} {First
  principles computational materials design for energy storage materials in
  lithium ion batteries},\ }\href {https://doi.org/10.1039/B901825E} {\bibfield
   {journal} {\bibinfo  {journal} {Energy \& Environmental Sciences}\ }\textbf
  {\bibinfo {volume} {2}},\ \bibinfo {pages} {589} (\bibinfo {year}
  {2009})}\BibitemShut {NoStop}%
\bibitem [{\citenamefont {Rahm}\ \emph {et~al.}(2021)\citenamefont {Rahm},
  \citenamefont {Löfgren}, \citenamefont {Fransson},\ and\ \citenamefont
  {Erhart}}]{RahLöfFraErh21}%
  \BibitemOpen
  \bibfield  {author} {\bibinfo {author} {\bibfnamefont {J.~M.}\ \bibnamefont
  {Rahm}}, \bibinfo {author} {\bibfnamefont {J.}~\bibnamefont {Löfgren}},
  \bibinfo {author} {\bibfnamefont {E.}~\bibnamefont {Fransson}},\ and\
  \bibinfo {author} {\bibfnamefont {P.}~\bibnamefont {Erhart}},\ }\bibfield
  {title} {\bibinfo {title} {A tale of two phase diagrams: {Interplay} of
  ordering and hydrogen uptake in {Pd–Au–H}},\ }\href
  {https://doi.org/10.1016/j.actamat.2021.116893} {\bibfield  {journal}
  {\bibinfo  {journal} {Acta Materialia}\ }\textbf {\bibinfo {volume} {211}},\
  \bibinfo {pages} {116893} (\bibinfo {year} {2021})}\BibitemShut {NoStop}%
\bibitem [{\citenamefont {Gren}\ \emph {et~al.}(2021)\citenamefont {Gren},
  \citenamefont {Fransson}, \citenamefont {\AA{}ngqvist}, \citenamefont
  {Erhart},\ and\ \citenamefont {Wahnstr\"om}}]{GreFraAngErhWah2021}%
  \BibitemOpen
  \bibfield  {author} {\bibinfo {author} {\bibfnamefont {M.}~\bibnamefont
  {Gren}}, \bibinfo {author} {\bibfnamefont {E.}~\bibnamefont {Fransson}},
  \bibinfo {author} {\bibfnamefont {M.}~\bibnamefont {\AA{}ngqvist}}, \bibinfo
  {author} {\bibfnamefont {P.}~\bibnamefont {Erhart}},\ and\ \bibinfo {author}
  {\bibfnamefont {G.}~\bibnamefont {Wahnstr\"om}},\ }\bibfield  {title}
  {\bibinfo {title} {Modeling of vibrational and configurational degrees of
  freedom in hexagonal and cubic tungsten carbide at high temperatures},\
  }\href {https://doi.org/10.1103/PhysRevMaterials.5.033804} {\bibfield
  {journal} {\bibinfo  {journal} {Physical Review Materials}\ }\textbf
  {\bibinfo {volume} {5}},\ \bibinfo {pages} {033804} (\bibinfo {year}
  {2021})}\BibitemShut {NoStop}%
\bibitem [{\citenamefont {Van~der Ven}\ \emph {et~al.}(1998)\citenamefont
  {Van~der Ven}, \citenamefont {Aydinol}, \citenamefont {Ceder}, \citenamefont
  {Kresse},\ and\ \citenamefont {Hafner}}]{VanAydCed98}%
  \BibitemOpen
  \bibfield  {author} {\bibinfo {author} {\bibfnamefont {A.}~\bibnamefont
  {Van~der Ven}}, \bibinfo {author} {\bibfnamefont {M.~K.}\ \bibnamefont
  {Aydinol}}, \bibinfo {author} {\bibfnamefont {G.}~\bibnamefont {Ceder}},
  \bibinfo {author} {\bibfnamefont {G.}~\bibnamefont {Kresse}},\ and\ \bibinfo
  {author} {\bibfnamefont {J.}~\bibnamefont {Hafner}},\ }\bibfield  {title}
  {\bibinfo {title} {First-principles investigation of phase stability in
  \ce{Li_xCoO2}},\ }\href {https://doi.org/10.1103/PhysRevB.58.2975} {\bibfield
   {journal} {\bibinfo  {journal} {Physical Review B}\ }\textbf {\bibinfo
  {volume} {58}},\ \bibinfo {pages} {2975} (\bibinfo {year}
  {1998})}\BibitemShut {NoStop}%
\bibitem [{\citenamefont {Ceder}\ \emph {et~al.}(2000)\citenamefont {Ceder},
  \citenamefont {Van~der Ven}, \citenamefont {Marianetti},\ and\ \citenamefont
  {Morgan}}]{CedVenMar00}%
  \BibitemOpen
  \bibfield  {author} {\bibinfo {author} {\bibfnamefont {G.}~\bibnamefont
  {Ceder}}, \bibinfo {author} {\bibfnamefont {A.}~\bibnamefont {Van~der Ven}},
  \bibinfo {author} {\bibfnamefont {C.}~\bibnamefont {Marianetti}},\ and\
  \bibinfo {author} {\bibfnamefont {D.}~\bibnamefont {Morgan}},\ }\bibfield
  {title} {\bibinfo {title} {First-principles alloy theory in oxides},\ }\href
  {https://doi.org/10.1088/0965-0393/8/3/311} {\bibfield  {journal} {\bibinfo
  {journal} {Modelling and Simulation in Materials Science and Engineering}\
  }\textbf {\bibinfo {volume} {8}},\ \bibinfo {pages} {311} (\bibinfo {year}
  {2000})}\BibitemShut {NoStop}%
\bibitem [{\citenamefont {Zhou}\ \emph {et~al.}(2006)\citenamefont {Zhou},
  \citenamefont {Maxisch},\ and\ \citenamefont {Ceder}}]{ZhoMaxCed06}%
  \BibitemOpen
  \bibfield  {author} {\bibinfo {author} {\bibfnamefont {F.}~\bibnamefont
  {Zhou}}, \bibinfo {author} {\bibfnamefont {T.}~\bibnamefont {Maxisch}},\ and\
  \bibinfo {author} {\bibfnamefont {G.}~\bibnamefont {Ceder}},\ }\bibfield
  {title} {\bibinfo {title} {Configurational electronic entropy and the phase
  diagram of mixed-valence oxides: {The} case of \ce{Li_xFeP4}},\ }\href
  {https://doi.org/10.1103/PhysRevLett.97.155704} {\bibfield  {journal}
  {\bibinfo  {journal} {Physical Review Letters}\ }\textbf {\bibinfo {volume}
  {97}},\ \bibinfo {pages} {155704} (\bibinfo {year} {2006})}\BibitemShut
  {NoStop}%
\bibitem [{\citenamefont {Bechtel}\ and\ \citenamefont {Van~der
  Ven}(2018)}]{BecVan18}%
  \BibitemOpen
  \bibfield  {author} {\bibinfo {author} {\bibfnamefont {J.~S.}\ \bibnamefont
  {Bechtel}}\ and\ \bibinfo {author} {\bibfnamefont {A.}~\bibnamefont {Van~der
  Ven}},\ }\bibfield  {title} {\bibinfo {title} {First-principles
  thermodynamics study of phase stability in inorganic halide perovskite solid
  solutions},\ }\bibfield  {journal} {\bibinfo  {journal} {Physical Review
  Materials}\ }\textbf {\bibinfo {volume} {2}},\ \href
  {https://doi.org/10.1103/physrevmaterials.2.045401}
  {10.1103/physrevmaterials.2.045401} (\bibinfo {year} {2018})\BibitemShut
  {NoStop}%
\bibitem [{\citenamefont {Linder{\"a}lv}\ \emph {et~al.}(2022)\citenamefont
  {Linder{\"a}lv}, \citenamefont {Rahm},\ and\ \citenamefont
  {Erhart}}]{LinRahErh22}%
  \BibitemOpen
  \bibfield  {author} {\bibinfo {author} {\bibfnamefont {C.}~\bibnamefont
  {Linder{\"a}lv}}, \bibinfo {author} {\bibfnamefont {J.~M.}\ \bibnamefont
  {Rahm}},\ and\ \bibinfo {author} {\bibfnamefont {P.}~\bibnamefont {Erhart}},\
  }\bibfield  {title} {\bibinfo {title} {High-throughput characterization of
  transition metal dichalcogenide alloys: Thermodynamic stability and
  electronic band alignment},\ }\href
  {https://doi.org/10.1021/acs.chemmater.2c01176} {\bibfield  {journal}
  {\bibinfo  {journal} {Chemistry of Materials}\ }\textbf {\bibinfo {volume}
  {34}},\ \bibinfo {pages} {9364} (\bibinfo {year} {2022})}\BibitemShut
  {NoStop}%
\bibitem [{\citenamefont {Kim}\ \emph {et~al.}(2010)\citenamefont {Kim},
  \citenamefont {Kaviany}, \citenamefont {Thomas}, \citenamefont {Van~der Ven},
  \citenamefont {Uher},\ and\ \citenamefont {Huang}}]{KimKavTho10}%
  \BibitemOpen
  \bibfield  {author} {\bibinfo {author} {\bibfnamefont {H.}~\bibnamefont
  {Kim}}, \bibinfo {author} {\bibfnamefont {M.}~\bibnamefont {Kaviany}},
  \bibinfo {author} {\bibfnamefont {J.~C.}\ \bibnamefont {Thomas}}, \bibinfo
  {author} {\bibfnamefont {A.}~\bibnamefont {Van~der Ven}}, \bibinfo {author}
  {\bibfnamefont {C.}~\bibnamefont {Uher}},\ and\ \bibinfo {author}
  {\bibfnamefont {B.}~\bibnamefont {Huang}},\ }\bibfield  {title} {\bibinfo
  {title} {Structural order-disorder transitions and phonon conductivity of
  partially filled skutterudites},\ }\href
  {https://doi.org/10.1103/PhysRevLett.105.265901} {\bibfield  {journal}
  {\bibinfo  {journal} {Physical Review Letters}\ }\textbf {\bibinfo {volume}
  {105}},\ \bibinfo {pages} {265901} (\bibinfo {year} {2010})}\BibitemShut
  {NoStop}%
\bibitem [{\citenamefont {Burton}\ \emph {et~al.}(2011)\citenamefont {Burton},
  \citenamefont {van~de Walle},\ and\ \citenamefont {Stokes}}]{BurWalSto11}%
  \BibitemOpen
  \bibfield  {author} {\bibinfo {author} {\bibfnamefont {B.~P.}\ \bibnamefont
  {Burton}}, \bibinfo {author} {\bibfnamefont {A.}~\bibnamefont {van~de
  Walle}},\ and\ \bibinfo {author} {\bibfnamefont {H.~T.}\ \bibnamefont
  {Stokes}},\ }\bibfield  {title} {\bibinfo {title} {First principles phase
  diagram calculations for the octahedral-interstitial system \ce{ZrO_x}, $0
  \leq x \leq 1/2$},\ }\href {https://doi.org/10.1143/JPSJ.81.014004}
  {\bibfield  {journal} {\bibinfo  {journal} {Journal of the Physical Society
  of Japan}\ }\textbf {\bibinfo {volume} {81}},\ \bibinfo {pages} {014004}
  (\bibinfo {year} {2011})}\BibitemShut {NoStop}%
\bibitem [{\citenamefont {Burton}\ and\ \citenamefont {van~de
  Walle}(2012)}]{BurWal12}%
  \BibitemOpen
  \bibfield  {author} {\bibinfo {author} {\bibfnamefont {B.~P.}\ \bibnamefont
  {Burton}}\ and\ \bibinfo {author} {\bibfnamefont {A.}~\bibnamefont {van~de
  Walle}},\ }\bibfield  {title} {\bibinfo {title} {First principles phase
  diagram calculations for the {Octahedral-Interstitial} system \ce{ZrO_x}, $0
  \leq x \leq 1/2$},\ }\href {https://doi.org/10.1016/j.calphad.2011.12.011}
  {\bibfield  {journal} {\bibinfo  {journal} {Calphad}\ }\textbf {\bibinfo
  {volume} {37}},\ \bibinfo {pages} {151} (\bibinfo {year} {2012})}\BibitemShut
  {NoStop}%
\bibitem [{\citenamefont {{\AA}ngqvist}\ \emph {et~al.}(2016)\citenamefont
  {{\AA}ngqvist}, \citenamefont {Lindroth},\ and\ \citenamefont
  {Erhart}}]{AngLinErh16}%
  \BibitemOpen
  \bibfield  {author} {\bibinfo {author} {\bibfnamefont {M.}~\bibnamefont
  {{\AA}ngqvist}}, \bibinfo {author} {\bibfnamefont {D.~O.}\ \bibnamefont
  {Lindroth}},\ and\ \bibinfo {author} {\bibfnamefont {P.}~\bibnamefont
  {Erhart}},\ }\bibfield  {title} {\bibinfo {title} {Optimization of the
  thermoelectric power factor: {Coupling} between chemical order and transport
  properties},\ }\href {https://doi.org/10.1021/acs.chemmater.6b02117}
  {\bibfield  {journal} {\bibinfo  {journal} {Chemistry of Materials}\ }\textbf
  {\bibinfo {volume} {28}},\ \bibinfo {pages} {6877} (\bibinfo {year}
  {2016})}\BibitemShut {NoStop}%
\bibitem [{\citenamefont {{\AA}ngqvist}\ and\ \citenamefont
  {Erhart}(2017)}]{AngErh17}%
  \BibitemOpen
  \bibfield  {author} {\bibinfo {author} {\bibfnamefont {M.}~\bibnamefont
  {{\AA}ngqvist}}\ and\ \bibinfo {author} {\bibfnamefont {P.}~\bibnamefont
  {Erhart}},\ }\bibfield  {title} {\bibinfo {title} {Understanding chemical
  ordering in intermetallic clathrates from atomic scale simulations},\ }\href
  {https://doi.org/10.1021/acs.chemmater.7b02686} {\bibfield  {journal}
  {\bibinfo  {journal} {Chemistry of Materials}\ }\textbf {\bibinfo {volume}
  {29}},\ \bibinfo {pages} {7554} (\bibinfo {year} {2017})}\BibitemShut
  {NoStop}%
\bibitem [{\citenamefont {Troppenz}\ \emph {et~al.}(2017)\citenamefont
  {Troppenz}, \citenamefont {Rigamonti},\ and\ \citenamefont
  {Draxl}}]{TroRigDra17}%
  \BibitemOpen
  \bibfield  {author} {\bibinfo {author} {\bibfnamefont {M.}~\bibnamefont
  {Troppenz}}, \bibinfo {author} {\bibfnamefont {S.}~\bibnamefont
  {Rigamonti}},\ and\ \bibinfo {author} {\bibfnamefont {C.}~\bibnamefont
  {Draxl}},\ }\bibfield  {title} {\bibinfo {title} {Predicting ground-state
  configurations and electronic properties of the thermoelectric clathrates
  \ce{Ba8Al_xSi_{46-x}} and \ce{Sr8Al_xSi_{46-x}}},\ }\href
  {https://doi.org/10.1021/acs.chemmater.6b05027} {\bibfield  {journal}
  {\bibinfo  {journal} {Chemistry of Materials}\ }\textbf {\bibinfo {volume}
  {29}},\ \bibinfo {pages} {2414} (\bibinfo {year} {2017})}\BibitemShut
  {NoStop}%
\bibitem [{\citenamefont {Gunda}\ \emph {et~al.}(2018)\citenamefont {Gunda},
  \citenamefont {Puchala},\ and\ \citenamefont {Van~der Ven}}]{GunPucVan18}%
  \BibitemOpen
  \bibfield  {author} {\bibinfo {author} {\bibfnamefont {N.~S.~H.}\
  \bibnamefont {Gunda}}, \bibinfo {author} {\bibfnamefont {B.}~\bibnamefont
  {Puchala}},\ and\ \bibinfo {author} {\bibfnamefont {A.}~\bibnamefont {Van~der
  Ven}},\ }\bibfield  {title} {\bibinfo {title} {Resolving phase stability in
  the {Ti}-{O} binary with first-principles statistical mechanics methods},\
  }\href {https://doi.org/10.1103/PhysRevMaterials.2.033604} {\bibfield
  {journal} {\bibinfo  {journal} {Physical Review Materials}\ }\textbf
  {\bibinfo {volume} {2}},\ \bibinfo {pages} {033604} (\bibinfo {year}
  {2018})}\BibitemShut {NoStop}%
\bibitem [{\citenamefont {Drautz}\ \emph {et~al.}(2001)\citenamefont {Drautz},
  \citenamefont {Reichert}, \citenamefont {F\"ahnle}, \citenamefont {Dosch},\
  and\ \citenamefont {Sanchez}}]{DraReiFah01}%
  \BibitemOpen
  \bibfield  {author} {\bibinfo {author} {\bibfnamefont {R.}~\bibnamefont
  {Drautz}}, \bibinfo {author} {\bibfnamefont {H.}~\bibnamefont {Reichert}},
  \bibinfo {author} {\bibfnamefont {M.}~\bibnamefont {F\"ahnle}}, \bibinfo
  {author} {\bibfnamefont {H.}~\bibnamefont {Dosch}},\ and\ \bibinfo {author}
  {\bibfnamefont {J.~M.}\ \bibnamefont {Sanchez}},\ }\bibfield  {title}
  {\bibinfo {title} {Spontaneous {L1$_2$} order at \ce{Ni_{90}Al_{10}}(110)
  surfaces: {An} x-ray and first-principles-calculation study},\ }\href
  {https://doi.org/10.1103/PhysRevLett.87.236102} {\bibfield  {journal}
  {\bibinfo  {journal} {Physical Review Letters}\ }\textbf {\bibinfo {volume}
  {87}},\ \bibinfo {pages} {236102} (\bibinfo {year} {2001})}\BibitemShut
  {NoStop}%
\bibitem [{\citenamefont {Sluiter}\ and\ \citenamefont
  {Kawazoe}(2003)}]{SluKaw03}%
  \BibitemOpen
  \bibfield  {author} {\bibinfo {author} {\bibfnamefont {M.~H.~F.}\
  \bibnamefont {Sluiter}}\ and\ \bibinfo {author} {\bibfnamefont
  {Y.}~\bibnamefont {Kawazoe}},\ }\bibfield  {title} {\bibinfo {title} {Cluster
  expansion method for adsorption: {Application} to hydrogen chemisorption on
  graphene},\ }\href {https://doi.org/10.1103/PhysRevB.68.085410} {\bibfield
  {journal} {\bibinfo  {journal} {Physical Review B}\ }\textbf {\bibinfo
  {volume} {68}},\ \bibinfo {pages} {085410} (\bibinfo {year}
  {2003})}\BibitemShut {NoStop}%
\bibitem [{\citenamefont {Welker}\ \emph {et~al.}(2010)\citenamefont {Welker},
  \citenamefont {Wieckhorst}, \citenamefont {Kerscher},\ and\ \citenamefont
  {M\"uller}}]{WelWieKer10}%
  \BibitemOpen
  \bibfield  {author} {\bibinfo {author} {\bibfnamefont {P.}~\bibnamefont
  {Welker}}, \bibinfo {author} {\bibfnamefont {O.}~\bibnamefont {Wieckhorst}},
  \bibinfo {author} {\bibfnamefont {T.~C.}\ \bibnamefont {Kerscher}},\ and\
  \bibinfo {author} {\bibfnamefont {S.}~\bibnamefont {M\"uller}},\ }\bibfield
  {title} {\bibinfo {title} {Predicting the segregation profile of the
  \ce{Pt_{25}Rh_{75}} (100) surface from first-principles},\ }\href
  {https://doi.org/10.1088/0953-8984/22/38/384203} {\bibfield  {journal}
  {\bibinfo  {journal} {Journal of Physics: Condensed Matter}\ }\textbf
  {\bibinfo {volume} {22}},\ \bibinfo {pages} {384203} (\bibinfo {year}
  {2010})}\BibitemShut {NoStop}%
\bibitem [{\citenamefont {Stephens}\ \emph {et~al.}(2010)\citenamefont
  {Stephens}, \citenamefont {Ham},\ and\ \citenamefont {Hwang}}]{SteHamHwa10}%
  \BibitemOpen
  \bibfield  {author} {\bibinfo {author} {\bibfnamefont {J.~A.}\ \bibnamefont
  {Stephens}}, \bibinfo {author} {\bibfnamefont {H.~C.}\ \bibnamefont {Ham}},\
  and\ \bibinfo {author} {\bibfnamefont {G.~S.}\ \bibnamefont {Hwang}},\
  }\bibfield  {title} {\bibinfo {title} {Atomic arrangements of {AuPt/Pt(111)}
  and {AuPd/Pd(111)} surface alloys: {A} combined density functional theory and
  {Monte Carlo} study},\ }\href {https://doi.org/10.1021/jp1074384} {\bibfield
  {journal} {\bibinfo  {journal} {Journal of Physical Chemistry C}\ }\textbf
  {\bibinfo {volume} {114}},\ \bibinfo {pages} {21516} (\bibinfo {year}
  {2010})}\BibitemShut {NoStop}%
\bibitem [{\citenamefont {Chen}\ \emph {et~al.}(2011)\citenamefont {Chen},
  \citenamefont {Schmidt}, \citenamefont {Schneider},\ and\ \citenamefont
  {Wolverton}}]{CheSchSch11}%
  \BibitemOpen
  \bibfield  {author} {\bibinfo {author} {\bibfnamefont {W.}~\bibnamefont
  {Chen}}, \bibinfo {author} {\bibfnamefont {D.}~\bibnamefont {Schmidt}},
  \bibinfo {author} {\bibfnamefont {W.~F.}\ \bibnamefont {Schneider}},\ and\
  \bibinfo {author} {\bibfnamefont {C.}~\bibnamefont {Wolverton}},\ }\bibfield
  {title} {\bibinfo {title} {Ordering and oxygen adsorption in
  {Au–Pt/Pt(111)} surface alloys},\ }\href
  {https://doi.org/10.1021/jp205995j} {\bibfield  {journal} {\bibinfo
  {journal} {The Journal of Physical Chemistry C}\ }\textbf {\bibinfo {volume}
  {115}},\ \bibinfo {pages} {17915} (\bibinfo {year} {2011})}\BibitemShut
  {NoStop}%
\bibitem [{\citenamefont {Cao}\ and\ \citenamefont {Mueller}(2015)}]{CaoMue15}%
  \BibitemOpen
  \bibfield  {author} {\bibinfo {author} {\bibfnamefont {L.}~\bibnamefont
  {Cao}}\ and\ \bibinfo {author} {\bibfnamefont {T.}~\bibnamefont {Mueller}},\
  }\bibfield  {title} {\bibinfo {title} {Rational design of \ce{Pt3Ni} surface
  structures for the oxygen reduction reaction},\ }\href
  {https://doi.org/10.1021/acs.jpcc.5b04951} {\bibfield  {journal} {\bibinfo
  {journal} {Journal of Physical Chemistry C}\ }\textbf {\bibinfo {volume}
  {119}},\ \bibinfo {pages} {17735} (\bibinfo {year} {2015})}\BibitemShut
  {NoStop}%
\bibitem [{\citenamefont {Herder}\ \emph {et~al.}(2015)\citenamefont {Herder},
  \citenamefont {Bray},\ and\ \citenamefont {Schneider}}]{HerBraSch15}%
  \BibitemOpen
  \bibfield  {author} {\bibinfo {author} {\bibfnamefont {L.~M.}\ \bibnamefont
  {Herder}}, \bibinfo {author} {\bibfnamefont {J.~M.}\ \bibnamefont {Bray}},\
  and\ \bibinfo {author} {\bibfnamefont {W.~F.}\ \bibnamefont {Schneider}},\
  }\bibfield  {title} {\bibinfo {title} {Comparison of cluster expansion
  fitting algorithms for interactions at surfaces},\ }\href
  {https://doi.org/10.1016/j.susc.2015.02.017} {\bibfield  {journal} {\bibinfo
  {journal} {Surface Science}\ }\textbf {\bibinfo {volume} {640}},\ \bibinfo
  {pages} {104} (\bibinfo {year} {2015})}\BibitemShut {NoStop}%
\bibitem [{\citenamefont {Fransson}\ \emph
  {et~al.}(2021{\natexlab{a}})\citenamefont {Fransson}, \citenamefont {Gren},\
  and\ \citenamefont {Wahnström}}]{FraGreWah2021}%
  \BibitemOpen
  \bibfield  {author} {\bibinfo {author} {\bibfnamefont {E.}~\bibnamefont
  {Fransson}}, \bibinfo {author} {\bibfnamefont {M.}~\bibnamefont {Gren}},\
  and\ \bibinfo {author} {\bibfnamefont {G.}~\bibnamefont {Wahnström}},\
  }\bibfield  {title} {\bibinfo {title} {Complexions and grain growth
  retardation: First-principles modeling of phase boundaries in {WC-Co}
  cemented carbides at elevated temperatures},\ }\href
  {https://doi.org/10.1016/j.actamat.2021.117128} {\bibfield  {journal}
  {\bibinfo  {journal} {Acta Materialia}\ }\textbf {\bibinfo {volume} {216}},\
  \bibinfo {pages} {117128} (\bibinfo {year} {2021}{\natexlab{a}})}\BibitemShut
  {NoStop}%
\bibitem [{\citenamefont {Fransson}\ \emph
  {et~al.}(2021{\natexlab{b}})\citenamefont {Fransson}, \citenamefont {Gren},
  \citenamefont {Larsson},\ and\ \citenamefont
  {Wahnstr\"om}}]{FraGreLarWah2021}%
  \BibitemOpen
  \bibfield  {author} {\bibinfo {author} {\bibfnamefont {E.}~\bibnamefont
  {Fransson}}, \bibinfo {author} {\bibfnamefont {M.}~\bibnamefont {Gren}},
  \bibinfo {author} {\bibfnamefont {H.}~\bibnamefont {Larsson}},\ and\ \bibinfo
  {author} {\bibfnamefont {G.}~\bibnamefont {Wahnstr\"om}},\ }\bibfield
  {title} {\bibinfo {title} {First-principles modeling of complexions at the
  phase boundaries in {Ti}-doped {WC-Co} cemented carbides at finite
  temperatures},\ }\href {https://doi.org/10.1103/PhysRevMaterials.5.093801}
  {\bibfield  {journal} {\bibinfo  {journal} {Physical Review Materials}\
  }\textbf {\bibinfo {volume} {5}},\ \bibinfo {pages} {093801} (\bibinfo {year}
  {2021}{\natexlab{b}})}\BibitemShut {NoStop}%
\bibitem [{\citenamefont {Ekborg-Tanner}\ and\ \citenamefont
  {Erhart}(2021)}]{EkbErh21}%
  \BibitemOpen
  \bibfield  {author} {\bibinfo {author} {\bibfnamefont {P.}~\bibnamefont
  {Ekborg-Tanner}}\ and\ \bibinfo {author} {\bibfnamefont {P.}~\bibnamefont
  {Erhart}},\ }\bibfield  {title} {\bibinfo {title} {Hydrogen-driven surface
  segregation in {Pd} alloys from atomic-scale simulations},\ }\href
  {https://doi.org/10.1021/acs.jpcc.1c00575} {\bibfield  {journal} {\bibinfo
  {journal} {The Journal of Physical Chemistry C}\ }\textbf {\bibinfo {volume}
  {125}},\ \bibinfo {pages} {17248} (\bibinfo {year} {2021})}\BibitemShut
  {NoStop}%
\bibitem [{\citenamefont {Xie}\ and\ \citenamefont {Jiang}(2023)}]{XieJia23}%
  \BibitemOpen
  \bibfield  {author} {\bibinfo {author} {\bibfnamefont {J.-Z.}\ \bibnamefont
  {Xie}}\ and\ \bibinfo {author} {\bibfnamefont {H.}~\bibnamefont {Jiang}},\
  }\bibfield  {title} {\bibinfo {title} {Revealing carbon vacancy distribution
  on $\alpha$-\ce{MoC_{1–x}} surfaces by machine-learning force-field-aided
  cluster expansion approach},\ }\href
  {https://doi.org/10.1021/acs.jpcc.3c01941} {\bibfield  {journal} {\bibinfo
  {journal} {The Journal of Physical Chemistry C}\ }\textbf {\bibinfo {volume}
  {127}},\ \bibinfo {pages} {13228–13237} (\bibinfo {year}
  {2023})}\BibitemShut {NoStop}%
\bibitem [{\citenamefont {Mueller}\ and\ \citenamefont
  {Ceder}(2010)}]{MueCed10}%
  \BibitemOpen
  \bibfield  {author} {\bibinfo {author} {\bibfnamefont {T.}~\bibnamefont
  {Mueller}}\ and\ \bibinfo {author} {\bibfnamefont {G.}~\bibnamefont
  {Ceder}},\ }\bibfield  {title} {\bibinfo {title} {Effect of particle size on
  hydrogen release from sodium alanate nanoparticles},\ }\href
  {https://doi.org/10.1021/nn101224j} {\bibfield  {journal} {\bibinfo
  {journal} {ACS Nano}\ }\textbf {\bibinfo {volume} {4}},\ \bibinfo {pages}
  {5647} (\bibinfo {year} {2010})}\BibitemShut {NoStop}%
\bibitem [{\citenamefont {Chepulskii}\ \emph {et~al.}(2010)\citenamefont
  {Chepulskii}, \citenamefont {Butler}, \citenamefont {van~de Walle},\ and\
  \citenamefont {Curtarolo}}]{CheButWal10}%
  \BibitemOpen
  \bibfield  {author} {\bibinfo {author} {\bibfnamefont {R.~V.}\ \bibnamefont
  {Chepulskii}}, \bibinfo {author} {\bibfnamefont {W.~H.}\ \bibnamefont
  {Butler}}, \bibinfo {author} {\bibfnamefont {A.}~\bibnamefont {van~de
  Walle}},\ and\ \bibinfo {author} {\bibfnamefont {S.}~\bibnamefont
  {Curtarolo}},\ }\bibfield  {title} {\bibinfo {title} {Surface segregation in
  nanoparticles from first principles: {The} case of {FePt}},\ }\href
  {https://doi.org/10.1016/j.scriptamat.2009.10.019} {\bibfield  {journal}
  {\bibinfo  {journal} {Scripta Materialia}\ }\textbf {\bibinfo {volume}
  {62}},\ \bibinfo {pages} {179} (\bibinfo {year} {2010})}\BibitemShut
  {NoStop}%
\bibitem [{\citenamefont {Yuge}(2011)}]{Yug11}%
  \BibitemOpen
  \bibfield  {author} {\bibinfo {author} {\bibfnamefont {K.}~\bibnamefont
  {Yuge}},\ }\bibfield  {title} {\bibinfo {title} {Concentration effects on
  segregation behavior of {Pt-Rh} nanoparticles},\ }\href
  {https://doi.org/10.1103/PhysRevB.84.085451} {\bibfield  {journal} {\bibinfo
  {journal} {Physical Review B}\ }\textbf {\bibinfo {volume} {84}},\ \bibinfo
  {pages} {1} (\bibinfo {year} {2011})}\BibitemShut {NoStop}%
\bibitem [{\citenamefont {Mueller}(2012)}]{Mue12}%
  \BibitemOpen
  \bibfield  {author} {\bibinfo {author} {\bibfnamefont {T.}~\bibnamefont
  {Mueller}},\ }\bibfield  {title} {\bibinfo {title} {Ab initio determination
  of structure-property relationships in alloy nanoparticles},\ }\href
  {https://doi.org/10.1103/PhysRevB.86.144201} {\bibfield  {journal} {\bibinfo
  {journal} {Physical Review B}\ }\textbf {\bibinfo {volume} {86}},\ \bibinfo
  {pages} {144201} (\bibinfo {year} {2012})}\BibitemShut {NoStop}%
\bibitem [{\citenamefont {Tan}\ \emph {et~al.}(2012)\citenamefont {Tan},
  \citenamefont {Wang}, \citenamefont {Johnson},\ and\ \citenamefont
  {Bai}}]{TanWanJoh12}%
  \BibitemOpen
  \bibfield  {author} {\bibinfo {author} {\bibfnamefont {T.~L.}\ \bibnamefont
  {Tan}}, \bibinfo {author} {\bibfnamefont {L.-L.}\ \bibnamefont {Wang}},
  \bibinfo {author} {\bibfnamefont {D.~D.}\ \bibnamefont {Johnson}},\ and\
  \bibinfo {author} {\bibfnamefont {K.}~\bibnamefont {Bai}},\ }\bibfield
  {title} {\bibinfo {title} {A comprehensive search for stable {Pt-Pd}
  nanoalloy configurations and their use as tunable catalysts},\ }\href
  {https://doi.org/10.1021/nl302405k} {\bibfield  {journal} {\bibinfo
  {journal} {Nano Letters}\ }\textbf {\bibinfo {volume} {12}},\ \bibinfo
  {pages} {4875} (\bibinfo {year} {2012})}\BibitemShut {NoStop}%
\bibitem [{\citenamefont {Wang}\ \emph {et~al.}(2014)\citenamefont {Wang},
  \citenamefont {Tan},\ and\ \citenamefont {Johnson}}]{WanTanJoh14}%
  \BibitemOpen
  \bibfield  {author} {\bibinfo {author} {\bibfnamefont {L.-L.}\ \bibnamefont
  {Wang}}, \bibinfo {author} {\bibfnamefont {T.~L.}\ \bibnamefont {Tan}},\ and\
  \bibinfo {author} {\bibfnamefont {D.~D.}\ \bibnamefont {Johnson}},\
  }\bibfield  {title} {\bibinfo {title} {Configurational thermodynamics of
  alloyed nanoparticles with adsorbates},\ }\href
  {https://doi.org/10.1021/nl503519m} {\bibfield  {journal} {\bibinfo
  {journal} {Nano Letters}\ }\textbf {\bibinfo {volume} {14}},\ \bibinfo
  {pages} {7077} (\bibinfo {year} {2014})}\BibitemShut {NoStop}%
\bibitem [{\citenamefont {Li}\ \emph {et~al.}(2018)\citenamefont {Li},
  \citenamefont {Raciti}, \citenamefont {Pu}, \citenamefont {Cao},
  \citenamefont {He}, \citenamefont {Wang},\ and\ \citenamefont
  {Mueller}}]{LiRacPu18}%
  \BibitemOpen
  \bibfield  {author} {\bibinfo {author} {\bibfnamefont {C.}~\bibnamefont
  {Li}}, \bibinfo {author} {\bibfnamefont {D.}~\bibnamefont {Raciti}}, \bibinfo
  {author} {\bibfnamefont {T.}~\bibnamefont {Pu}}, \bibinfo {author}
  {\bibfnamefont {L.}~\bibnamefont {Cao}}, \bibinfo {author} {\bibfnamefont
  {C.}~\bibnamefont {He}}, \bibinfo {author} {\bibfnamefont {C.}~\bibnamefont
  {Wang}},\ and\ \bibinfo {author} {\bibfnamefont {T.}~\bibnamefont
  {Mueller}},\ }\bibfield  {title} {\bibinfo {title} {Improved prediction of
  nanoalloy structures by the explicit inclusion of adsorbates in cluster
  expansions},\ }\href {https://doi.org/10.1021/acs.jpcc.8b03868} {\bibfield
  {journal} {\bibinfo  {journal} {The Journal of Physical Chemistry C}\
  }\textbf {\bibinfo {volume} {122}},\ \bibinfo {pages} {18040} (\bibinfo
  {year} {2018})}\BibitemShut {NoStop}%
\bibitem [{\citenamefont {Cao}\ \emph {et~al.}(2018)\citenamefont {Cao},
  \citenamefont {Li},\ and\ \citenamefont {Mueller}}]{CaoLiMue18}%
  \BibitemOpen
  \bibfield  {author} {\bibinfo {author} {\bibfnamefont {L.}~\bibnamefont
  {Cao}}, \bibinfo {author} {\bibfnamefont {C.}~\bibnamefont {Li}},\ and\
  \bibinfo {author} {\bibfnamefont {T.}~\bibnamefont {Mueller}},\ }\bibfield
  {title} {\bibinfo {title} {The use of cluster expansions to predict the
  structures and properties of surfaces and nanostructured materials},\ }\href
  {https://doi.org/10.1021/acs.jcim.8b00413} {\bibfield  {journal} {\bibinfo
  {journal} {Journal of Chemical Information and Modeling}\ }\textbf {\bibinfo
  {volume} {58}},\ \bibinfo {pages} {2401–2413} (\bibinfo {year}
  {2018})}\BibitemShut {NoStop}%
\bibitem [{\citenamefont {Van~der Ven}\ \emph {et~al.}(2001)\citenamefont
  {Van~der Ven}, \citenamefont {Ceder}, \citenamefont {Asta},\ and\
  \citenamefont {Tepesch}}]{VanCedAst01}%
  \BibitemOpen
  \bibfield  {author} {\bibinfo {author} {\bibfnamefont {A.}~\bibnamefont
  {Van~der Ven}}, \bibinfo {author} {\bibfnamefont {G.}~\bibnamefont {Ceder}},
  \bibinfo {author} {\bibfnamefont {M.}~\bibnamefont {Asta}},\ and\ \bibinfo
  {author} {\bibfnamefont {P.~D.}\ \bibnamefont {Tepesch}},\ }\bibfield
  {title} {\bibinfo {title} {First-principles theory of ionic diffusion with
  nondilute carriers},\ }\href {https://doi.org/10.1103/PhysRevB.64.184307}
  {\bibfield  {journal} {\bibinfo  {journal} {Physical Review B}\ }\textbf
  {\bibinfo {volume} {64}},\ \bibinfo {pages} {184307} (\bibinfo {year}
  {2001})}\BibitemShut {NoStop}%
\bibitem [{\citenamefont {Morgan}\ \emph {et~al.}(2000)\citenamefont {Morgan},
  \citenamefont {van~de Walle}, \citenamefont {Ceder}, \citenamefont
  {Althoff},\ and\ \citenamefont {de~Fontaine}}]{MorWalCed00}%
  \BibitemOpen
  \bibfield  {author} {\bibinfo {author} {\bibfnamefont {D.}~\bibnamefont
  {Morgan}}, \bibinfo {author} {\bibfnamefont {A.}~\bibnamefont {van~de
  Walle}}, \bibinfo {author} {\bibfnamefont {G.}~\bibnamefont {Ceder}},
  \bibinfo {author} {\bibfnamefont {J.~D.}\ \bibnamefont {Althoff}},\ and\
  \bibinfo {author} {\bibfnamefont {D.}~\bibnamefont {de~Fontaine}},\
  }\bibfield  {title} {\bibinfo {title} {Vibrational thermodynamics: coupling
  of chemical order and size effects},\ }\href
  {https://doi.org/10.1088/0965-0393/8/3/310} {\bibfield  {journal} {\bibinfo
  {journal} {Modelling and Simulation in Materials Science and Engineering}\
  }\textbf {\bibinfo {volume} {8}},\ \bibinfo {pages} {295} (\bibinfo {year}
  {2000})}\BibitemShut {NoStop}%
\bibitem [{\citenamefont {van~de Walle}\ and\ \citenamefont
  {Ceder}(2002{\natexlab{a}})}]{WalCed02b}%
  \BibitemOpen
  \bibfield  {author} {\bibinfo {author} {\bibfnamefont {A.}~\bibnamefont
  {van~de Walle}}\ and\ \bibinfo {author} {\bibfnamefont {G.}~\bibnamefont
  {Ceder}},\ }\bibfield  {title} {\bibinfo {title} {The effect of lattice
  vibrations on substitutional alloy thermodynamics},\ }\href
  {https://doi.org/10.1103/RevModPhys.74.11} {\bibfield  {journal} {\bibinfo
  {journal} {Review of Modern Physics}\ }\textbf {\bibinfo {volume} {74}},\
  \bibinfo {pages} {11} (\bibinfo {year} {2002}{\natexlab{a}})}\BibitemShut
  {NoStop}%
\bibitem [{\citenamefont {Brorsson}\ \emph {et~al.}(2021)\citenamefont
  {Brorsson}, \citenamefont {Zhang}, \citenamefont {Palmqvist},\ and\
  \citenamefont {Erhart}}]{BroZhaPal21}%
  \BibitemOpen
  \bibfield  {author} {\bibinfo {author} {\bibfnamefont {J.}~\bibnamefont
  {Brorsson}}, \bibinfo {author} {\bibfnamefont {Y.}~\bibnamefont {Zhang}},
  \bibinfo {author} {\bibfnamefont {A.~E.~C.}\ \bibnamefont {Palmqvist}},\ and\
  \bibinfo {author} {\bibfnamefont {P.}~\bibnamefont {Erhart}},\ }\bibfield
  {title} {\bibinfo {title} {Order-disorder transition in inorganic clathrates
  controls electrical transport properties},\ }\href
  {https://doi.org/10.1021/acs.chemmater.1c00731} {\bibfield  {journal}
  {\bibinfo  {journal} {Chemistry of Materials}\ }\textbf {\bibinfo {volume}
  {33}},\ \bibinfo {pages} {4500} (\bibinfo {year} {2021})}\BibitemShut
  {NoStop}%
\bibitem [{\citenamefont {{van de Walle}}\ \emph {et~al.}(2002)\citenamefont
  {{van de Walle}}, \citenamefont {Asta},\ and\ \citenamefont {Ceder}}]{ATAT}%
  \BibitemOpen
  \bibfield  {author} {\bibinfo {author} {\bibfnamefont {A.}~\bibnamefont {{van
  de Walle}}}, \bibinfo {author} {\bibfnamefont {M.}~\bibnamefont {Asta}},\
  and\ \bibinfo {author} {\bibfnamefont {G.}~\bibnamefont {Ceder}},\ }\bibfield
   {title} {\bibinfo {title} {The alloy theoretic automated toolkit: {A} user
  guide},\ }\href {https://doi.org/10.1016/S0364-5916(02)80006-2} {\bibfield
  {journal} {\bibinfo  {journal} {Calphad}\ }\textbf {\bibinfo {volume} {26}},\
  \bibinfo {pages} {539} (\bibinfo {year} {2002})}\BibitemShut {NoStop}%
\bibitem [{\citenamefont {Lerch}\ \emph {et~al.}(2009)\citenamefont {Lerch},
  \citenamefont {Wieckhorst}, \citenamefont {Hart}, \citenamefont {Forcade},\
  and\ \citenamefont {M\"uller}}]{LerWieHar09}%
  \BibitemOpen
  \bibfield  {author} {\bibinfo {author} {\bibfnamefont {D.}~\bibnamefont
  {Lerch}}, \bibinfo {author} {\bibfnamefont {O.}~\bibnamefont {Wieckhorst}},
  \bibinfo {author} {\bibfnamefont {G.~L.~W.}\ \bibnamefont {Hart}}, \bibinfo
  {author} {\bibfnamefont {R.~W.}\ \bibnamefont {Forcade}},\ and\ \bibinfo
  {author} {\bibfnamefont {S.}~\bibnamefont {M\"uller}},\ }\bibfield  {title}
  {\bibinfo {title} {{UNCLE}: a code for constructing cluster expansions for
  arbitrary lattices with minimal user-input},\ }\href
  {https://doi.org/10.1088/0965-0393/17/5/055003} {\bibfield  {journal}
  {\bibinfo  {journal} {Modelling and Simulation in Materials Science and
  Engineering}\ }\textbf {\bibinfo {volume} {17}},\ \bibinfo {pages} {055003}
  (\bibinfo {year} {2009})}\BibitemShut {NoStop}%
\bibitem [{\citenamefont {Chang}\ \emph {et~al.}(2019)\citenamefont {Chang},
  \citenamefont {Kleiven}, \citenamefont {Melander}, \citenamefont {Akola},
  \citenamefont {Garcia-Lastra},\ and\ \citenamefont {Vegge}}]{CLEASE}%
  \BibitemOpen
  \bibfield  {author} {\bibinfo {author} {\bibfnamefont {J.~H.}\ \bibnamefont
  {Chang}}, \bibinfo {author} {\bibfnamefont {D.}~\bibnamefont {Kleiven}},
  \bibinfo {author} {\bibfnamefont {M.}~\bibnamefont {Melander}}, \bibinfo
  {author} {\bibfnamefont {J.}~\bibnamefont {Akola}}, \bibinfo {author}
  {\bibfnamefont {J.~M.}\ \bibnamefont {Garcia-Lastra}},\ and\ \bibinfo
  {author} {\bibfnamefont {T.}~\bibnamefont {Vegge}},\ }\bibfield  {title}
  {\bibinfo {title} {{CLEASE}: a versatile and user-friendly implementation of
  cluster expansion method},\ }\href {https://doi.org/10.1088/1361-648X/ab1bbc}
  {\bibfield  {journal} {\bibinfo  {journal} {Journal of Physics: Condensed
  Matter}\ }\textbf {\bibinfo {volume} {31}},\ \bibinfo {pages} {325901}
  (\bibinfo {year} {2019})}\BibitemShut {NoStop}%
\bibitem [{\citenamefont {Puchala}\ \emph {et~al.}(2023)\citenamefont
  {Puchala}, \citenamefont {Thomas}, \citenamefont {Natarajan}, \citenamefont
  {Goiri}, \citenamefont {Behara}, \citenamefont {Kaufman},\ and\ \citenamefont
  {{Van der Ven}}}]{CASM}%
  \BibitemOpen
  \bibfield  {author} {\bibinfo {author} {\bibfnamefont {B.}~\bibnamefont
  {Puchala}}, \bibinfo {author} {\bibfnamefont {J.~C.}\ \bibnamefont {Thomas}},
  \bibinfo {author} {\bibfnamefont {A.~R.}\ \bibnamefont {Natarajan}}, \bibinfo
  {author} {\bibfnamefont {J.~G.}\ \bibnamefont {Goiri}}, \bibinfo {author}
  {\bibfnamefont {S.~S.}\ \bibnamefont {Behara}}, \bibinfo {author}
  {\bibfnamefont {J.~L.}\ \bibnamefont {Kaufman}},\ and\ \bibinfo {author}
  {\bibfnamefont {A.}~\bibnamefont {{Van der Ven}}},\ }\bibfield  {title}
  {\bibinfo {title} {{{CASM}} --- {{A}} software package for first-principles
  based study of multicomponent crystalline solids},\ }\href
  {https://doi.org/10.1016/j.commatsci.2022.111897} {\bibfield  {journal}
  {\bibinfo  {journal} {Computational Materials Science}\ }\textbf {\bibinfo
  {volume} {217}},\ \bibinfo {pages} {111897} (\bibinfo {year}
  {2023})}\BibitemShut {NoStop}%
\bibitem [{\citenamefont {{\AA}ngqvist}\ \emph {et~al.}(2019)\citenamefont
  {{\AA}ngqvist}, \citenamefont {Mu{\~{n}}oz}, \citenamefont {Rahm},
  \citenamefont {Fransson}, \citenamefont {Durniak}, \citenamefont {Rozyczko},
  \citenamefont {Rod},\ and\ \citenamefont {Erhart}}]{Angqvist2019}%
  \BibitemOpen
  \bibfield  {author} {\bibinfo {author} {\bibfnamefont {M.}~\bibnamefont
  {{\AA}ngqvist}}, \bibinfo {author} {\bibfnamefont {W.~A.}\ \bibnamefont
  {Mu{\~{n}}oz}}, \bibinfo {author} {\bibfnamefont {J.~M.}\ \bibnamefont
  {Rahm}}, \bibinfo {author} {\bibfnamefont {E.}~\bibnamefont {Fransson}},
  \bibinfo {author} {\bibfnamefont {C.}~\bibnamefont {Durniak}}, \bibinfo
  {author} {\bibfnamefont {P.}~\bibnamefont {Rozyczko}}, \bibinfo {author}
  {\bibfnamefont {T.~H.}\ \bibnamefont {Rod}},\ and\ \bibinfo {author}
  {\bibfnamefont {P.}~\bibnamefont {Erhart}},\ }\bibfield  {title} {\bibinfo
  {title} {{ICET} {\textendash} a python library for constructing and sampling
  alloy cluster expansions},\ }\href {https://doi.org/10.1002/adts.201900015}
  {\bibfield  {journal} {\bibinfo  {journal} {Advanced Theory and Simulations}\
  }\textbf {\bibinfo {volume} {2}},\ \bibinfo {pages} {1900015} (\bibinfo
  {year} {2019})},\ \bibinfo {note}
  {\url{https://icet.materialsmodeling.org/}}\BibitemShut {NoStop}%
\bibitem [{\citenamefont {Barroso-Luque}\ \emph {et~al.}(2022)\citenamefont
  {Barroso-Luque}, \citenamefont {Yang}, \citenamefont {Xie}, \citenamefont
  {Chen}, \citenamefont {Kam}, \citenamefont {Jadidi}, \citenamefont {Zhong},\
  and\ \citenamefont {Ceder}}]{SMOL}%
  \BibitemOpen
  \bibfield  {author} {\bibinfo {author} {\bibfnamefont {L.}~\bibnamefont
  {Barroso-Luque}}, \bibinfo {author} {\bibfnamefont {J.~H.}\ \bibnamefont
  {Yang}}, \bibinfo {author} {\bibfnamefont {F.}~\bibnamefont {Xie}}, \bibinfo
  {author} {\bibfnamefont {T.}~\bibnamefont {Chen}}, \bibinfo {author}
  {\bibfnamefont {R.~L.}\ \bibnamefont {Kam}}, \bibinfo {author} {\bibfnamefont
  {Z.}~\bibnamefont {Jadidi}}, \bibinfo {author} {\bibfnamefont
  {P.}~\bibnamefont {Zhong}},\ and\ \bibinfo {author} {\bibfnamefont
  {G.}~\bibnamefont {Ceder}},\ }\bibfield  {title} {\bibinfo {title} {smol: A
  python package for cluster expansions and beyond},\ }\href
  {https://doi.org/10.21105/joss.04504} {\bibfield  {journal} {\bibinfo
  {journal} {Journal of Open Source Software}\ }\textbf {\bibinfo {volume}
  {7}},\ \bibinfo {pages} {4504} (\bibinfo {year} {2022})}\BibitemShut
  {NoStop}%
\bibitem [{url()}]{url-link}%
  \BibitemOpen
  \href@noop {} {}\bibinfo {note} {Cluster expansion tutorials,
  \url{https://ce-tutorials.materialsmodeling.org}, accessed
  2024-04-23.}\BibitemShut {Stop}%
\bibitem [{zen()}]{zenodo-link}%
  \BibitemOpen
  \href@noop {} {}\bibinfo {note} {The jupyter notebooks associated with this
  tutorial as well as underlying data are available at
  \href{https://doi.org/10.5281/zenodo.10997198}{doi:10.5281/zenodo.10997198}}\BibitemShut
  {NoStop}%
\bibitem [{\citenamefont {Sanchez}\ \emph {et~al.}(1984)\citenamefont
  {Sanchez}, \citenamefont {Ducastelle},\ and\ \citenamefont
  {Gratias}}]{SanDucGra84}%
  \BibitemOpen
  \bibfield  {author} {\bibinfo {author} {\bibfnamefont {J.~M.}\ \bibnamefont
  {Sanchez}}, \bibinfo {author} {\bibfnamefont {F.}~\bibnamefont
  {Ducastelle}},\ and\ \bibinfo {author} {\bibfnamefont {D.}~\bibnamefont
  {Gratias}},\ }\bibfield  {title} {\bibinfo {title} {Generalized cluster
  description of multicomponent systems},\ }\href
  {https://doi.org/10.1016/0378-4371(84)90096-7} {\bibfield  {journal}
  {\bibinfo  {journal} {Physica A: Statistical Mechanics and its Applications}\
  }\textbf {\bibinfo {volume} {128}},\ \bibinfo {pages} {334} (\bibinfo {year}
  {1984})}\BibitemShut {NoStop}%
\bibitem [{\citenamefont {Sanchez}(2010)}]{Sanchez2010}%
  \BibitemOpen
  \bibfield  {author} {\bibinfo {author} {\bibfnamefont {J.~M.}\ \bibnamefont
  {Sanchez}},\ }\bibfield  {title} {\bibinfo {title} {Cluster expansion and the
  configurational theory of alloys},\ }\href
  {https://doi.org/10.1103/PhysRevB.81.224202} {\bibfield  {journal} {\bibinfo
  {journal} {Physical Review B}\ }\textbf {\bibinfo {volume} {81}},\ \bibinfo
  {pages} {224202} (\bibinfo {year} {2010})}\BibitemShut {NoStop}%
\bibitem [{\citenamefont {de~Fontaine}(1994)}]{Fon94}%
  \BibitemOpen
  \bibfield  {author} {\bibinfo {author} {\bibfnamefont {D.}~\bibnamefont
  {de~Fontaine}},\ }\bibfield  {title} {\bibinfo {title} {Cluster approach to
  order-disorder transformations in alloys},\ }in\ \href
  {https://doi.org/10.1016/S0081-1947(08)60639-6} {\emph {\bibinfo {booktitle}
  {Solid {State} {Physics}}}},\ Vol.~\bibinfo {volume} {47},\ \bibinfo {editor}
  {edited by\ \bibinfo {editor} {\bibfnamefont {H.}~\bibnamefont {Ehrenreich}}\
  and\ \bibinfo {editor} {\bibfnamefont {D.}~\bibnamefont {Turnbull}}}\
  (\bibinfo  {publisher} {Academic Press},\ \bibinfo {year} {1994})\
  p.~\bibinfo {pages} {33}\BibitemShut {NoStop}%
\bibitem [{\citenamefont {Zunger}\ \emph {et~al.}(2002)\citenamefont {Zunger},
  \citenamefont {Wang}, \citenamefont {Hart},\ and\ \citenamefont
  {Sanati}}]{Zunger2002}%
  \BibitemOpen
  \bibfield  {author} {\bibinfo {author} {\bibfnamefont {A.}~\bibnamefont
  {Zunger}}, \bibinfo {author} {\bibfnamefont {L.~G.}\ \bibnamefont {Wang}},
  \bibinfo {author} {\bibfnamefont {G.~L.~W.}\ \bibnamefont {Hart}},\ and\
  \bibinfo {author} {\bibfnamefont {M.}~\bibnamefont {Sanati}},\ }\bibfield
  {title} {\bibinfo {title} {Obtaining {Ising}-like expansions for binary
  alloys from first principles},\ }\href
  {https://doi.org/10.1088/0965-0393/10/6/306} {\bibfield  {journal} {\bibinfo
  {journal} {Modelling and Simulation in Materials Science and Engineering}\
  }\textbf {\bibinfo {volume} {10}},\ \bibinfo {pages} {685} (\bibinfo {year}
  {2002})}\BibitemShut {NoStop}%
\bibitem [{\citenamefont {van~de Walle}(2009)}]{Wal09}%
  \BibitemOpen
  \bibfield  {author} {\bibinfo {author} {\bibfnamefont {A.}~\bibnamefont
  {van~de Walle}},\ }\bibfield  {title} {\bibinfo {title} {Multicomponent
  multisublattice alloys, nonconfigurational entropy and other additions to the
  {Alloy} {Theoretic} {Automated} {Toolkit}},\ }\href
  {https://doi.org/10.1016/j.calphad.2008.12.005} {\bibfield  {journal}
  {\bibinfo  {journal} {Calphad}\ }\bibinfo {series} {Tools for {Computational}
  {Thermodynamics}},\ \textbf {\bibinfo {volume} {33}},\ \bibinfo {pages} {266}
  (\bibinfo {year} {2009})}\BibitemShut {NoStop}%
\bibitem [{\citenamefont {Xie}\ \emph {et~al.}(2022)\citenamefont {Xie},
  \citenamefont {Zhou},\ and\ \citenamefont {Jiang}}]{XieZhoJia22}%
  \BibitemOpen
  \bibfield  {author} {\bibinfo {author} {\bibfnamefont {J.-Z.}\ \bibnamefont
  {Xie}}, \bibinfo {author} {\bibfnamefont {X.-Y.}\ \bibnamefont {Zhou}},\ and\
  \bibinfo {author} {\bibfnamefont {H.}~\bibnamefont {Jiang}},\ }\bibfield
  {title} {\bibinfo {title} {Perspective on optimal strategies of building
  cluster expansion models for configurationally disordered materials},\ }\href
  {https://doi.org/10.1063/5.0106788} {\bibfield  {journal} {\bibinfo
  {journal} {The Journal of Chemical Physics}\ }\textbf {\bibinfo {volume}
  {157}},\ \bibinfo {pages} {200901} (\bibinfo {year} {2022})}\BibitemShut
  {NoStop}%
\bibitem [{\citenamefont {Mueller}(2017)}]{Mueller2010}%
  \BibitemOpen
  \bibfield  {author} {\bibinfo {author} {\bibfnamefont {T.}~\bibnamefont
  {Mueller}},\ }\bibfield  {title} {\bibinfo {title} {{Comment on ``Cluster
  expansion and the configurational theory of alloys''}},\ }\href
  {https://doi.org/10.1103/PhysRevB.95.216201} {\bibfield  {journal} {\bibinfo
  {journal} {Physical Review B}\ }\textbf {\bibinfo {volume} {95}},\ \bibinfo
  {pages} {216201} (\bibinfo {year} {2017})}\BibitemShut {NoStop}%
\bibitem [{\citenamefont {Sanchez}(2017)}]{Sanchez2017}%
  \BibitemOpen
  \bibfield  {author} {\bibinfo {author} {\bibfnamefont {J.~M.}\ \bibnamefont
  {Sanchez}},\ }\bibfield  {title} {\bibinfo {title} {{Reply to ``Comment on
  `Cluster expansion and the configurational theory of alloys'''}},\ }\href
  {https://doi.org/10.1103/PhysRevB.95.216202} {\bibfield  {journal} {\bibinfo
  {journal} {Physical Review B}\ }\textbf {\bibinfo {volume} {95}},\ \bibinfo
  {pages} {216202} (\bibinfo {year} {2017})}\BibitemShut {NoStop}%
\bibitem [{\citenamefont {{Barroso-Luque}}\ \emph {et~al.}(2022)\citenamefont
  {{Barroso-Luque}}, \citenamefont {Zhong}, \citenamefont {Yang}, \citenamefont
  {Xie}, \citenamefont {Chen}, \citenamefont {Ouyang},\ and\ \citenamefont
  {Ceder}}]{BarZhoYan22}%
  \BibitemOpen
  \bibfield  {author} {\bibinfo {author} {\bibfnamefont {L.}~\bibnamefont
  {{Barroso-Luque}}}, \bibinfo {author} {\bibfnamefont {P.}~\bibnamefont
  {Zhong}}, \bibinfo {author} {\bibfnamefont {J.~H.}\ \bibnamefont {Yang}},
  \bibinfo {author} {\bibfnamefont {F.}~\bibnamefont {Xie}}, \bibinfo {author}
  {\bibfnamefont {T.}~\bibnamefont {Chen}}, \bibinfo {author} {\bibfnamefont
  {B.}~\bibnamefont {Ouyang}},\ and\ \bibinfo {author} {\bibfnamefont
  {G.}~\bibnamefont {Ceder}},\ }\bibfield  {title} {\bibinfo {title} {Cluster
  expansions of multicomponent ionic materials: {{Formalism}} and
  methodology},\ }\href {https://doi.org/10.1103/PhysRevB.106.144202}
  {\bibfield  {journal} {\bibinfo  {journal} {Physical Review B}\ }\textbf
  {\bibinfo {volume} {106}},\ \bibinfo {pages} {144202} (\bibinfo {year}
  {2022})}\BibitemShut {NoStop}%
\bibitem [{\citenamefont {Fant}\ \emph {et~al.}(2021)\citenamefont {Fant},
  \citenamefont {Ångqvist}, \citenamefont {Hellman},\ and\ \citenamefont
  {Erhart}}]{FanAngHel21}%
  \BibitemOpen
  \bibfield  {author} {\bibinfo {author} {\bibfnamefont {M.}~\bibnamefont
  {Fant}}, \bibinfo {author} {\bibfnamefont {M.}~\bibnamefont {Ångqvist}},
  \bibinfo {author} {\bibfnamefont {A.}~\bibnamefont {Hellman}},\ and\ \bibinfo
  {author} {\bibfnamefont {P.}~\bibnamefont {Erhart}},\ }\bibfield  {title}
  {\bibinfo {title} {To every rule there is an exception: A rational extension
  of loewenstein’s rule},\ }\href {https://doi.org/10.1002/anie.202013256}
  {\bibfield  {journal} {\bibinfo  {journal} {Angewandte Chemie - International
  Edition}\ }\textbf {\bibinfo {volume} {60}},\ \bibinfo {pages} {5132}
  (\bibinfo {year} {2021})}\BibitemShut {NoStop}%
\bibitem [{\citenamefont {Laks}\ \emph {et~al.}(1992)\citenamefont {Laks},
  \citenamefont {Ferreira}, \citenamefont {Froyen},\ and\ \citenamefont
  {Zunger}}]{Laks1992}%
  \BibitemOpen
  \bibfield  {author} {\bibinfo {author} {\bibfnamefont {D.~B.}\ \bibnamefont
  {Laks}}, \bibinfo {author} {\bibfnamefont {L.~G.}\ \bibnamefont {Ferreira}},
  \bibinfo {author} {\bibfnamefont {S.}~\bibnamefont {Froyen}},\ and\ \bibinfo
  {author} {\bibfnamefont {A.}~\bibnamefont {Zunger}},\ }\bibfield  {title}
  {\bibinfo {title} {Efficient cluster expansion for substitutional systems},\
  }\href {https://doi.org/10.1103/PhysRevB.46.12587} {\bibfield  {journal}
  {\bibinfo  {journal} {Physical Review B}\ }\textbf {\bibinfo {volume} {46}},\
  \bibinfo {pages} {12587} (\bibinfo {year} {1992})}\BibitemShut {NoStop}%
\bibitem [{\citenamefont {Rahm}\ \emph {et~al.}(2022)\citenamefont {Rahm},
  \citenamefont {Löfgren},\ and\ \citenamefont {Erhart}}]{Rahm2022}%
  \BibitemOpen
  \bibfield  {author} {\bibinfo {author} {\bibfnamefont {J.~M.}\ \bibnamefont
  {Rahm}}, \bibinfo {author} {\bibfnamefont {J.}~\bibnamefont {Löfgren}},\
  and\ \bibinfo {author} {\bibfnamefont {P.}~\bibnamefont {Erhart}},\
  }\bibfield  {title} {\bibinfo {title} {Quantitative predictions of
  thermodynamic hysteresis: Temperature-dependent character of the phase
  transition in {Pd–H}},\ }\href
  {https://doi.org/10.1016/j.actamat.2022.117697} {\bibfield  {journal}
  {\bibinfo  {journal} {Acta Materialia}\ }\textbf {\bibinfo {volume} {227}},\
  \bibinfo {pages} {117697} (\bibinfo {year} {2022})}\BibitemShut {NoStop}%
\bibitem [{\citenamefont {Nelson}\ \emph
  {et~al.}(2013{\natexlab{a}})\citenamefont {Nelson}, \citenamefont
  {Ozoli\c{n}\v{s}}, \citenamefont {Reese}, \citenamefont {Zhou},\ and\
  \citenamefont {Hart}}]{NelOzoRee13}%
  \BibitemOpen
  \bibfield  {author} {\bibinfo {author} {\bibfnamefont {L.~J.}\ \bibnamefont
  {Nelson}}, \bibinfo {author} {\bibfnamefont {V.}~\bibnamefont
  {Ozoli\c{n}\v{s}}}, \bibinfo {author} {\bibfnamefont {C.~S.}\ \bibnamefont
  {Reese}}, \bibinfo {author} {\bibfnamefont {F.}~\bibnamefont {Zhou}},\ and\
  \bibinfo {author} {\bibfnamefont {G.~L.~W.}\ \bibnamefont {Hart}},\
  }\bibfield  {title} {\bibinfo {title} {Cluster expansion made easy with
  {Bayesian} compressive sensing},\ }\href
  {https://doi.org/10.1103/PhysRevB.88.155105} {\bibfield  {journal} {\bibinfo
  {journal} {Physical Review B}\ }\textbf {\bibinfo {volume} {88}},\ \bibinfo
  {pages} {155105} (\bibinfo {year} {2013}{\natexlab{a}})}\BibitemShut
  {NoStop}%
\bibitem [{\citenamefont {Zhou}\ \emph {et~al.}(2014)\citenamefont {Zhou},
  \citenamefont {Nielson}, \citenamefont {Xia},\ and\ \citenamefont
  {Ozoli\c{n}\v{s}}}]{Zhou2014}%
  \BibitemOpen
  \bibfield  {author} {\bibinfo {author} {\bibfnamefont {F.}~\bibnamefont
  {Zhou}}, \bibinfo {author} {\bibfnamefont {W.}~\bibnamefont {Nielson}},
  \bibinfo {author} {\bibfnamefont {Y.}~\bibnamefont {Xia}},\ and\ \bibinfo
  {author} {\bibfnamefont {V.}~\bibnamefont {Ozoli\c{n}\v{s}}},\ }\bibfield
  {title} {\bibinfo {title} {Lattice anharmonicity and thermal conductivity
  from compressive sensing of first-principles calculations},\ }\href
  {https://doi.org/10.1103/PhysRevLett.113.185501} {\bibfield  {journal}
  {\bibinfo  {journal} {Physical Review Letters}\ }\textbf {\bibinfo {volume}
  {113}},\ \bibinfo {pages} {185501} (\bibinfo {year} {2014})}\BibitemShut
  {NoStop}%
\bibitem [{\citenamefont {Fransson}\ \emph {et~al.}(2020)\citenamefont
  {Fransson}, \citenamefont {Eriksson},\ and\ \citenamefont
  {Erhart}}]{FraEriErh20}%
  \BibitemOpen
  \bibfield  {author} {\bibinfo {author} {\bibfnamefont {E.}~\bibnamefont
  {Fransson}}, \bibinfo {author} {\bibfnamefont {F.}~\bibnamefont {Eriksson}},\
  and\ \bibinfo {author} {\bibfnamefont {P.}~\bibnamefont {Erhart}},\
  }\bibfield  {title} {\bibinfo {title} {Efficient construction of linear
  models in materials modeling and applications to force constant expansions},\
  }\href {https://doi.org/10.1038/s41524-020-00404-5} {\bibfield  {journal}
  {\bibinfo  {journal} {npj Computational Materials}\ }\textbf {\bibinfo
  {volume} {6}},\ \bibinfo {pages} {135} (\bibinfo {year} {2020})},\ \bibinfo
  {note} {\url{https://trainstation.materialsmodeling.org/}}\BibitemShut
  {NoStop}%
\bibitem [{\citenamefont {Cheng}\ \emph {et~al.}(2020)\citenamefont {Cheng},
  \citenamefont {Zhu}, \citenamefont {Zhou},\ and\ \citenamefont
  {Sun}}]{Cheng2020}%
  \BibitemOpen
  \bibfield  {author} {\bibinfo {author} {\bibfnamefont {Y.}~\bibnamefont
  {Cheng}}, \bibinfo {author} {\bibfnamefont {L.}~\bibnamefont {Zhu}}, \bibinfo
  {author} {\bibfnamefont {J.}~\bibnamefont {Zhou}},\ and\ \bibinfo {author}
  {\bibfnamefont {Z.}~\bibnamefont {Sun}},\ }\bibfield  {title} {\bibinfo
  {title} {{pyGACE}: {Combining} the genetic algorithm and cluster expansion
  methods to predict the ground-state structure of systems containing point
  defects},\ }\href {https://doi.org/10.1016/j.commatsci.2019.109482}
  {\bibfield  {journal} {\bibinfo  {journal} {Computational Materials Science}\
  }\textbf {\bibinfo {volume} {174}},\ \bibinfo {pages} {109482} (\bibinfo
  {year} {2020})}\BibitemShut {NoStop}%
\bibitem [{\citenamefont {MacKay}(1992)}]{MacKay1992}%
  \BibitemOpen
  \bibfield  {author} {\bibinfo {author} {\bibfnamefont {D.~J.~C.}\
  \bibnamefont {MacKay}},\ }\bibfield  {title} {\bibinfo {title} {Bayesian
  interpolation},\ }\href {https://doi.org/10.1162/neco.1992.4.3.415}
  {\bibfield  {journal} {\bibinfo  {journal} {Neural Computation}\ }\textbf
  {\bibinfo {volume} {4}},\ \bibinfo {pages} {415–447} (\bibinfo {year}
  {1992})}\BibitemShut {NoStop}%
\bibitem [{\citenamefont {Pedregosa}\ \emph {et~al.}(2011)\citenamefont
  {Pedregosa}, \citenamefont {Varoquaux}, \citenamefont {Gramfort},
  \citenamefont {Michel}, \citenamefont {Thirion}, \citenamefont {Grisel},
  \citenamefont {Blondel}, \citenamefont {Prettenhofer}, \citenamefont {Weiss},
  \citenamefont {Dubourg}, \citenamefont {Vanderplas}, \citenamefont {Passos},
  \citenamefont {Cournapeau}, \citenamefont {Brucher}, \citenamefont {Perrot},\
  and\ \citenamefont {Duchesnay}}]{scikit-learn}%
  \BibitemOpen
  \bibfield  {author} {\bibinfo {author} {\bibfnamefont {F.}~\bibnamefont
  {Pedregosa}}, \bibinfo {author} {\bibfnamefont {G.}~\bibnamefont
  {Varoquaux}}, \bibinfo {author} {\bibfnamefont {A.}~\bibnamefont {Gramfort}},
  \bibinfo {author} {\bibfnamefont {V.}~\bibnamefont {Michel}}, \bibinfo
  {author} {\bibfnamefont {B.}~\bibnamefont {Thirion}}, \bibinfo {author}
  {\bibfnamefont {O.}~\bibnamefont {Grisel}}, \bibinfo {author} {\bibfnamefont
  {M.}~\bibnamefont {Blondel}}, \bibinfo {author} {\bibfnamefont
  {P.}~\bibnamefont {Prettenhofer}}, \bibinfo {author} {\bibfnamefont
  {R.}~\bibnamefont {Weiss}}, \bibinfo {author} {\bibfnamefont
  {V.}~\bibnamefont {Dubourg}}, \bibinfo {author} {\bibfnamefont
  {J.}~\bibnamefont {Vanderplas}}, \bibinfo {author} {\bibfnamefont
  {A.}~\bibnamefont {Passos}}, \bibinfo {author} {\bibfnamefont
  {D.}~\bibnamefont {Cournapeau}}, \bibinfo {author} {\bibfnamefont
  {M.}~\bibnamefont {Brucher}}, \bibinfo {author} {\bibfnamefont
  {M.}~\bibnamefont {Perrot}},\ and\ \bibinfo {author} {\bibfnamefont
  {E.}~\bibnamefont {Duchesnay}},\ }\bibfield  {title} {\bibinfo {title}
  {Scikit-learn: Machine learning in {P}ython},\ }\href@noop {} {\bibfield
  {journal} {\bibinfo  {journal} {Journal of Machine Learning Research}\
  }\textbf {\bibinfo {volume} {12}},\ \bibinfo {pages} {2825} (\bibinfo {year}
  {2011})}\BibitemShut {NoStop}%
\bibitem [{\citenamefont {Akaike}(1974)}]{Aka74}%
  \BibitemOpen
  \bibfield  {author} {\bibinfo {author} {\bibfnamefont {H.}~\bibnamefont
  {Akaike}},\ }\bibfield  {title} {\bibinfo {title} {A new look at the
  statistical model identification},\ }\href
  {https://doi.org/10.1109/TAC.1974.1100705} {\bibfield  {journal} {\bibinfo
  {journal} {IEEE Transactions on Automatic Control}\ }\textbf {\bibinfo
  {volume} {19}},\ \bibinfo {pages} {716} (\bibinfo {year} {1974})}\BibitemShut
  {NoStop}%
\bibitem [{\citenamefont {Schwarz}(1978)}]{Sch78}%
  \BibitemOpen
  \bibfield  {author} {\bibinfo {author} {\bibfnamefont {G.}~\bibnamefont
  {Schwarz}},\ }\bibfield  {title} {\bibinfo {title} {Estimating the dimension
  of a model},\ }\href {https://doi.org/10.1214/aos/1176344136} {\bibfield
  {journal} {\bibinfo  {journal} {The Annals of Statistics}\ }\textbf {\bibinfo
  {volume} {6}},\ \bibinfo {pages} {461} (\bibinfo {year} {1978})}\BibitemShut
  {NoStop}%
\bibitem [{\citenamefont {Aho}\ \emph {et~al.}(2014)\citenamefont {Aho},
  \citenamefont {Derryberry},\ and\ \citenamefont {Peterson}}]{AhoDerPet14}%
  \BibitemOpen
  \bibfield  {author} {\bibinfo {author} {\bibfnamefont {K.}~\bibnamefont
  {Aho}}, \bibinfo {author} {\bibfnamefont {D.}~\bibnamefont {Derryberry}},\
  and\ \bibinfo {author} {\bibfnamefont {T.}~\bibnamefont {Peterson}},\
  }\bibfield  {title} {\bibinfo {title} {Model selection for ecologists: The
  worldviews of {AIC} and {BIC}},\ }\href {https://doi.org/10.1890/13-1452.1}
  {\bibfield  {journal} {\bibinfo  {journal} {Ecology}\ }\textbf {\bibinfo
  {volume} {95}},\ \bibinfo {pages} {631} (\bibinfo {year} {2014})}\BibitemShut
  {NoStop}%
\bibitem [{\citenamefont {Murphy}(2013)}]{Mur13}%
  \BibitemOpen
  \bibfield  {author} {\bibinfo {author} {\bibfnamefont {K.~P.}\ \bibnamefont
  {Murphy}},\ }\href {https://doi.org/10.1137/1.9781421407944} {\emph {\bibinfo
  {title} {Machine Learning: A Probabilistic Perspective}}}\ (\bibinfo
  {publisher} {MIT Press},\ \bibinfo {address} {Cambridge, MA},\ \bibinfo
  {year} {2013})\BibitemShut {NoStop}%
\bibitem [{\citenamefont {Mueller}\ and\ \citenamefont
  {Ceder}(2009)}]{MueCed09}%
  \BibitemOpen
  \bibfield  {author} {\bibinfo {author} {\bibfnamefont {T.}~\bibnamefont
  {Mueller}}\ and\ \bibinfo {author} {\bibfnamefont {G.}~\bibnamefont
  {Ceder}},\ }\bibfield  {title} {\bibinfo {title} {Bayesian approach to
  cluster expansions},\ }\href {https://doi.org/10.1103/PhysRevB.80.024103}
  {\bibfield  {journal} {\bibinfo  {journal} {Physical Review B}\ }\textbf
  {\bibinfo {volume} {80}},\ \bibinfo {pages} {024103} (\bibinfo {year}
  {2009})}\BibitemShut {NoStop}%
\bibitem [{\citenamefont {Zou}(2006)}]{Zou06}%
  \BibitemOpen
  \bibfield  {author} {\bibinfo {author} {\bibfnamefont {H.}~\bibnamefont
  {Zou}},\ }\bibfield  {title} {\bibinfo {title} {The adaptive {Lasso} and its
  oracle properties},\ }\href {https://doi.org/10.1198/016214506000000735}
  {\bibfield  {journal} {\bibinfo  {journal} {Journal of the American
  Statistical Association}\ }\textbf {\bibinfo {volume} {101}},\ \bibinfo
  {pages} {1418} (\bibinfo {year} {2006})}\BibitemShut {NoStop}%
\bibitem [{\citenamefont {Candes}\ and\ \citenamefont {Tao}(2005)}]{CanTao05}%
  \BibitemOpen
  \bibfield  {author} {\bibinfo {author} {\bibfnamefont {E.}~\bibnamefont
  {Candes}}\ and\ \bibinfo {author} {\bibfnamefont {T.}~\bibnamefont {Tao}},\
  }\bibfield  {title} {\bibinfo {title} {Decoding by linear programming},\
  }\href {https://doi.org/10.1109/TIT.2005.858979} {\bibfield  {journal}
  {\bibinfo  {journal} {IEEE Transactions on Information Theory}\ }\textbf
  {\bibinfo {volume} {51}},\ \bibinfo {pages} {4203} (\bibinfo {year}
  {2005})}\BibitemShut {NoStop}%
\bibitem [{\citenamefont {van~de Walle}\ and\ \citenamefont
  {Ceder}(2002{\natexlab{b}})}]{WalCed02}%
  \BibitemOpen
  \bibfield  {author} {\bibinfo {author} {\bibfnamefont {A.}~\bibnamefont
  {van~de Walle}}\ and\ \bibinfo {author} {\bibfnamefont {G.}~\bibnamefont
  {Ceder}},\ }\bibfield  {title} {\bibinfo {title} {Automating first-principles
  phase diagram calculations},\ }\href
  {https://doi.org/10.1361/105497102770331596} {\bibfield  {journal} {\bibinfo
  {journal} {Journal of Phase Equilibria}\ }\textbf {\bibinfo {volume} {23}},\
  \bibinfo {pages} {348} (\bibinfo {year} {2002}{\natexlab{b}})}\BibitemShut
  {NoStop}%
\bibitem [{\citenamefont {Seko}\ \emph {et~al.}(2009)\citenamefont {Seko},
  \citenamefont {Koyama},\ and\ \citenamefont {Tanaka}}]{Sek2009}%
  \BibitemOpen
  \bibfield  {author} {\bibinfo {author} {\bibfnamefont {A.}~\bibnamefont
  {Seko}}, \bibinfo {author} {\bibfnamefont {Y.}~\bibnamefont {Koyama}},\ and\
  \bibinfo {author} {\bibfnamefont {I.}~\bibnamefont {Tanaka}},\ }\bibfield
  {title} {\bibinfo {title} {Cluster expansion method for multicomponent
  systems based on optimal selection of structures for density-functional
  theory calculations},\ }\href {https://doi.org/10.1103/PhysRevB.80.165122}
  {\bibfield  {journal} {\bibinfo  {journal} {Phys. Rev. B}\ }\textbf {\bibinfo
  {volume} {80}},\ \bibinfo {pages} {165122} (\bibinfo {year}
  {2009})}\BibitemShut {NoStop}%
\bibitem [{\citenamefont {Kleiven}\ \emph {et~al.}(2021)\citenamefont
  {Kleiven}, \citenamefont {Akola}, \citenamefont {Peterson}, \citenamefont
  {Vegge},\ and\ \citenamefont {Chang}}]{Kleiven2021}%
  \BibitemOpen
  \bibfield  {author} {\bibinfo {author} {\bibfnamefont {D.}~\bibnamefont
  {Kleiven}}, \bibinfo {author} {\bibfnamefont {J.}~\bibnamefont {Akola}},
  \bibinfo {author} {\bibfnamefont {A.~A.}\ \bibnamefont {Peterson}}, \bibinfo
  {author} {\bibfnamefont {T.}~\bibnamefont {Vegge}},\ and\ \bibinfo {author}
  {\bibfnamefont {J.~H.}\ \bibnamefont {Chang}},\ }\bibfield  {title} {\bibinfo
  {title} {Training sets based on uncertainty estimates in the
  cluster-expansion method},\ }\href {https://doi.org/10.1088/2515-7655/abf9ef}
  {\bibfield  {journal} {\bibinfo  {journal} {Journal of Physics: Energy}\
  }\textbf {\bibinfo {volume} {3}},\ \bibinfo {pages} {034012} (\bibinfo {year}
  {2021})}\BibitemShut {NoStop}%
\bibitem [{\citenamefont {Golub}\ and\ \citenamefont
  {Van~Loan}(2013)}]{GolGenLoaCha13}%
  \BibitemOpen
  \bibfield  {author} {\bibinfo {author} {\bibfnamefont {G.~H.}\ \bibnamefont
  {Golub}}\ and\ \bibinfo {author} {\bibfnamefont {C.~F.}\ \bibnamefont
  {Van~Loan}},\ }\href {https://doi.org/10.1137/1.9781421407944} {\emph
  {\bibinfo {title} {Matrix Computations}}},\ \bibinfo {edition} {4th}\ ed.\
  (\bibinfo  {publisher} {Johns Hopkins University Press},\ \bibinfo {address}
  {Philadelphia, PA},\ \bibinfo {year} {2013})\BibitemShut {NoStop}%
\bibitem [{\citenamefont {Nelson}\ \emph
  {et~al.}(2013{\natexlab{b}})\citenamefont {Nelson}, \citenamefont {Hart},
  \citenamefont {Zhou},\ and\ \citenamefont {Ozoli\c{n}\v{s}}}]{NelHarZho13}%
  \BibitemOpen
  \bibfield  {author} {\bibinfo {author} {\bibfnamefont {L.~J.}\ \bibnamefont
  {Nelson}}, \bibinfo {author} {\bibfnamefont {G.~L.~W.}\ \bibnamefont {Hart}},
  \bibinfo {author} {\bibfnamefont {F.}~\bibnamefont {Zhou}},\ and\ \bibinfo
  {author} {\bibfnamefont {V.}~\bibnamefont {Ozoli\c{n}\v{s}}},\ }\bibfield
  {title} {\bibinfo {title} {Compressive sensing as a paradigm for building
  physics models},\ }\href {https://doi.org/10.1103/PhysRevB.87.035125}
  {\bibfield  {journal} {\bibinfo  {journal} {Physical Review B}\ }\textbf
  {\bibinfo {volume} {87}},\ \bibinfo {pages} {035125} (\bibinfo {year}
  {2013}{\natexlab{b}})}\BibitemShut {NoStop}%
\bibitem [{\citenamefont {Hart}\ and\ \citenamefont
  {Forcade}(2008)}]{HarFor08}%
  \BibitemOpen
  \bibfield  {author} {\bibinfo {author} {\bibfnamefont {G.~L.~W.}\
  \bibnamefont {Hart}}\ and\ \bibinfo {author} {\bibfnamefont {R.~W.}\
  \bibnamefont {Forcade}},\ }\bibfield  {title} {\bibinfo {title} {Algorithm
  for generating derivative structures},\ }\href
  {https://doi.org/10.1103/PhysRevB.77.224115} {\bibfield  {journal} {\bibinfo
  {journal} {Physical Review B}\ }\textbf {\bibinfo {volume} {77}},\ \bibinfo
  {pages} {224115} (\bibinfo {year} {2008})}\BibitemShut {NoStop}%
\bibitem [{\citenamefont {Cockayne}\ and\ \citenamefont {van~de
  Walle}(2010)}]{CocWal10}%
  \BibitemOpen
  \bibfield  {author} {\bibinfo {author} {\bibfnamefont {E.}~\bibnamefont
  {Cockayne}}\ and\ \bibinfo {author} {\bibfnamefont {A.}~\bibnamefont {van~de
  Walle}},\ }\bibfield  {title} {\bibinfo {title} {Building effective models
  from sparse but precise data: {Application} to an alloy cluster expansion
  model},\ }\href {https://doi.org/10.1103/PhysRevB.81.012104} {\bibfield
  {journal} {\bibinfo  {journal} {Physical Review B}\ }\textbf {\bibinfo
  {volume} {81}},\ \bibinfo {pages} {012104} (\bibinfo {year}
  {2010})}\BibitemShut {NoStop}%
\bibitem [{\citenamefont {Frenkel}\ and\ \citenamefont {Smit}(2023)}]{Fre23}%
  \BibitemOpen
  \bibfield  {author} {\bibinfo {author} {\bibfnamefont {D.}~\bibnamefont
  {Frenkel}}\ and\ \bibinfo {author} {\bibfnamefont {B.}~\bibnamefont {Smit}},\
  }\bibfield  {title} {\bibinfo {title} {{Monte Carlo} simulations in various
  ensembles},\ }in\ \href {https://doi.org/10.1016/B978-0-32-390292-2.00015-5}
  {\emph {\bibinfo {booktitle} {Understanding Molecular Simulation}}},\
  \bibinfo {editor} {edited by\ \bibinfo {editor} {\bibfnamefont
  {D.}~\bibnamefont {Frenkel}}\ and\ \bibinfo {editor} {\bibfnamefont
  {B.}~\bibnamefont {Smit}}}\ (\bibinfo  {publisher} {Academic Press},\
  \bibinfo {year} {2023})\ \bibinfo {edition} {3rd}\ ed.,\ Chap.~\bibinfo
  {chapter} {6}, pp.\ \bibinfo {pages} {181--232}\BibitemShut {NoStop}%
\bibitem [{\citenamefont {Pathria}\ and\ \citenamefont
  {Beale}(2022)}]{Pathria22}%
  \BibitemOpen
  \bibfield  {author} {\bibinfo {author} {\bibfnamefont {R.}~\bibnamefont
  {Pathria}}\ and\ \bibinfo {author} {\bibfnamefont {P.~D.}\ \bibnamefont
  {Beale}},\ }\bibfield  {title} {\bibinfo {title} {{4 - The grand canonical
  ensemble}},\ }in\ \href
  {https://doi.org/https://doi.org/10.1016/B978-0-08-102692-2.00013-2} {\emph
  {\bibinfo {booktitle} {Statistical Mechanics (Fourth Edition)}}},\ \bibinfo
  {editor} {edited by\ \bibinfo {editor} {\bibfnamefont {R.}~\bibnamefont
  {Pathria}}\ and\ \bibinfo {editor} {\bibfnamefont {P.~D.}\ \bibnamefont
  {Beale}}}\ (\bibinfo  {publisher} {Academic Press},\ \bibinfo {year} {2022})\
  \bibinfo {edition} {fourth edition}\ ed.,\ pp.\ \bibinfo {pages}
  {93--116}\BibitemShut {NoStop}%
\bibitem [{\citenamefont {Sadigh}\ and\ \citenamefont
  {Erhart}(2012)}]{SadErh12}%
  \BibitemOpen
  \bibfield  {author} {\bibinfo {author} {\bibfnamefont {B.}~\bibnamefont
  {Sadigh}}\ and\ \bibinfo {author} {\bibfnamefont {P.}~\bibnamefont
  {Erhart}},\ }\bibfield  {title} {\bibinfo {title} {Calculation of excess free
  energies of precipitates via direct thermodynamic integration across phase
  boundaries},\ }\href {https://doi.org/10.1103/PhysRevB.86.134204} {\bibfield
  {journal} {\bibinfo  {journal} {Physical Review B}\ }\textbf {\bibinfo
  {volume} {86}},\ \bibinfo {pages} {134204} (\bibinfo {year}
  {2012})}\BibitemShut {NoStop}%
\bibitem [{\citenamefont {Ma}(1985)}]{Ma_StatPhys}%
  \BibitemOpen
  \bibfield  {author} {\bibinfo {author} {\bibfnamefont {S.-K.}\ \bibnamefont
  {Ma}},\ }\href@noop {} {\emph {\bibinfo {title} {Statistical Mechanics}}}\
  (\bibinfo  {publisher} {World Scientific Publishing},\ \bibinfo {address}
  {Singapore, Singapore},\ \bibinfo {year} {1985})\BibitemShut {NoStop}%
\bibitem [{\citenamefont {van~de Walle}\ and\ \citenamefont
  {Asta}(2002)}]{WalAst02}%
  \BibitemOpen
  \bibfield  {author} {\bibinfo {author} {\bibfnamefont {A.}~\bibnamefont
  {van~de Walle}}\ and\ \bibinfo {author} {\bibfnamefont {M.}~\bibnamefont
  {Asta}},\ }\bibfield  {title} {\bibinfo {title} {Self-driven lattice-model
  monte carlo simulations of alloy thermodynamic properties and phase
  diagrams},\ }\href {https://doi.org/10.1088/0965-0393/10/5/304} {\bibfield
  {journal} {\bibinfo  {journal} {Modelling and Simulation in Materials Science
  and Engineering}\ }\textbf {\bibinfo {volume} {10}},\ \bibinfo {pages} {521}
  (\bibinfo {year} {2002})}\BibitemShut {NoStop}%
\bibitem [{\citenamefont {Sadigh}\ \emph {et~al.}(2012)\citenamefont {Sadigh},
  \citenamefont {Erhart}, \citenamefont {Stukowski}, \citenamefont {Caro},
  \citenamefont {Martinez},\ and\ \citenamefont {Zepeda-Ruiz}}]{SadErhStu12}%
  \BibitemOpen
  \bibfield  {author} {\bibinfo {author} {\bibfnamefont {B.}~\bibnamefont
  {Sadigh}}, \bibinfo {author} {\bibfnamefont {P.}~\bibnamefont {Erhart}},
  \bibinfo {author} {\bibfnamefont {A.}~\bibnamefont {Stukowski}}, \bibinfo
  {author} {\bibfnamefont {A.}~\bibnamefont {Caro}}, \bibinfo {author}
  {\bibfnamefont {E.}~\bibnamefont {Martinez}},\ and\ \bibinfo {author}
  {\bibfnamefont {L.}~\bibnamefont {Zepeda-Ruiz}},\ }\bibfield  {title}
  {\bibinfo {title} {Scalable parallel monte carlo algorithm for atomistic
  simulations of precipitation in alloys},\ }\href
  {https://doi.org/10.1103/PhysRevB.85.184203} {\bibfield  {journal} {\bibinfo
  {journal} {Physical Review B}\ }\textbf {\bibinfo {volume} {85}},\ \bibinfo
  {pages} {184203} (\bibinfo {year} {2012})}\BibitemShut {NoStop}%
\bibitem [{\citenamefont {Van~der Ven}\ and\ \citenamefont
  {Ceder}(2004)}]{VanCed04}%
  \BibitemOpen
  \bibfield  {author} {\bibinfo {author} {\bibfnamefont {A.}~\bibnamefont
  {Van~der Ven}}\ and\ \bibinfo {author} {\bibfnamefont {G.}~\bibnamefont
  {Ceder}},\ }\bibfield  {title} {\bibinfo {title} {Ordering in
  \ce{Li_x(Ni_{0.5}Mn_{0.5})O2} and its relation to charge capacity and
  electrochemical behavior in rechargeable lithium batteries},\ }\href
  {https://doi.org/10.1016/j.elecom.2004.07.018} {\bibfield  {journal}
  {\bibinfo  {journal} {Electrochemistry Communications}\ }\textbf {\bibinfo
  {volume} {6}},\ \bibinfo {pages} {1045–1050} (\bibinfo {year}
  {2004})}\BibitemShut {NoStop}%
\bibitem [{\citenamefont {Chen}\ \emph {et~al.}(2017)\citenamefont {Chen},
  \citenamefont {Sai~Gautam}, \citenamefont {Huang},\ and\ \citenamefont
  {Ceder}}]{CheSaiHua17}%
  \BibitemOpen
  \bibfield  {author} {\bibinfo {author} {\bibfnamefont {T.}~\bibnamefont
  {Chen}}, \bibinfo {author} {\bibfnamefont {G.}~\bibnamefont {Sai~Gautam}},
  \bibinfo {author} {\bibfnamefont {W.}~\bibnamefont {Huang}},\ and\ \bibinfo
  {author} {\bibfnamefont {G.}~\bibnamefont {Ceder}},\ }\bibfield  {title}
  {\bibinfo {title} {First-principles study of the voltage profile and mobility
  of {Mg} intercalation in a chromium oxide spinel},\ }\href
  {https://doi.org/10.1021/acs.chemmater.7b04038} {\bibfield  {journal}
  {\bibinfo  {journal} {Chemistry of Materials}\ }\textbf {\bibinfo {volume}
  {30}},\ \bibinfo {pages} {153–162} (\bibinfo {year} {2017})}\BibitemShut
  {NoStop}%
\bibitem [{\citenamefont {Yu}\ \emph {et~al.}(2014)\citenamefont {Yu},
  \citenamefont {Ling}, \citenamefont {Bhattacharya}, \citenamefont {Thomas},
  \citenamefont {Thornton},\ and\ \citenamefont {Van~der Ven}}]{YuLinTho14}%
  \BibitemOpen
  \bibfield  {author} {\bibinfo {author} {\bibfnamefont {H.-C.}\ \bibnamefont
  {Yu}}, \bibinfo {author} {\bibfnamefont {C.}~\bibnamefont {Ling}}, \bibinfo
  {author} {\bibfnamefont {J.}~\bibnamefont {Bhattacharya}}, \bibinfo {author}
  {\bibfnamefont {J.~C.}\ \bibnamefont {Thomas}}, \bibinfo {author}
  {\bibfnamefont {K.}~\bibnamefont {Thornton}},\ and\ \bibinfo {author}
  {\bibfnamefont {A.}~\bibnamefont {Van~der Ven}},\ }\bibfield  {title}
  {\bibinfo {title} {Designing the next generation high capacity battery
  electrodes},\ }\href {https://doi.org/10.1039/c3ee43154a} {\bibfield
  {journal} {\bibinfo  {journal} {Energy \& Environmental Science}\ }\textbf
  {\bibinfo {volume} {7}},\ \bibinfo {pages} {1760} (\bibinfo {year}
  {2014})}\BibitemShut {NoStop}%
\bibitem [{\citenamefont {Chang}\ \emph {et~al.}(2021)\citenamefont {Chang},
  \citenamefont {Jørgensen}, \citenamefont {Loftager}, \citenamefont
  {Bhowmik}, \citenamefont {Lastra},\ and\ \citenamefont
  {Vegge}}]{ChaJorLof21}%
  \BibitemOpen
  \bibfield  {author} {\bibinfo {author} {\bibfnamefont {J.~H.}\ \bibnamefont
  {Chang}}, \bibinfo {author} {\bibfnamefont {P.~B.}\ \bibnamefont
  {Jørgensen}}, \bibinfo {author} {\bibfnamefont {S.}~\bibnamefont
  {Loftager}}, \bibinfo {author} {\bibfnamefont {A.}~\bibnamefont {Bhowmik}},
  \bibinfo {author} {\bibfnamefont {J.~M.~G.}\ \bibnamefont {Lastra}},\ and\
  \bibinfo {author} {\bibfnamefont {T.}~\bibnamefont {Vegge}},\ }\bibfield
  {title} {\bibinfo {title} {On-the-fly assessment of diffusion barriers of
  disordered transition metal oxyfluorides using local descriptors},\ }\href
  {https://doi.org/10.1016/j.electacta.2021.138551} {\bibfield  {journal}
  {\bibinfo  {journal} {Electrochimica Acta}\ }\textbf {\bibinfo {volume}
  {388}},\ \bibinfo {pages} {138551} (\bibinfo {year} {2021})}\BibitemShut
  {NoStop}%
\bibitem [{\citenamefont {Yamamoto}\ \emph {et~al.}(2017)\citenamefont
  {Yamamoto}, \citenamefont {Iikubo}, \citenamefont {Yamasaki}, \citenamefont
  {Ogomi},\ and\ \citenamefont {Hayase}}]{YamIikYam17}%
  \BibitemOpen
  \bibfield  {author} {\bibinfo {author} {\bibfnamefont {K.}~\bibnamefont
  {Yamamoto}}, \bibinfo {author} {\bibfnamefont {S.}~\bibnamefont {Iikubo}},
  \bibinfo {author} {\bibfnamefont {J.}~\bibnamefont {Yamasaki}}, \bibinfo
  {author} {\bibfnamefont {Y.}~\bibnamefont {Ogomi}},\ and\ \bibinfo {author}
  {\bibfnamefont {S.}~\bibnamefont {Hayase}},\ }\bibfield  {title} {\bibinfo
  {title} {Structural stability of iodide perovskite: A combined cluster
  expansion method and first-principles study},\ }\href
  {https://doi.org/10.1021/acs.jpcc.7b07910} {\bibfield  {journal} {\bibinfo
  {journal} {The Journal of Physical Chemistry C}\ }\textbf {\bibinfo {volume}
  {121}},\ \bibinfo {pages} {27797–27804} (\bibinfo {year}
  {2017})}\BibitemShut {NoStop}%
\bibitem [{\citenamefont {Xu}\ and\ \citenamefont {Jiang}(2019)}]{XuJia19}%
  \BibitemOpen
  \bibfield  {author} {\bibinfo {author} {\bibfnamefont {X.}~\bibnamefont
  {Xu}}\ and\ \bibinfo {author} {\bibfnamefont {H.}~\bibnamefont {Jiang}},\
  }\bibfield  {title} {\bibinfo {title} {Cluster expansion based
  configurational averaging approach to bandgaps of semiconductor alloys},\
  }\href {https://doi.org/10.1063/1.5078399} {\bibfield  {journal} {\bibinfo
  {journal} {The Journal of Chemical Physics}\ }\textbf {\bibinfo {volume}
  {150}},\ \bibinfo {pages} {034102} (\bibinfo {year} {2019})}\BibitemShut
  {NoStop}%
\bibitem [{\citenamefont {Han}\ \emph {et~al.}(2022)\citenamefont {Han},
  \citenamefont {Yeu}, \citenamefont {Ye}, \citenamefont {Hwang},\ and\
  \citenamefont {Choi}}]{HanYeuYe22}%
  \BibitemOpen
  \bibfield  {author} {\bibinfo {author} {\bibfnamefont {G.}~\bibnamefont
  {Han}}, \bibinfo {author} {\bibfnamefont {I.~W.}\ \bibnamefont {Yeu}},
  \bibinfo {author} {\bibfnamefont {K.~H.}\ \bibnamefont {Ye}}, \bibinfo
  {author} {\bibfnamefont {C.~S.}\ \bibnamefont {Hwang}},\ and\ \bibinfo
  {author} {\bibfnamefont {J.-H.}\ \bibnamefont {Choi}},\ }\bibfield  {title}
  {\bibinfo {title} {Atomistic prediction on the composition- and
  configuration-dependent bandgap of {Ga(As,Sb)} using cluster expansion and ab
  initio thermodynamics},\ }\href {https://doi.org/10.1016/j.mseb.2022.115713}
  {\bibfield  {journal} {\bibinfo  {journal} {Materials Science and
  Engineering: B}\ }\textbf {\bibinfo {volume} {280}},\ \bibinfo {pages}
  {115713} (\bibinfo {year} {2022})}\BibitemShut {NoStop}%
\bibitem [{\citenamefont {Yu}\ and\ \citenamefont {Carter}(2016)}]{YuCar16}%
  \BibitemOpen
  \bibfield  {author} {\bibinfo {author} {\bibfnamefont {K.}~\bibnamefont
  {Yu}}\ and\ \bibinfo {author} {\bibfnamefont {E.~A.}\ \bibnamefont
  {Carter}},\ }\bibfield  {title} {\bibinfo {title} {Elucidating structural
  disorder and the effects of {Cu} vacancies on the electronic properties of
  \ce{Cu2ZnSnS4}},\ }\href {https://doi.org/10.1021/acs.chemmater.5b04351}
  {\bibfield  {journal} {\bibinfo  {journal} {Chemistry of Materials}\ }\textbf
  {\bibinfo {volume} {28}},\ \bibinfo {pages} {864–869} (\bibinfo {year}
  {2016})}\BibitemShut {NoStop}%
\bibitem [{\citenamefont {Eom}\ \emph {et~al.}(2018)\citenamefont {Eom},
  \citenamefont {Kim}, \citenamefont {Lim}, \citenamefont {Han}, \citenamefont
  {Han}, \citenamefont {Lee}, \citenamefont {Lebègue}, \citenamefont {Hwang},
  \citenamefont {Min},\ and\ \citenamefont {Kim}}]{EomKimLim18}%
  \BibitemOpen
  \bibfield  {author} {\bibinfo {author} {\bibfnamefont {T.}~\bibnamefont
  {Eom}}, \bibinfo {author} {\bibfnamefont {W.~J.}\ \bibnamefont {Kim}},
  \bibinfo {author} {\bibfnamefont {H.-K.}\ \bibnamefont {Lim}}, \bibinfo
  {author} {\bibfnamefont {M.~H.}\ \bibnamefont {Han}}, \bibinfo {author}
  {\bibfnamefont {K.~H.}\ \bibnamefont {Han}}, \bibinfo {author} {\bibfnamefont
  {E.-K.}\ \bibnamefont {Lee}}, \bibinfo {author} {\bibfnamefont
  {S.}~\bibnamefont {Lebègue}}, \bibinfo {author} {\bibfnamefont {Y.~J.}\
  \bibnamefont {Hwang}}, \bibinfo {author} {\bibfnamefont {B.~K.}\ \bibnamefont
  {Min}},\ and\ \bibinfo {author} {\bibfnamefont {H.}~\bibnamefont {Kim}},\
  }\bibfield  {title} {\bibinfo {title} {Cluster expansion method for
  simulating realistic size of nanoparticle catalysts with an application in
  \ce{CO2} electroreduction},\ }\href
  {https://doi.org/10.1021/acs.jpcc.8b02886} {\bibfield  {journal} {\bibinfo
  {journal} {The Journal of Physical Chemistry C}\ }\textbf {\bibinfo {volume}
  {122}},\ \bibinfo {pages} {9245–9254} (\bibinfo {year} {2018})}\BibitemShut
  {NoStop}%
\bibitem [{\citenamefont {Ai}\ \emph {et~al.}(2023{\natexlab{a}})\citenamefont
  {Ai}, \citenamefont {Chang}, \citenamefont {Tygesen}, \citenamefont {Vegge},\
  and\ \citenamefont {Hansen}}]{AiChaTyg23a}%
  \BibitemOpen
  \bibfield  {author} {\bibinfo {author} {\bibfnamefont {C.}~\bibnamefont
  {Ai}}, \bibinfo {author} {\bibfnamefont {J.~H.}\ \bibnamefont {Chang}},
  \bibinfo {author} {\bibfnamefont {A.~S.}\ \bibnamefont {Tygesen}}, \bibinfo
  {author} {\bibfnamefont {T.}~\bibnamefont {Vegge}},\ and\ \bibinfo {author}
  {\bibfnamefont {H.~A.}\ \bibnamefont {Hansen}},\ }\bibfield  {title}
  {\bibinfo {title} {Impact of hydrogen concentration for \ce{CO2} reduction on
  \ce{PdH_x}: A combination study of cluster expansion and kinetics analysis},\
  }\href {https://doi.org/10.1016/j.jcat.2023.115188} {\bibfield  {journal}
  {\bibinfo  {journal} {Journal of Catalysis}\ }\textbf {\bibinfo {volume}
  {428}},\ \bibinfo {pages} {115188} (\bibinfo {year}
  {2023}{\natexlab{a}})}\BibitemShut {NoStop}%
\bibitem [{\citenamefont {Yang}\ \emph {et~al.}(2020)\citenamefont {Yang},
  \citenamefont {Tan},\ and\ \citenamefont {Saidi}}]{YanTanSai20}%
  \BibitemOpen
  \bibfield  {author} {\bibinfo {author} {\bibfnamefont {T.~T.}\ \bibnamefont
  {Yang}}, \bibinfo {author} {\bibfnamefont {T.~L.}\ \bibnamefont {Tan}},\ and\
  \bibinfo {author} {\bibfnamefont {W.~A.}\ \bibnamefont {Saidi}},\ }\bibfield
  {title} {\bibinfo {title} {High activity toward the hydrogen evolution
  reaction on the edges of \ce{MoS2}-supported platinum nanoclusters using
  cluster expansion and electrochemical modeling},\ }\href
  {https://doi.org/10.1021/acs.chemmater.9b05244} {\bibfield  {journal}
  {\bibinfo  {journal} {Chemistry of Materials}\ }\textbf {\bibinfo {volume}
  {32}},\ \bibinfo {pages} {1315–1321} (\bibinfo {year} {2020})}\BibitemShut
  {NoStop}%
\bibitem [{\citenamefont {Ai}\ \emph {et~al.}(2023{\natexlab{b}})\citenamefont
  {Ai}, \citenamefont {Chang}, \citenamefont {Tygesen}, \citenamefont {Vegge},\
  and\ \citenamefont {Hansen}}]{AiChaTyg23b}%
  \BibitemOpen
  \bibfield  {author} {\bibinfo {author} {\bibfnamefont {C.}~\bibnamefont
  {Ai}}, \bibinfo {author} {\bibfnamefont {J.~H.}\ \bibnamefont {Chang}},
  \bibinfo {author} {\bibfnamefont {A.~S.}\ \bibnamefont {Tygesen}}, \bibinfo
  {author} {\bibfnamefont {T.}~\bibnamefont {Vegge}},\ and\ \bibinfo {author}
  {\bibfnamefont {H.~A.}\ \bibnamefont {Hansen}},\ }\bibfield  {title}
  {\bibinfo {title} {High‐throughput compositional screening of
  \ce{Pd_xTi_{1‐x}H_y} and \ce{Pd_xNb_{1‐x}H_y} hydrides for \ce{CO2}
  reduction},\ }\href {https://doi.org/10.1002/cssc.202301277} {\bibfield
  {journal} {\bibinfo  {journal} {ChemSusChem}\ }\textbf {\bibinfo {volume}
  {17}},\ \bibinfo {pages} {e202301277} (\bibinfo {year}
  {2023}{\natexlab{b}})}\BibitemShut {NoStop}%
\bibitem [{\citenamefont {Chan}\ \emph {et~al.}(2010)\citenamefont {Chan},
  \citenamefont {Reed}, \citenamefont {Donadio}, \citenamefont {Mueller},
  \citenamefont {Meng}, \citenamefont {Galli},\ and\ \citenamefont
  {Ceder}}]{ChaReeDon10}%
  \BibitemOpen
  \bibfield  {author} {\bibinfo {author} {\bibfnamefont {M.~K.~Y.}\
  \bibnamefont {Chan}}, \bibinfo {author} {\bibfnamefont {J.}~\bibnamefont
  {Reed}}, \bibinfo {author} {\bibfnamefont {D.}~\bibnamefont {Donadio}},
  \bibinfo {author} {\bibfnamefont {T.}~\bibnamefont {Mueller}}, \bibinfo
  {author} {\bibfnamefont {Y.~S.}\ \bibnamefont {Meng}}, \bibinfo {author}
  {\bibfnamefont {G.}~\bibnamefont {Galli}},\ and\ \bibinfo {author}
  {\bibfnamefont {G.}~\bibnamefont {Ceder}},\ }\bibfield  {title} {\bibinfo
  {title} {Cluster expansion and optimization of thermal conductivity in {SiGe}
  nanowires},\ }\href {https://doi.org/10.1103/physrevb.81.174303} {\bibfield
  {journal} {\bibinfo  {journal} {Physical Review B}\ }\textbf {\bibinfo
  {volume} {81}},\ \bibinfo {pages} {174303} (\bibinfo {year}
  {2010})}\BibitemShut {NoStop}%
\bibitem [{\citenamefont {Isaacs}\ and\ \citenamefont
  {Wolverton}(2019)}]{IsaWol19}%
  \BibitemOpen
  \bibfield  {author} {\bibinfo {author} {\bibfnamefont {E.~B.}\ \bibnamefont
  {Isaacs}}\ and\ \bibinfo {author} {\bibfnamefont {C.}~\bibnamefont
  {Wolverton}},\ }\bibfield  {title} {\bibinfo {title} {Electronic structure
  and phase stability of {Yb}-filled \ce{CoSb3} skutterudite thermoelectrics
  from first-principles},\ }\href@noop {} {\bibfield  {journal} {\bibinfo
  {journal} {Chemistry of Materials}\ }\textbf {\bibinfo {volume} {31}},\
  \bibinfo {pages} {6154} (\bibinfo {year} {2019})}\BibitemShut {NoStop}%
\bibitem [{\citenamefont {Maisel}\ \emph {et~al.}(2016)\citenamefont {Maisel},
  \citenamefont {H\"{o}fler},\ and\ \citenamefont {M\"{u}ller}}]{MaiHofMul16}%
  \BibitemOpen
  \bibfield  {author} {\bibinfo {author} {\bibfnamefont {S.~B.}\ \bibnamefont
  {Maisel}}, \bibinfo {author} {\bibfnamefont {M.}~\bibnamefont {H\"{o}fler}},\
  and\ \bibinfo {author} {\bibfnamefont {S.}~\bibnamefont {M\"{u}ller}},\
  }\bibfield  {title} {\bibinfo {title} {Configurationally exhaustive
  first-principles study of a quaternary superalloy with a vast configuration
  space},\ }\href {https://doi.org/10.1103/physrevb.94.014116} {\bibfield
  {journal} {\bibinfo  {journal} {Physical Review B}\ }\textbf {\bibinfo
  {volume} {94}},\ \bibinfo {pages} {014116} (\bibinfo {year}
  {2016})}\BibitemShut {NoStop}%
\bibitem [{\citenamefont {Sharma}\ \emph {et~al.}(2016)\citenamefont {Sharma},
  \citenamefont {Singh}, \citenamefont {Johnson}, \citenamefont {Liaw},\ and\
  \citenamefont {Balasubramanian}}]{ShaSinJoh16}%
  \BibitemOpen
  \bibfield  {author} {\bibinfo {author} {\bibfnamefont {A.}~\bibnamefont
  {Sharma}}, \bibinfo {author} {\bibfnamefont {P.}~\bibnamefont {Singh}},
  \bibinfo {author} {\bibfnamefont {D.~D.}\ \bibnamefont {Johnson}}, \bibinfo
  {author} {\bibfnamefont {P.~K.}\ \bibnamefont {Liaw}},\ and\ \bibinfo
  {author} {\bibfnamefont {G.}~\bibnamefont {Balasubramanian}},\ }\bibfield
  {title} {\bibinfo {title} {Atomistic clustering-ordering and high-strain
  deformation of an \ce{Al_{0.1}CrCoFeNi} high-entropy alloy},\ }\href
  {https://doi.org/10.1038/srep31028} {\bibfield  {journal} {\bibinfo
  {journal} {Scientific Reports}\ }\textbf {\bibinfo {volume} {6}},\ \bibinfo
  {pages} {31028} (\bibinfo {year} {2016})}\BibitemShut {NoStop}%
\bibitem [{\citenamefont {Alidoust}\ \emph {et~al.}(2020)\citenamefont
  {Alidoust}, \citenamefont {Kleiven},\ and\ \citenamefont
  {Akola}}]{AliKleAko20}%
  \BibitemOpen
  \bibfield  {author} {\bibinfo {author} {\bibfnamefont {M.}~\bibnamefont
  {Alidoust}}, \bibinfo {author} {\bibfnamefont {D.}~\bibnamefont {Kleiven}},\
  and\ \bibinfo {author} {\bibfnamefont {J.}~\bibnamefont {Akola}},\ }\bibfield
   {title} {\bibinfo {title} {Density functional simulations of pressurized
  \ce{Mg-Zn} and \ce{Al-Zn} alloys},\ }\href
  {https://doi.org/10.1103/physrevmaterials.4.045002} {\bibfield  {journal}
  {\bibinfo  {journal} {Physical Review Materials}\ }\textbf {\bibinfo {volume}
  {4}},\ \bibinfo {pages} {045002} (\bibinfo {year} {2020})}\BibitemShut
  {NoStop}%
\bibitem [{\citenamefont {Natarajan}\ and\ \citenamefont {Van~der
  Ven}(2017)}]{NatVan17}%
  \BibitemOpen
  \bibfield  {author} {\bibinfo {author} {\bibfnamefont {A.~R.}\ \bibnamefont
  {Natarajan}}\ and\ \bibinfo {author} {\bibfnamefont {A.}~\bibnamefont
  {Van~der Ven}},\ }\bibfield  {title} {\bibinfo {title} {First-principles
  investigation of phase stability in the \ce{Mg-Sc} binary alloy},\ }\href
  {https://doi.org/10.1103/physrevb.95.214107} {\bibfield  {journal} {\bibinfo
  {journal} {Physical Review B}\ }\textbf {\bibinfo {volume} {95}},\ \bibinfo
  {pages} {214107} (\bibinfo {year} {2017})}\BibitemShut {NoStop}%
\bibitem [{\citenamefont {Eriksson}\ \emph {et~al.}(2019)\citenamefont
  {Eriksson}, \citenamefont {Fransson},\ and\ \citenamefont
  {Erhart}}]{EriFraErh19}%
  \BibitemOpen
  \bibfield  {author} {\bibinfo {author} {\bibfnamefont {F.}~\bibnamefont
  {Eriksson}}, \bibinfo {author} {\bibfnamefont {E.}~\bibnamefont {Fransson}},\
  and\ \bibinfo {author} {\bibfnamefont {P.}~\bibnamefont {Erhart}},\
  }\bibfield  {title} {\bibinfo {title} {The hiphive package for the extraction
  of high-order force constants by machine learning},\ }\href
  {https://doi.org/10.1002/adts.201800184} {\bibfield  {journal} {\bibinfo
  {journal} {Advanced Theory and Simulations}\ }\textbf {\bibinfo {volume}
  {2}},\ \bibinfo {pages} {1800184} (\bibinfo {year} {2019})}\BibitemShut
  {NoStop}%
\bibitem [{\citenamefont {Huang}\ \emph {et~al.}(2017)\citenamefont {Huang},
  \citenamefont {Urban}, \citenamefont {Rong}, \citenamefont {Ding},
  \citenamefont {Luo},\ and\ \citenamefont {Ceder}}]{HuaUrbRon17}%
  \BibitemOpen
  \bibfield  {author} {\bibinfo {author} {\bibfnamefont {W.}~\bibnamefont
  {Huang}}, \bibinfo {author} {\bibfnamefont {A.}~\bibnamefont {Urban}},
  \bibinfo {author} {\bibfnamefont {Z.}~\bibnamefont {Rong}}, \bibinfo {author}
  {\bibfnamefont {Z.}~\bibnamefont {Ding}}, \bibinfo {author} {\bibfnamefont
  {C.}~\bibnamefont {Luo}},\ and\ \bibinfo {author} {\bibfnamefont
  {G.}~\bibnamefont {Ceder}},\ }\bibfield  {title} {\bibinfo {title}
  {Construction of ground-state preserving sparse lattice models for predictive
  materials simulations},\ }\bibfield  {journal} {\bibinfo  {journal} {npj
  Computational Materials}\ }\textbf {\bibinfo {volume} {3}},\ \href
  {https://doi.org/10.1038/s41524-017-0032-0} {10.1038/s41524-017-0032-0}
  (\bibinfo {year} {2017})\BibitemShut {NoStop}%
\bibitem [{\citenamefont {Zhong}\ \emph {et~al.}(2022)\citenamefont {Zhong},
  \citenamefont {Chen}, \citenamefont {{Barroso-Luque}}, \citenamefont {Xie},\
  and\ \citenamefont {Ceder}}]{ZhoCheBar22}%
  \BibitemOpen
  \bibfield  {author} {\bibinfo {author} {\bibfnamefont {P.}~\bibnamefont
  {Zhong}}, \bibinfo {author} {\bibfnamefont {T.}~\bibnamefont {Chen}},
  \bibinfo {author} {\bibfnamefont {L.}~\bibnamefont {{Barroso-Luque}}},
  \bibinfo {author} {\bibfnamefont {F.}~\bibnamefont {Xie}},\ and\ \bibinfo
  {author} {\bibfnamefont {G.}~\bibnamefont {Ceder}},\ }\bibfield  {title}
  {\bibinfo {title} {An $\ell_0\ell_2$-norm regularized regression model for
  construction of robust cluster expansions in multicomponent systems},\ }\href
  {https://doi.org/10.1103/PhysRevB.106.024203} {\bibfield  {journal} {\bibinfo
   {journal} {Physical Review B}\ }\textbf {\bibinfo {volume} {106}},\ \bibinfo
  {pages} {024203} (\bibinfo {year} {2022})}\BibitemShut {NoStop}%
\bibitem [{\citenamefont {Larsen}\ \emph {et~al.}(2018)\citenamefont {Larsen},
  \citenamefont {Jacobsen},\ and\ \citenamefont {Schiøtz}}]{LarJacSch18}%
  \BibitemOpen
  \bibfield  {author} {\bibinfo {author} {\bibfnamefont {P.~M.}\ \bibnamefont
  {Larsen}}, \bibinfo {author} {\bibfnamefont {K.~W.}\ \bibnamefont
  {Jacobsen}},\ and\ \bibinfo {author} {\bibfnamefont {J.}~\bibnamefont
  {Schiøtz}},\ }\bibfield  {title} {\bibinfo {title} {Rich ground-state
  chemical ordering in nanoparticles: Exact solution of a model for \ce{Ag-Au}
  clusters},\ }\href {https://doi.org/10.1103/physrevlett.120.256101}
  {\bibfield  {journal} {\bibinfo  {journal} {Physical Review Letters}\
  }\textbf {\bibinfo {volume} {120}},\ \bibinfo {pages} {256101} (\bibinfo
  {year} {2018})}\BibitemShut {NoStop}%
\bibitem [{\citenamefont {Huang}\ \emph {et~al.}(2016)\citenamefont {Huang},
  \citenamefont {Kitchaev}, \citenamefont {Dacek}, \citenamefont {Rong},
  \citenamefont {Urban}, \citenamefont {Cao}, \citenamefont {Luo},\ and\
  \citenamefont {Ceder}}]{HuaKit2016}%
  \BibitemOpen
  \bibfield  {author} {\bibinfo {author} {\bibfnamefont {W.}~\bibnamefont
  {Huang}}, \bibinfo {author} {\bibfnamefont {D.~A.}\ \bibnamefont {Kitchaev}},
  \bibinfo {author} {\bibfnamefont {S.~T.}\ \bibnamefont {Dacek}}, \bibinfo
  {author} {\bibfnamefont {Z.}~\bibnamefont {Rong}}, \bibinfo {author}
  {\bibfnamefont {A.}~\bibnamefont {Urban}}, \bibinfo {author} {\bibfnamefont
  {S.}~\bibnamefont {Cao}}, \bibinfo {author} {\bibfnamefont {C.}~\bibnamefont
  {Luo}},\ and\ \bibinfo {author} {\bibfnamefont {G.}~\bibnamefont {Ceder}},\
  }\bibfield  {title} {\bibinfo {title} {Finding and proving the exact ground
  state of a generalized ising model by convex optimization and max-sat},\
  }\href {https://doi.org/10.1103/PhysRevB.94.134424} {\bibfield  {journal}
  {\bibinfo  {journal} {Phys. Rev. B}\ }\textbf {\bibinfo {volume} {94}},\
  \bibinfo {pages} {134424} (\bibinfo {year} {2016})}\BibitemShut {NoStop}%
\bibitem [{\citenamefont {Kleiven}\ and\ \citenamefont
  {Akola}(2020)}]{KleAko20}%
  \BibitemOpen
  \bibfield  {author} {\bibinfo {author} {\bibfnamefont {D.}~\bibnamefont
  {Kleiven}}\ and\ \bibinfo {author} {\bibfnamefont {J.}~\bibnamefont
  {Akola}},\ }\bibfield  {title} {\bibinfo {title} {Precipitate formation in
  aluminium alloys: Multi-scale modelling approach},\ }\href
  {https://doi.org/10.1016/j.actamat.2020.05.050} {\bibfield  {journal}
  {\bibinfo  {journal} {Acta Materialia}\ }\textbf {\bibinfo {volume} {195}},\
  \bibinfo {pages} {123–131} (\bibinfo {year} {2020})}\BibitemShut {NoStop}%
\bibitem [{\citenamefont {Deng}\ \emph {et~al.}(2023)\citenamefont {Deng},
  \citenamefont {Mishra}, \citenamefont {Xie}, \citenamefont {Saeed},
  \citenamefont {Gautam},\ and\ \citenamefont {Canepa}}]{DenMisXie23}%
  \BibitemOpen
  \bibfield  {author} {\bibinfo {author} {\bibfnamefont {Z.}~\bibnamefont
  {Deng}}, \bibinfo {author} {\bibfnamefont {T.~P.}\ \bibnamefont {Mishra}},
  \bibinfo {author} {\bibfnamefont {W.}~\bibnamefont {Xie}}, \bibinfo {author}
  {\bibfnamefont {D.~A.}\ \bibnamefont {Saeed}}, \bibinfo {author}
  {\bibfnamefont {G.~S.}\ \bibnamefont {Gautam}},\ and\ \bibinfo {author}
  {\bibfnamefont {P.}~\bibnamefont {Canepa}},\ }\bibfield  {title} {\bibinfo
  {title} {kmcpy: A python package to simulate transport properties in solids
  with kinetic monte carlo},\ }\href
  {https://doi.org/10.1016/j.commatsci.2023.112394} {\bibfield  {journal}
  {\bibinfo  {journal} {Computational Materials Science}\ }\textbf {\bibinfo
  {volume} {229}},\ \bibinfo {pages} {112394} (\bibinfo {year}
  {2023})}\BibitemShut {NoStop}%
\bibitem [{\citenamefont {Bl\"ochl}(1994)}]{Blo94}%
  \BibitemOpen
  \bibfield  {author} {\bibinfo {author} {\bibfnamefont {P.~E.}\ \bibnamefont
  {Bl\"ochl}},\ }\bibfield  {title} {\bibinfo {title} {Projector augmented-wave
  method},\ }\href {https://doi.org/10.1103/PhysRevB.50.17953} {\bibfield
  {journal} {\bibinfo  {journal} {Physical Review B}\ }\textbf {\bibinfo
  {volume} {50}},\ \bibinfo {pages} {17953} (\bibinfo {year}
  {1994})}\BibitemShut {NoStop}%
\bibitem [{\citenamefont {Kresse}\ and\ \citenamefont
  {Joubert}(1999)}]{KreJou99}%
  \BibitemOpen
  \bibfield  {author} {\bibinfo {author} {\bibfnamefont {G.}~\bibnamefont
  {Kresse}}\ and\ \bibinfo {author} {\bibfnamefont {D.}~\bibnamefont
  {Joubert}},\ }\bibfield  {title} {\bibinfo {title} {From ultrasoft
  pseudopotentials to the projector augmented-wave method},\ }\href
  {https://doi.org/10.1103/PhysRevB.59.1758} {\bibfield  {journal} {\bibinfo
  {journal} {Physical Review B}\ }\textbf {\bibinfo {volume} {59}},\ \bibinfo
  {pages} {1758} (\bibinfo {year} {1999})}\BibitemShut {NoStop}%
\bibitem [{\citenamefont {Kresse}\ and\ \citenamefont
  {Hafner}(1993)}]{KreHaf93}%
  \BibitemOpen
  \bibfield  {author} {\bibinfo {author} {\bibfnamefont {G.}~\bibnamefont
  {Kresse}}\ and\ \bibinfo {author} {\bibfnamefont {J.}~\bibnamefont
  {Hafner}},\ }\bibfield  {title} {\bibinfo {title} {Ab initio molecular
  dynamics for liquid metals},\ }\href
  {https://doi.org/10.1103/PhysRevB.47.558} {\bibfield  {journal} {\bibinfo
  {journal} {Physical Review B}\ }\textbf {\bibinfo {volume} {47}},\ \bibinfo
  {pages} {558} (\bibinfo {year} {1993})}\BibitemShut {NoStop}%
\bibitem [{\citenamefont {Kresse}\ and\ \citenamefont
  {Furthm\"uller}(1996{\natexlab{a}})}]{KreFur1996-1}%
  \BibitemOpen
  \bibfield  {author} {\bibinfo {author} {\bibfnamefont {G.}~\bibnamefont
  {Kresse}}\ and\ \bibinfo {author} {\bibfnamefont {J.}~\bibnamefont
  {Furthm\"uller}},\ }\bibfield  {title} {\bibinfo {title} {Efficient iterative
  schemes for ab initio total-energy calculations using a plane-wave basis
  set},\ }\href {https://doi.org/10.1103/PhysRevB.54.11169} {\bibfield
  {journal} {\bibinfo  {journal} {Physical Review B}\ }\textbf {\bibinfo
  {volume} {54}},\ \bibinfo {pages} {11169} (\bibinfo {year}
  {1996}{\natexlab{a}})}\BibitemShut {NoStop}%
\bibitem [{\citenamefont {Kresse}\ and\ \citenamefont
  {Furthm\"uller}(1996{\natexlab{b}})}]{KreFur1996-2}%
  \BibitemOpen
  \bibfield  {author} {\bibinfo {author} {\bibfnamefont {G.}~\bibnamefont
  {Kresse}}\ and\ \bibinfo {author} {\bibfnamefont {J.}~\bibnamefont
  {Furthm\"uller}},\ }\bibfield  {title} {\bibinfo {title} {Efficiency of
  ab-initio total energy calculations for metals and semiconductors using a
  plane-wave basis set},\ }\href
  {https://doi.org/https://doi.org/10.1016/0927-0256(96)00008-0} {\bibfield
  {journal} {\bibinfo  {journal} {Computational Materials Science}\ }\textbf
  {\bibinfo {volume} {6}},\ \bibinfo {pages} {15} (\bibinfo {year}
  {1996}{\natexlab{b}})}\BibitemShut {NoStop}%
\bibitem [{\citenamefont {Dion}\ \emph {et~al.}(2004)\citenamefont {Dion},
  \citenamefont {Rydberg}, \citenamefont {Schr\"oder}, \citenamefont
  {Langreth},\ and\ \citenamefont {Lundqvist}}]{DioRydSch04}%
  \BibitemOpen
  \bibfield  {author} {\bibinfo {author} {\bibfnamefont {M.}~\bibnamefont
  {Dion}}, \bibinfo {author} {\bibfnamefont {H.}~\bibnamefont {Rydberg}},
  \bibinfo {author} {\bibfnamefont {E.}~\bibnamefont {Schr\"oder}}, \bibinfo
  {author} {\bibfnamefont {D.~C.}\ \bibnamefont {Langreth}},\ and\ \bibinfo
  {author} {\bibfnamefont {B.~I.}\ \bibnamefont {Lundqvist}},\ }\bibfield
  {title} {\bibinfo {title} {Van der {Waals} density functional for general
  geometries},\ }\href {https://doi.org/10.1103/PhysRevLett.92.246401}
  {\bibfield  {journal} {\bibinfo  {journal} {Physical Review Letters}\
  }\textbf {\bibinfo {volume} {92}},\ \bibinfo {pages} {246401} (\bibinfo
  {year} {2004})}\BibitemShut {NoStop}%
\bibitem [{\citenamefont {Berland}\ and\ \citenamefont
  {Hyldgaard}(2014)}]{BerHyl2014}%
  \BibitemOpen
  \bibfield  {author} {\bibinfo {author} {\bibfnamefont {K.}~\bibnamefont
  {Berland}}\ and\ \bibinfo {author} {\bibfnamefont {P.}~\bibnamefont
  {Hyldgaard}},\ }\bibfield  {title} {\bibinfo {title} {Exchange functional
  that tests the robustness of the plasmon description of the {van der Waals}
  density functional},\ }\href {https://doi.org/10.1103/PhysRevB.89.035412}
  {\bibfield  {journal} {\bibinfo  {journal} {Physical Review B}\ }\textbf
  {\bibinfo {volume} {89}},\ \bibinfo {pages} {035412} (\bibinfo {year}
  {2014})}\BibitemShut {NoStop}%
\bibitem [{\citenamefont {Cowley}(1950)}]{Cowley1950}%
  \BibitemOpen
  \bibfield  {author} {\bibinfo {author} {\bibfnamefont {J.~M.}\ \bibnamefont
  {Cowley}},\ }\bibfield  {title} {\bibinfo {title} {An approximate theory of
  order in alloys},\ }\href {https://doi.org/10.1103/PhysRev.77.669} {\bibfield
   {journal} {\bibinfo  {journal} {Physical Review}\ }\textbf {\bibinfo
  {volume} {77}},\ \bibinfo {pages} {669} (\bibinfo {year} {1950})}\BibitemShut
  {NoStop}%
\end{thebibliography}
\end{document}